%% file: arxiv-version.tex
\documentclass{vldb}
\usepackage{balance}

\usepackage{epstopdf}
\usepackage{graphicx}

\usepackage{enumitem}
\setlist{noitemsep} 

\usepackage[pdftex,dvipsnames]{xcolor}  

\newif\iftechreport
\techreporttrue

\usepackage{graphics}
\usepackage{epstopdf}
\usepackage{pstricks}
\usepackage{multirow}
\usepackage{tablefootnote}
\usepackage{ifthen}
\usepackage{hyperref}

\usepackage{amsmath,amssymb,amstext} 

\usepackage[noend]{algorithmic}
\usepackage{algorithm2e}
\SetKwProg{Fn}{Function}{}{}

\definecolor{annotationColor}{RGB}{0,0,139}

\usepackage{tikz}
\usetikzlibrary{positioning}
\usetikzlibrary{shapes.multipart}

\usepackage{caption}
\usepackage{subcaption}
\usepackage{hhline}
\usepackage{pifont}
\usepackage{array}
\newcolumntype{L}[1]{>{\raggedright\let\newline\\\arraybackslash\hspace{-3pt}}m{#1}}
\newcolumntype{C}[1]{>{\centering\let\newline\\\arraybackslash\hspace{0pt}}m{#1}}
\newcolumntype{R}[1]{>{\raggedleft\let\newline\\\arraybackslash\hspace{0pt}}m{#1}}

\usepackage{chngcntr}

\usepackage[font=small,skip=0pt]{caption}

\usepackage{pdflscape}

\usepackage{lipsum}                     
\usepackage{xargs}                      
\usepackage{graphicx} 
\usepackage[colorinlistoftodos,prependcaption,textsize=tiny]{todonotes}

\usepackage{pgfplots}
\usetikzlibrary{pgfplots.groupplots}
\pgfplotsset{compat=newest}
\usepackage{tikz}
\usetikzlibrary{patterns}
\usetikzlibrary{pgfplots.groupplots}

\usepackage{balance}

\begin{document}


\title{Experimental Analysis of Distributed Graph Systems\thanks{A shorter version of this paper has been accepted for publication in Volume 11 of \emph{Proc. VLDB Endowment}.}}

\numberofauthors{1} 
\author{
	\alignauthor
	Khaled Ammar, M. Tamer \"{O}zsu\\
	\affaddr{David R. Cheriton School of Computer Science}\\
	\affaddr{University of Waterloo, Waterloo, Ontario, Canada}\\
	\email{\{khaled.ammar, tamer.ozsu\}@uwaterloo.ca}
}

\maketitle

\begin{abstract}
	This paper evaluates eight parallel graph processing systems: Hadoop, HaLoop, Vertica, Giraph, GraphLab (PowerGraph), Blogel, Flink Gelly, and GraphX (SPARK) over four very large datasets (Twitter, World Road Network, UK 200705, and ClueWeb) using four workloads (PageRank, WCC, SSSP and K-hop). The main objective is to perform an independent scale-out study by experimentally analyzing the performance, usability, and scalability (using up to 128 machines) of these systems. In addition to performance results, we discuss our experiences in using these systems and suggest some system tuning heuristics that lead to better performance.
\end{abstract}

\input{introduction}
\input{systems}

\input{workloads}
\input{design}

\input{results}
\input{related}
\input{conclusions}

\balance

\section{Acknowledgments}
We  thank Semih Salihoglu and Khuzaima Daudjee for their feedback on the paper. We also acknowledge the contribution of our undergraduate research assistants: Heli Wang, Runsheng Guo, and Anselme Goetschmann. This research was supported in part by Natural Sciences and Engineering Research Council (NSERC) of Canada.

\bibliographystyle{abbrv}
\bibliography{publications,references,uw-ethesis}

\balancecolumns
\end{document}

%% file: introduction.tex
\section{Introduction}
\label{Sec-exp:intro}
In the last decade, a number of graph processing systems have been developed. These  are typically divided into graph analytics systems (e.g. Giraph) and graph database systems (e.g. Neo4j) based on the workloads they process. In this paper we focus on graph analytics systems. Many of these use parallel processing to scale-out to a high number of computing nodes to accommodate very large graphs and high computation costs. Single machine solutions have also been proposed, but our focus in this paper is on scale-out systems. Although each of the proposals are accompanied by a performance study, objective, independent, and comprehensive evaluation of the proposed systems is not widely available. This paper reports the results of our extensive and systematic performance evaluation  of eight graph analytics systems over four real datasets with different characteristics. The choice of these eight systems  is based on a classification discussed in Section~\ref{LR:ComputingParadigm} and include:
\begin{itemize}
	\item Vertex-centric:
	\begin{itemize}
		\item Synchronous: Giraph~\cite{Giraph}, GraphLab~\cite{powerGraph}, Blogel-Vertex (Blogel-V)~\cite{Blogel}
		\item Asynchronous: GraphLab~\cite{powerGraph}
	\end{itemize}
	\item Block-centric: Blogel-Block (Blogel-B)~\cite{Blogel}
	\item MapReduce: Hadoop~\cite{Hadoop}
	\item MapReduce extensions: HaLoop~\cite{Haloop}, Spark/GraphX~\cite{GraphX}
	\item Relational: Vertica~\cite{Vertica4Graph}
	\item Stream: Flink Gelly~\cite{Gelly}
\end{itemize}

The experiments generated more than 20 GB of log files that were used for analysis. The novel aspects of this study are the following:

\begin{itemize}

	\item We study a comprehensive set of systems that cover most computation models ($\S$~\ref{LR:ComputingParadigm}). Previous studies (e.g.,~\cite{Sakr-GraphExperiments, han2014experimental, experiments}) consider only vertex-centric systems.

	\item Compared to previous studies, we use a wider set of real datasets: web graphs (UK200705, ClueWeb), social networks (Twitter), and road networks (world road network). Although, web graphs and social networks share some common properties, such as power-law distribution~\cite{ChapterLawsGenerators} and shrinking diameter~\cite{DynamicGraphProperties}, road networks are different, for example with their very large diameters ($48$K in our dataset). 
			
	\item We suggest several system tuning heuristics and a number of enhancements to existing systems for improved performance and usability.

	\item This is the first study that considers the COST metric for parallelization ($\S$~\ref{Cost-experiments}).

	\item We develop a visualization tool\footnote{\url{
			https://tinyurl.com/ya5plcr3
			%https://github.com/khaledammar/System\_performance\_demo
			}} that processes different system log files, extracts interesting information, and displays several types of figures for comparisons. Independent of the performance results we report, this tool is itself useful for experimental evaluation.
	
\end{itemize}

The major findings of our study are the following:

\begin{itemize}
	
\item Blogel is the overall winner. The execution time of Blogel-B is shortest, but Blogel-V is faster when we consider the end-to-end processing including data loading and partitioning ($\S$~\ref{BlogelDiscussion}). 
	
	\item Existing graph processing systems are inefficient over graphs with large diameters, such as the road network ($\S$~\ref{Sec:SyncVSasyncPAGERANK},~\ref{GraphXdiscussion},~\ref{wcc}).
	
	\item GraphLab performance is sensitive to cluster size (Section~\ref{results:graphlab_partitioning}). 
	
	\item Giraph has a similar performance to GraphLab when both systems use random partitioning ($\S$~\ref{results:GraphlabGirap}).
		
	\item GraphX is not suitable for graph workloads or datasets that require large number of iterations ($\S$~\ref{GraphXdiscussion}).

	\item General data parallelization frameworks such as Hadoop and Spark have additional computation overhead  ($\S$~\ref{Overhead-discussion}) that carry over to their graph systems (HaLoop, GraphX).  However, they could be useful when processing large graphs over shared resources or when resources are limited ($\S$~\ref{HaLoop-results}).
		
	\item Vertica is significantly slower than native graph processing systems. Although, its memory footprint is small, its I/O wait time and network communication is significantly high ($\S$~\ref{sec-vertica-results}).

\end{itemize}

It can be claimed that some of the performance differences could be due to the choice of the implementation language (Java or C++). It is common knowledge that C++ has better overall performance than Java for multiple reasons. Although we are not aware of a system that has both C++ and Java implementations to conduct a more controlled experiment,  the fact that GraphLab and Giraph have similar performance when they use the same partitioning algorithm (random) suggests that implementation language may not be a main factor. Nevertheless, this point requires further study.

We introduce the systems under investigation in Section~\ref{Sec-exp:sys} and the workloads in Section~\ref{Sec-exp:workload}. Section~\ref{Sec-exp:design} explains the experimental setup while Section~\ref{Sec-exp:results} presents our quantitative results. Section~\ref{relatedStudies} compares our work with related works, and we conclude in Section~\ref{Sec-exp:conclusion}.

%% file: systems.tex
\section{Systems}
\label{LR:ComputingParadigm}
\label{Sec-exp:sys}

We evaluate seven graph processing systems in this study. 
All systems, except Vertica, read datasets from and write results to a distributed file system such as HDFS (Hadoop Distributed File System). Vertica is a relational database system and it uses its own distributed storage. It is included in this study because of a recent claim that it performs  comparable to native graph systems~\cite{Vertica4Graph}. Hadoop, HaLoop, and Giraph are developed in Java and utilize the Hadoop MapReduce framework to execute all workloads. Flink Gelly has Scala and Java APIs, both use existing libraries to read and write data from HDFS. Blogel and GraphLab are developed in C++; they use libraries to read and write from HDFS. Finally, GraphX is developed using Scala and run on top of Apache Spark. We comment on the differences in programming languages in Section \ref{Sec-exp:conclusion}.

We categorize parallel graph systems based on their computational model, which explains our choice of the systems under study. A summary of the features of these systems is given in Table \ref{graphSystemsTable}. We also describe the special configurations used in this study. 

\begin{table*}[ht]
	\center
	\tiny
	\begin{tabular}{|C{2cm}|| C{1cm}| C{1.3cm} | C{1.5cm} | C{1.1cm} |C{1.1cm}|C{1.6cm}|C{1.35cm}|}
		\hline \textbf{System} & Memory /Disk & Architecture & Computing paradigm & Declarative Language & Partitioning& Synchronization& Fault Tolerance \\ 
		\hline
		
		\hline Hadoop~\cite{Dean2004, Hadoop} &  Disk & Parallel & BSP & \ding{53}& Random & Synchronous & re-execution \\ 
		
		\hline HaLoop~\cite{Haloop} & Disk & Parallel & BSP-extension & \ding{53}  & Random & Synchronous & re-execution \\

		\hline Pregel/Giraph/GPS\newline\cite{pregel, Giraph, GPS} & Memory & Parallel &  Vertex-Centric & \ding{53}& Random & Synchronous & global checkpoint \\ 
		
		\hline GraphLab~\cite{GraphLab} & Memory & Parallel & Vertex-Centric & \ding{53}& Random\newline Vertex-cut & (A)synchronous& global checkpoint\\
		
		\hline Spark/GraphX~\cite{Spark, GraphX} & Memory/Disk & Parallel & BSP-extension & \ding{53}& Random\newline Vertex-cut & Synchronous& global checkpoint\\
		
		\hline Giraph++~\cite{Giraph++} & Memory & Parallel & Block-Centric & \ding{53}& METIS & (A)synchronous & global checkpoint\\

		\hline Blogel~\cite{Blogel} & Memory & Parallel & Block-Centric & \ding{53}& Voronoi\newline 2D& Synchronous & global checkpoint\\
		
		\hline Vertica~\cite{Vertica4Graph} & Disk & Parallel & Relational & \ding{51}\newline(SQL)& Random& Synchronous & N/A\\
		
		\hline 
	\end{tabular}
	\caption{Graph processing systems} 
	\label{graphSystemsTable}
	\vspace{-5mm}
\end{table*}

\subsection{Vertex-Centric BSP}
\label{LR:VertexCentric}
Vertex-centric systems are also known as ``think-as-a-vertex''. Each vertex computes its new state based on its current state and the messages it receives from its neighbors. Each vertex then sends its new state to its neighbors using message passing. 
Synchronous versions follow the Bulk Synchronous Parallel (BSP) model that performs parallel computations in iterative steps, and synchronizes among machines at the end of each step. This means that messages sent in one iteration are not accessible by recipients in the same iteration; a recipient vertex receives its messages in the subsequent iteration. The computation stops when all vertices converge to a fixpoint or after a predefined number of iterations. This has been the most popular approach, and we study three systems in this category: Giraph, GraphLab, and Blogel-V.

\subsubsection{Giraph}
Giraph~\cite{Giraph} is the open source implementation of Pregel~\cite{pregel}, the prototypical vertex-centric BSP system. Giraph is implemented as a map-only application on Hadoop. It requires all data to be loaded in memory before starting the execution. Graph data is partitioned randomly using edge-cut approach, and each vertex is assigned to a partition.

Giraph API has one function, called \texttt{compute}. At every iteration, the \texttt{compute} function may update the vertex state based on its own data or based on its neighbors' data. The \texttt{compute} function may also send messages to other vertices. 

In our experiments, we use four mappers in each machine, and allow Hadoop to utilize 30GB memory in each machine. 

\subsubsection{GraphLab~/~PowerGraph}
\label{introduceGraphLab}
GraphLab~\cite{powerGraph} is a distributed graph processing system that is written in C++ and uses MPI for communication. Similar to Giraph, it keeps the graph in memory.  However, it does not depend on Hadoop and it introduces several modifications to the standard BSP model:

\begin{itemize}	
	\item Instead of using one \texttt{compute} function, it has three functions: \texttt{Gather}, \texttt{Apply}, and \texttt{Scatter} (GAS). The GAS model allows each vertex to \textit{gather} data from its neighbors, \textit{apply} the compute function on itself, and then \textit{scatter} relevant information to some neighbors if necessary. 
	
	\item It uses vertex-cut (i.e., edge-disjoint) partitioning instead of edge-cut. This replicates vertices and helps better distribute the work of vertices with very large degrees. These vertices exist in social network and web graphs, because they follow the power-law distribution~\cite{DynamicGraphsProperties}. Replication factor of a vertex refers to the number of machines on which that vertex is replicated. 
\end{itemize} 

GraphLab automatically uses all available cores and memory in the machine. It has multiple partitioning approaches that we study in further detail at Section~\ref{graphLabPartitioning}.

\subsubsection{Blogel-V}

Blogel~\cite{Blogel} adopts both vertex-centric and block-centric models (discussed in the next section). Blogel is implemented in C++ and uses MPI for communication between nodes. Blogel-V follows the standard BSP model. Its API has a \texttt{compute} function similar to Giraph.

\subsection{Vertex-Centric Asynchronous}

GraphLab~\cite{powerGraph} has an asynchronous mode where vertices can have access to the most recent data at other vertices within the same iteration. This avoids the overhead of waiting for all vertices to finish an iteration before starting a new one.  Synchronization is achieved by distributed locking. Both versions of GraphLab use the same configurations.

\subsection{Block-Centric BSP}
\label{LR:BlockCentric}
This category is also known as graph-centric. The main idea is to partition the graph into blocks of vertices, and run a serial algorithm within a block while synchronizing blocks on separate machines using BSP. The objective is to reduce the number of iterations, which leads to reducing the synchronization overhead. The number of blocks is expected to be significantly less than the number of vertices in a large graph, hence the performance gain from decreasing network communication. There are two prominent block-centric systems: Giraph++~\cite{Giraph++} and Blogel~\cite{Blogel}. Our study investigates Blogel, because Giraph++ is built on an earlier version of Giraph that does not implement the more recent optimizations proposed for Giraph~\cite{FB_trillionGraph, GPS}.

Blogel-B~\cite{Blogel} has a \texttt{compute} function for blocks, which typically includes a serial graph algorithm that runs within the block. Blogel-B partitions the dataset into multiple connected components using a partitioning algorithm based on Graph Voronoi Diagram (GVD)~\cite{GVD} partitioning. Additional partitioning techniques based on vertex properties in real graphs, such as 2-D coordinates (for road-network) or URL prefix (for web graph) have also been discussed, but we do not use these dataset-specific techniques in this study. We use the default parameters for Blogel-B's GVD partitioning~\cite{Blogel}.

\subsection{MapReduce}	
MapReduce~\cite{Dean2004} is a distributed BSP data processing framework whose goal is to simplify parallel processing by offering two simple interfaces: \texttt{map} and \texttt{reduce}. 
It achieves data-parallel computation by partitioning the data randomly to machines and executing the map and reduce functions on these partitions in parallel. Hadoop~\cite{Hadoop} is the most common open source implementation of MapReduce. It has been recognized that Hadoop is not suitable for graph algorithms that are iterative, due to excessive I/O with HDFS and data shuffling at every iteration~\cite{Haloop, Pegasus, GraphLab, pregel, Spark}. We nevertheless include Hadoop in this study, because there are cases where memory requirements will not allow other systems to run, and Hadoop is the only feasible alternative.  Hadoop is configured to use four mappers and two reducers in each machine. It is also granted $30$GB on each machine.  

\subsection{MapReduce Optimized} 
Modified MapReduce systems, such as HaLoop~\cite{Haloop} and  Spark~\cite{Spark}, address the shortcomings of MapReduce systems (in iterative workloads as graph processing) by caching reusable data between \texttt{map} and \texttt{reduce} steps and between iterations to avoid unnecessary scans of invariant data, and unnecessary data shuffling between machines.  

\subsubsection{HaLoop}
The main objective of HaLoop optimizations is to reduce data shuffling and reduce network usage after the first iteration. HaLoop proposes several modifications to enhance Hadoop's performance on iterative workloads:
\begin{itemize}
	\item A new programming model suitable for iterative programs, e.g., enabling loop control on the master node.
	
	\item Task scheduler in the master node is changed to be loop-aware. It keeps information about the location of sharded data, and tries to co-schedule tasks with data. This helps to decrease network communication.
	
	\item Slave nodes include a module for caching and indexing loop-invariant data that could be used in all iterations. The task tracker is modified to manage these modules.
	
	\item New support is introduced for fixpoint evaluation to optimize checking for convergence. The result of the last iteration is always locally cached to be used in the comparison instead of retrieving it again from HDFS. 
\end{itemize}

HaLoop configuration is very similar to Hadoop's: four mappers, two reducers, and $30$GB memory. However, in our environment HaLoop suffered from multiple errors because it keeps many files open. Therefore, we had to change the operating system's \texttt{nofile} limits.

\subsubsection{Spark/GraphX}
\label{sec:GraphX}
Similar to HaLoop, Spark caches dataset partitions for future use, but in memory instead of on local disk. The main feature of Spark is its fault tolerant in-memory abstraction, called resilient distributed datasets (RDD). GraphX~\cite{GraphX} is a graph library that extends Spark abstractions to implement graph operations. It uses vertex-cut partitioning (similar to GraphLab). Every iteration consists of multiple Spark jobs. A developer can decide what data portions should be cached for future use. However, cached data cannot change because they are used as RDDs, which are immutable. 

We run GraphX using the Spark standalone mode to eliminate any overhead or performance gain from Yarn, Mesos, or any other systems that facilitate resource sharing. GraphX has many configuration parameters. We configured its workers and reducers so that they can use all available memory in each machine. By default, Spark uses all available cores.

\subsection{Relational}
\label{LR:relationalDatabase}
\label{sec-vertica}
These systems~\cite{againstGraphProcessing, Vertica4Graph} use a relational database as a back-end storage and query engine for graph computations. 
A graph can be represented as an edge and a vertex table. 
Transferring information to neighbors is equivalent to joining these tables, and then updating the answer column in the vertex table. Each graph workload can be translated to a SQL query and executed on the tables.

Repeated joins over large vertex and edge tables is inefficient, and  several optimizations have been proposed for Vertica~\cite{Vertica4Graph}:

\begin{itemize}
	\item Instead of updating multiple values in the vertex table (which also means random access to data on disk), it may be more efficient to create a new table instead, and replace the old table with the new one (sequential disk access) if the number of updates is large. If the number  is very small, updating the table might be more efficient, but, it is not straightforward to estimate the number of updates beforehand. 
	
	\item In traversal workloads, such as Single Source Shortest Path (SSSP), it is common to only process a few vertices at every iteration. Instead of starting from the complete vertex table and filter these vertices, it is more efficient to keep active vertices in a temporary table and use it during the join operation. 

\end{itemize}

 Several changes were made to the cluster to satisfy all Vertica OS-level requirements. Before we start our experiments, instead of loading the data to HDFS, we load the data as a table of edges to Vertica.

\subsection{Stream Systems}
\label{Gelly}

There are a few systems in the literature, such as Timely and Differential Dataflow~\cite{Timely,DifferentialDataFlow}, Naiad~\cite{naiad}, Flink~\cite{Flink} and TensorFlow~\cite{TensorFlow} that model computations  as a series of operations. In these systems, a developer needs to define operators, and then connect them to describe  the data flow among operators. Data are streamed and processed through these operators. These are general data processing systems that sometimes support iteration in their data flow, therefore they can process graph algorithms. 

In our study, we consider Flink Gelly~\cite{Gelly} as a representative for this category. Gelly is the graph processing API built on top of Flink. It has two approaches: stream and batch. The stream reads data from an input stream and it pushes the received edges (or vertices) to the data flow as they arrive. The batch approach reads data from containers then process the whole dataset in the data flow operations described by the application developer. To be consistent with other systems, we use the batch approach in our experiments, which allow us to isolate the time required to read and prepare the graph from execution time.

%% file: workloads.tex
\section{Workloads}
\label{Sec-exp:workload}
In this study we consider four graph workloads: \texttt{PageRank}, \texttt{WCC} (weakly connected component), \texttt{SSSP} (single source shortest path) and \texttt{K-hop}. These are chosen because: (1) they are prominent in graph system studies, (2) they have different characteristics -- some (e.g, PageRank and WCC) are analytic workloads that involve iterative computation over all the vertices in the graph while others (e.g., SSSP and K-hop) are known as online workloads that operate on certain parts of the graph, and (3) there are implementations of each of them over the evaluated systems. Although every system offers its own implementation of these workloads, we made small changes to ensure uniformity of the algorithm and implementation across the systems.

\subsection{PageRank}
\label{pagerank}
PageRank has been the most popular workload for evaluating graph systems for iterative algorithms. In a nutshell, PageRank assigns an importance weight (rank) to each vertex based on its influence in the graph. This is achieved by an initial assignment of rank followed by iterative computation until a fixpoint of weights is found.

The iterative algorithm models a random walk agent that moves through the graph, such that when it is at vertex $u$, it may choose an outgoing edge from $u$ with probability $1-\delta$ or  jump to a random vertex with probability $\delta$. Therefore, a vertex PageRank value $pr(v)$, in a graph $G(V,E)$, follows the following equation:
\[
pr(v) = \delta + (1-\delta) \times \Sigma{\frac{pr(u)}{outDegree(u)}}~|~(u,v) \in E
\]
where $outDegree(u)$ is the number of directed edges from vertex $u$ to other vertices. Many implementations assume that $\delta = 0.15$ and start with an  initial rank of $1$ for each vertex.  Iteratively vertex ranks are computed using this formula until the rank of each vertex converges to a value.

The standard PageRank implementation follows syn\-chro\-nous computation, such that all vertices are involved in computation until convergence. In our experiments, convergence means the maximum change in any vertex rank is less than the initial value. This definition is more suitable than stopping after a fixed number of iterations, because it takes into consideration the properties of each dataset. Asynchronous or synchronous implementations that allow converged vertices to opt-out early from computation result in approximate answers. We will discuss this error and accuracy in the detailed experiments in Section~\ref{Sec:SyncVSasyncPAGERANK}.

\subsubsection{Self-edges issue in GraphLab}
GraphLab could not compute the correct PageRank values, because it does not support self-edges, which exists in the real graphs we use in this study. Changing the system to allow self-edges using flags or other potential implementations is possible, but is outside the scope of this study, as it requires significant changes to GraphLab code. This issue was communicated to GraphLab developers. 

\subsubsection{Block-centric implementation}
The block-centric implementation of PageRank in Blogel-B~\cite{Blogel} also did not generate accurate results. The proposed  algorithm has two steps: (1) Compute the initial PageRank value using block-computation and local PageRank; (2) Compute PageRank using vertex-computation. The first step constructs a graph of blocks, such that the weight of an edge between two blocks represents the number of graph edges between these two blocks. In the first iteration, each block runs local PageRank over local edges in the block, then it runs a vertex centric PageRank on the graph of blocks. This step continues until convergence.

The second step starts by initializing the PageRank of every vertex in a block $b$ as ($pr(v) \times pr(b)$) (where $pr(v)$ is the initial vertex pagerank and $pr(b)$ is the block pagerank after the first step converged)\footnote{There are other possible initialization functions that may include block and vertex pagerank values or degrees.}. The second step then runs until convergence. We also considered a version of this algorithm where each step runs for the number of iterations computed earlier to guarantee conversion. However, it is not clear how many iterations each step would need to guarantee the same results as the other computation models. Therefore, in our experiments we follow the version of block-computation PageRank proposed in the original Blogel paper~\cite{Blogel}.

\subsection{WCC}

The objective of a weakly connected component (WCC) algorithm is to find all subgraphs that are internally reachable regardless of the edge direction. HashMin~\cite{Pegasus} is the straightforward distributed implementation of WCC. It  labels each vertex in the input graph by the minimum vertex id reachable from the vertex, regardless of the edge directions. It starts out by considering each vertex to be in one component (i.e., each vertex id is its component id). Each vertex propagates its component id to its neighbors. The process terminates when a fixpoint is reached, i.e., no vertex changes its component id. This algorithm requires $O(d)$ iterations, where $d$ is the graph diameter.

HashMin algorithm has been implemented in all of the systems under consideration. However, we found that the result generated by some of these implementations are not correct, because of failure to process both directions of an edge. We corrected Blogel~\cite{Blogel} and Giraph~\cite{han2014experimental} implementations by adding an extra task to the first iteration: creating reverse edges, when necessary. Since GraphLab~\cite{GraphLab} allows vertices to access both ends of an edge regardless to the edge direction, it does not suffer from this overhead. However, as we will show later, memory requirements of GraphLab is typically larger than Giraph and Blogel for this very reason. 

\subsection{SSSP and K-hop}

The Single Source Shortest Path (\texttt{SSSP}) query is a graph traversal workload. It finds the shortest path from a given source vertex to every other vertex in the graph. Assuming the source node is $u$, a typical algorithm starts by initializing distance $dist(u,v)=\infty$ for any vertex $v \neq u$. 
Iteratively, using a breadth first search, the algorithm explores new vertices: at iteration $i$  new vertices that are $i$ hops away from the source vertex are considered. The number of iterations is $O(d)$. The algorithm stops when all reachable vertices are visited and $dist(u,v)$ is computed for all $v$.  

The \texttt{K-hop} query is very similar to \texttt{SSSP}, but it is bounded by $K$ hops. This query is relevant in evaluating graph systems because it is a traversal query, but its complexity (\#iteration) does not depend on the graph diameter. We fix $K$ to a small number, $3$, to reduce the impact of graph diameter on the performance, and to represent multiple use cases, such as the friends-of-friends query and its potential indexes.

In the results reported in this paper, to be consistent with other studies in the literature, we only use a random start vertex, which is chosen for each graph dataset, and used consistently in all experiments. 

%% file: design.tex
\section{Experiment Design}
\label{Sec-exp:design}

The experimental setting of this study is summarized in Table \ref{ExperimentDimensions}. 

\begin{table}[ht]
	\centering
	\small
	\begin{tabular}{l||m{6cm}}
		\hline
		\textbf{Dimension} & \textbf{potential values} \\
		\hline
		Systems   & Giraph, Blogel, Hadoop, HaLoop, GraphX, GraphLab, Vertica, Flink Gelly \\ 
		Workloads & WCC, PageRank, SSSP, K$-$hop \\ 
		Datasets  & Twitter, UK, ClueWeb,  WRN \\ 
		Cluster Size  & 16, 32, 64, 128 \\ 
		Instance type & r3.xlarge
	\end{tabular} 
	\caption{A summary of experiments dimensions}
	\label{ExperimentDimensions}
\end{table}

\subsection{Infrastructure}

All experiments are run on Amazon EC2 AWS r3.xlarge machines, each of which has $4$ cores and $30.5$ GB memory, Intel Xeon E5-2670 v2 (Ivy Bridge) processors, and SSD disks. They are optimized for memory-intensive applications and recommended for in-memory analytics. We test scalability over $16$, $32$, $64$, and $128$ machines (one master).

\subsection{Evaluated Metrics}
\label{metrics}
We measure two things: resource utilization and system performance. Each system is evaluated in isolation, with no resource sharing across systems or experiments. For resource utilization, we record CPU utilization for each process type (user, system, I/O, idle) and memory usage every second. We also record the total network traffic by measuring network card usage before and after workload execution. 
We report the following system performance metrics: (a) data-loading time, (b) result-saving time, (c) execution time, and (d) total response time (latency). 

Data-loading time includes reading data from HDFS and graph partitioning, when necessary. Ideally, total response time should equal load+execute+save. However, we report it separately, because it represents the end-to-end processing time, and occasionally includes some overhead that might not be explicitly reported by some systems, such as the time of repartitioning, networking, and synchronization.

\input{datasets}

\subsection{Configuration}

We report below the  experiments we performed to better understand and fix configurations of some of the systems. 

\subsubsection{Partitioning in GraphLab}
\label{graphLabPartitioning}

GraphLab has two main partitioning options: ``Random'' and ``Auto''. Random assigns edges to partitions using a hash function. Auto chooses between three partitioning algorithms (PDS, Grid and Oblivious), in order, based on the number of machines in the cluster. 

Typically, these partitioning methods try to decrease the replication factor (see Section \ref{introduceGraphLab}) by minimizing the number of machines at which each vertex is stored. This would decrease the network communication between machines. The details of these algorithms are as follows:
\begin{itemize}
	\item Grid assumes the cluster is a rectangle of height and width equal to $X$ and $Y$ and requires the number of machines $M=X \times Y$, such that $|X-Y| \leq 2$ for any positive numbers $X$ and $Y$. Using a hashing function to map each vertex to a machine $m$, the vertex could be replicated to any machine in the same column or same row that include $m$. An edge between two vertices can be assigned to any partition that can include a replica for both vertices. 
	
	\item PDS creates a perfect difference set~\cite{PartitioningStudy} $S$ of size $p+1$ if the number of machines $M=p^2+p+1$, where $p$ is a positive prime number. Then, for each value $i$ in $S$, it creates another set $S_i$ by adding this value to  $[0,M]$ mod $M$. Finally, using a hash function to map a vertex to a machine $m$, the vertex could be replicated to any machine in $S_i$ such that $m\in S_i$. Again, an edge could be placed in a machine that can include both of its vertices.
	
	\item Oblivious is a greedy heuristics for edge placement to reduce the partitioning factor. Given an edge $(u,v)$, such that $S_u$ is the set of machines that include replicas of $u$ and $S_v$ is the set of machines that include replicas of $v$, the edge will be placed in the least loaded machine in $S_e$, such that:
	\begin{itemize}
		\item if $S_u \cap S_v \neq \phi$ then $S_e = S_u \cap S_v$;
		\item if $S_u = \phi$ and $S_v \neq \phi$ then $S_e = S_u$;
		\item if $S_u = S_v = \phi$ then $S_e$ is the set of all machines;
		\item if $S_u \cap S_v = \phi$ then $S_e = S_u \cup S_v$.
	\end{itemize} 
\end{itemize}

Occasionally these algorithms do not reduce the replication factor when compared with random. For example, the difference between replication factor (Table~\ref{replicationFactor}) in random and auto for the Twitter dataset is not significant (less than $2\times$). Twitter dataset has some differences when compared with the the UK0705 dataset. For example, the maximum out-degree in the Twitter dataset is $3\times$ the maximum out-degree in the UK0705 dataset, despite the fact that Twitter dataset is $3\times$ smaller than UK0705. Unlike UK9795, the Twitter dataset has only one large component. Auto partitioning could not help GraphLab to enhance the efficiency of Twitter graph processing. At the same time, the auto replication factor for the UK0705 dataset is $5\times$ less than random replication factor in the $32$ cluster, though.

\begin{table}[b]
\tiny
	\centering
	\begin{tabular}{|c|c|c|c|}
		\hline \textbf{Dataset} & \textbf{Cluster Size} & \textbf{Random} & \textbf{Auto} \\ \hline
		\hline \multirow{4}{*}{Twitter} & 16 & 9.3 & 5.5 \\ 
		\cline{2-4} & 32  & 13.3 & 9.8 \\ 
		\cline{2-4} & 64  & 17.8 & 9.1 \\ 
		\cline{2-4} & 128  & 22.5 & 15.2 \\ 
		\hline \hline \multirow{4}{*}{WRN} & 16 & NA & NA \\ 
		\cline{2-4} & 32  & 3.0 & 2.2 \\ 
		\cline{2-4} & 64  & 3.0 & 3.0 \\ 
		\cline{2-4} & 128  & 3.0 & 2.3 \\ 
		\hline \hline \multirow{4}{*}{UK0705} & 16 & 5.7 & NA \\ 
		\cline{2-4} & 32  & 15.8 & 3.6 \\ 
		\cline{2-4} & 64  & 21.5 & 10.1 \\ 
		\cline{2-4} & 128  & 27.1 & 4.5 \\ 
		\hline 
	\end{tabular} 
	\caption{The replication factor in GraphLab.}
	\label{replicationFactor}
\end{table}

\subsubsection{CPU utilization in GraphLab}

\begin{figure}
	\centering
	\includegraphics[width=0.6\linewidth]{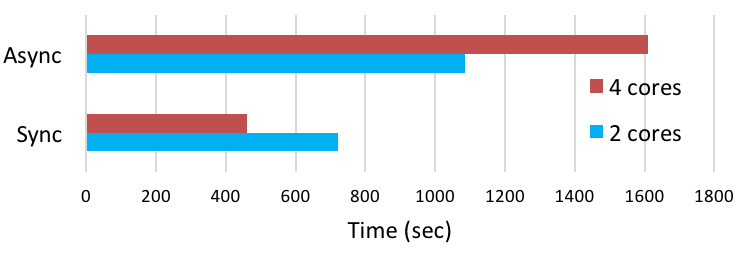}
	\caption{In a $16$-machines cluster, GraphLab synchronized mode benefits from using all 4 cores in computation, while asynchronous computation does not because vertices do computation and communication on the same time.}
	\label{GraphLab_number_cores} 
\end{figure}

GraphLab, by default, uses all the cores in every machine. It reserves two cores for networking and  overhead operations and uses the remaining cores for computations. Our experiments use GraphLab's default configuration. Nonetheless, we studied the value of this default configuration by changing the GraphLab code to use all available cores for computation (Figure~\ref{GraphLab_number_cores}). When we used all cores for computation, we obtained 40\% improvement (with the synchronous computation) on a 16-machine cluster over 30 iterations of PageRank computation using the Twitter dataset. On the other hand, asynchronous computation requires multiple communications while some vertices are still in the computation phase. Due to the expensive repetitive context switching, asynchronous does not benefit, and sometimes even under-performs, when all cores are used for computation.

\subsubsection{Number of Partitions in GraphX}

\begin{figure*}[ht]
	\centering
	\begin{subfigure}{.45\textwidth}
		\centering
		\includegraphics[width=0.8\linewidth]{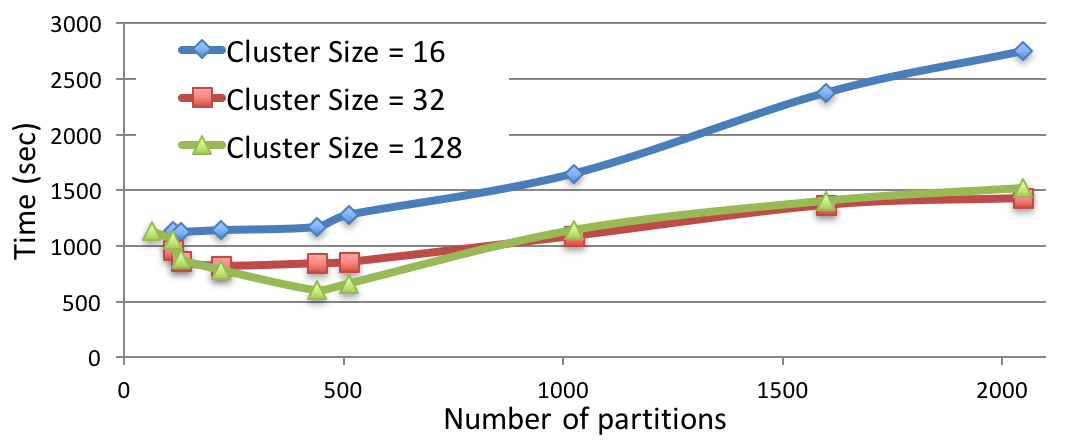}
		\caption{The default number of partitions in Twitter is $440$.}
		\label{GraphXpartitions_twitter}
	\end{subfigure}%
	~
	\begin{subfigure}{.45\textwidth}
		\centering
		\includegraphics[width=0.8\linewidth]{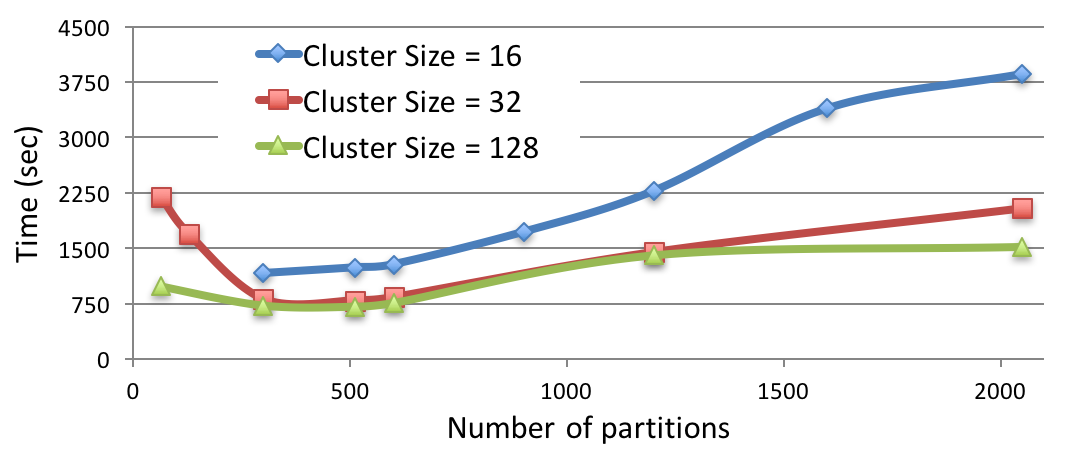}
		\caption{The default number of partitions in UK0705 is $1200$.}
		\label{GraphXpartitions_uk}
	\end{subfigure}
\normalsize
	\caption{Analysis of how number of partitions influence the performance of GraphX. The default number of partitions is not optimum.}
	\label{GraphXpartitions}
		\vspace{-5mm}
	
\end{figure*}

By default, the number of partitions is equal to the number of blocks\footnote{The default block size in HDFS is $64$ MB.} in the input file. Based on our communication with Spark engineers, this default value may not be optimum. We found that the default number of partitions may lead to reasonable performance in the case of small datasets. However, since this number does not consider the amount of available cores, it may lead to under utilization of the cluster computing power. Figure~\ref{GraphXpartitions} shows the influence of changing the number of partitions on two datasets (Twitter and UK0705) and three cluster sizes ($32$, $64$, $128$). The default number of partitions for the Twitter dataset is $440$, which also achieves the best performance among all clusters. However, the default number of partitions for the UK0705 dataset is $1200$, which is significantly larger than the number of cores\footnote{There are $4$ cores per machine. A $128$-machines cluster has $512$ cores.}. Therefore, the performance of GraphX using the default number of partitions is significantly worse than other options. 
Number of partitions used in our experiments are summarized in Table~\ref{ExperimentGraphXpartitions}.

\begin{table}[h!]
	\centering
	\tiny
	\begin{tabular}{|c|c||c|c|c|c|}
		\hline \multirow{2}{*}{\textbf{Dataset}} & \multirow{2}{*}{\textbf{\#blocks}} & \multicolumn{4}{c|}{\textbf{Cluster Size}} \\ \hhline{~~----}
		& & \textbf{16}   & \textbf{32}    & \textbf{64}     & \textbf{128} \\ \hline
		\hline Twitter & 440 & 128 & 256 & 440  & 440  \\ 
		\hline WRN     & 240 & 128 & 240 & 240  & 240  \\ 
		\hline UK200705& 1200& 128 & 256 & 512  & 1024 \\ 
		\hline 
	\end{tabular}
	\caption{Number of partitions GraphX in different cluster sizes.} 
	\label{ExperimentGraphXpartitions}
	\vspace{-0.7cm}
\end{table}

%% file: datasets.tex
\subsection{Datasets}
\label{Sec-exp:dataset}

Table~\ref{quantitative:realGraphs} summarizes the characteristics of the datasets used in this study: Twitter\footnote{http://law.di.unimi.it/webdata/twitter-2010}, World road network (WRN)\footnote{http://www.dis.uniroma1.it/challenge9/download.shtml}, UK200705\footnote{http://law.di.unimi.it/webdata/uk-2007-05/}, and ClueWeb\footnote{http://law.di.unimi.it/webdata/clueweb12/}. These are among the largest publicly available graph datasets, and they cover a wide range of graph characteristics. 

\begin{table}[ht]
	\centering
	\tiny
	\begin{tabular}{|l||c|c|c|c|}
		\hline Dataset& $|E|$ & Avg./Max. Degree & Diameter  \\ \hline
		\hline Twitter &  $1.46~B$ & $35$~/~$2.9M$ & $5.29$   \\ 
		\hline WRN &  $717~M$ & $1.05$~/~$9$ & $48~K$   \\ 
		\hline UK200705 & $3.7~B$ & $35.3$~/~$975K$ & $22.78$   \\ 
		\hline ClueWeb &  $42.5~B$ & $43.5$~/~$75M$ & $15.7$  \\ 
		\hline 
	\end{tabular}
	\caption{Real Graph Datasets} 
	\label{quantitative:realGraphs}
\end{table}

We partition all input graph datasets into chunks of similar sizes, and then load them to HDFS for all systems because this makes data loading more efficient for Blogel and GraphLab and has no impact on other systems. Note that the HDFS C++ library used in Blogel and GraphLab create a thread for each partition in the dataset. If there is only one data file, then only one thread executing on the master will read the entire graph, which significantly delays the loading process. In order to ensure a fair comparison, we prepared a dataset format that matches the typical requirement of each system. Specifically, we use three graph formats: adj, adj-long, and edge. The adjacency (adj) format is a typical adjacency list; each line includes a vertex id and then the ids of all vertices it is connected with. If a vertex does not have an out-edge, it does not need to have a line for itself. The adjacency-long (adj-long) format requires each vertex to have a line in the dataset input file. Moreover, the first value after the vertex id is the number of neighbor vertices. Edge format has a line for each edge in the graph. 

Hadoop, HaLoop, Giraph, and Graphlab use the adj format, which is the most concise format and significantly reduces the input size. Blogel needs to use the adj-long format for it to be able to create vertices that only have in-edges~\cite{BlogelCommunication}. This limitation could be fixed by adding an extra superstep in all computations to create missing vertices. However, this solution adds an overhead on the computation performance and was not preferred by Blogel developers when they were contacted. Finally, GraphX and Flink Gelly use edge-list.

%% file: results.tex
\section{Results \& Analysis}
\label{Sec-exp:results}

We summarize experimental parameters in Table~\ref{ExperimentDimensions}. The main experiment compares the performance of all systems with respect to all workloads, cluster sizes and datasets.  

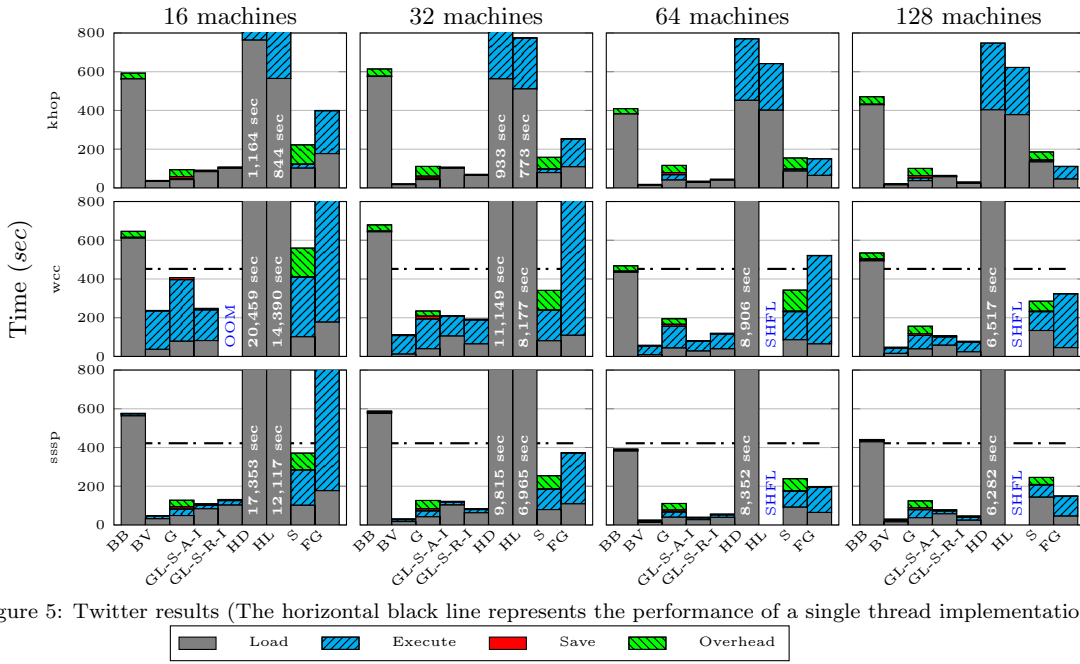
\begin{figure*}[th!]
	\centering
	\resizebox{15cm}{!}{\input{TekziStackGroupDataset-twitter-annotated-Gelly}}  

	\vspace{-2mm}
	\caption{Twitter results (The horizontal black line represents the performance of a single thread implementation)}
	\label{group-TW}
	\vspace{-5mm}
\end{figure*}

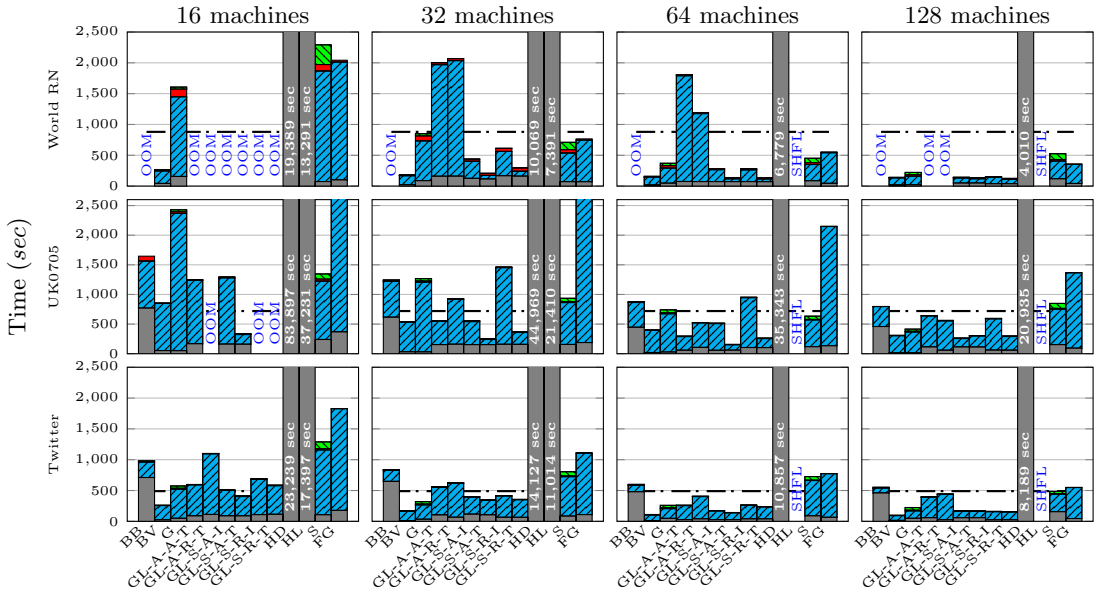
\begin{figure*}[th!]
	\centering
	\resizebox{15cm}{!}{\input{TekziStackGroupWorkload-pagerank-annotated-Gelly}}  

		\vspace{-2mm}
	\caption{PageRank query results (The horizontal black line represents the performance of a single thread implementation)}
	\label{group-PR}
		\vspace{-2mm}
\end{figure*}

\iftechreport
Figures ~\ref{group-PR},~\ref{KHOP},~\ref{SSSP},~\ref{WCC} show the detailed performance results of PageRank, Khop, SSSP, and WCC. Moreover, Figure ~\ref{group-TW} shows the results of Twitter dataset on all workloads and cluster sizes.
\fi

For all datasets except ClueWeb, we evaluate each system using all workloads across all cluster sizes; ClueWeb only fits in a cluster of $128$ machines and those results are reported separately in Table~\ref{ClueWeb-performance}. Empty entries in the result tables indicate that the execution did not successfully complete. There are multiple possible errors: timeout when an execution fails to complete in $24$ hours (TO), out-of-memory at any machine in the cluster (OOM), MPI error which only happens with Blogel-B (MPI), and shuffle error which only happens with HaLoop (SHFL). The following abbreviations are used for system names: \texttt{BV} and \texttt{BB} (Blogel -V and -B), \texttt{G} (Giraph), \texttt{S} (Spark/GraphX), \texttt{V} (Vertica), \texttt{HD} (Hadoop),  \texttt{HL} (HaLoop), \texttt{GL} (GraphLab), and \texttt{FG} (Flink Gelly). GraphLab experiments have six different versions identified by three symbols: (\texttt{A/S}) for asynchronous or synchronous computation, (\texttt{A/R}) for auto or random partitioning, and (\texttt{T/I}) for tolerance or iteration stopping criteria (discussed in PageRank workload in Section~\ref{pagerank}). For example, \texttt{GL-A-R-I} means GraphLab using asynchronous computation, random partitioning, and iteration stopping criteria.

\subsection{Blogel: The Overall Winner}
\label{BlogelDiscussion}
Vertex-centric Blogel (\texttt{BV}) has the best end-to-end performance. It is the only system that could finish the SSSP/WCC computation across all cluster sizes over WRN dataset, due to its large diameter. Moreover, it is the only system that could finish computations over ClueWeb in the $128$-machine cluster. It achieves this performance because it does not have an expensive infrastructure (such as Hadoop or Spark), uses efficient C++ libraries, utilizes all CPU cores, and has a small memory footprint.  

On the other hand, \texttt{BB} has the shortest execution time for queries that rely on reachability, such as WCC, SSSP, and K-hop for two reasons: (1) these queries benefit from Voronoi partitioning; and (2) block centric computation minimizes network overhead because it runs a serial algorithm with in each block before it communicates with other blocks. PageRank workload suffers from handling an awkward algorithm as discussed in Section~\ref{pagerank}. The purpose of running PageRank internally in each block is to start the global algorithm (considering all vertices in the graph) with a better initialization than a  straightforward initialization of equal PageRank value for each vertex. However, it turns out that the  algorithm used for this purpose does not generate good initial values, which hurt the overall performance.  This causes the block-centric version to take more iterations and more execution time after running the local PageRank. This result matches the original Blogel results and was discussed with Blogel developers. 

\begin{figure}[th!]
	\centering
	\resizebox{80mm}{!}{\includegraphics{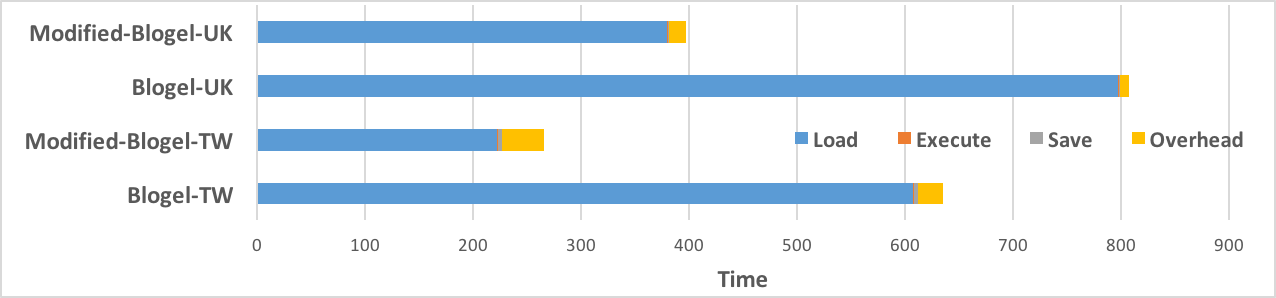}}
	\caption{Performance of modified-Blogel in computing WCC using a cluster of 16 machines, without HDFS overhead between partitioning and workload execution. The load time, which represents data reading, partitioning, and shuffling before execution, has been significantly reduced.}
	\label{BB_with_without_HDFS}
\end{figure}

The existing version of \texttt{BB}, reads data from HDFS, runs Voronoi partitioning, stores partitions in HDFS, reads these partitions again, and then runs a workload. We found that the end-to-end performance of block-centric computation has a significant overhead, due to the partitioning phase and the I/O overhead of writing and reading from HDFS. Removing the I/O overhead between partitioning and workload execution results in $50$\% reduction of the overall end-to-end response time (Figure~\ref{BB_with_without_HDFS}). 

Finally, \texttt{BB} could not process the WRN and ClueWeb graph because the GVD partitioning phase failed. After each sampling round, the Voronoi partitioner uses the master to aggregate block assignment data from each worker to count the size of each block. During this process the MPI interface uses an integer buffer to store data offset at a different location for each worker. Since WRN has a large number of vertices, the number of bytes received is larger than the maximum integer leading to an overflow and system crash. This issue can happen when MPI library is used to aggregate a large number of data items. The issue is known to the MPI community\footnote{\url{https://tinyurl.com/ybb9uu84
		%https://lists.mcs.anl.gov/pipermail/mpich-discuss/2011-March/009545.html
		}} and is a problem with MPI rather than Blogel. 

\begin{figure}[t]
	\centering
	\resizebox{80mm}{!}{\includegraphics{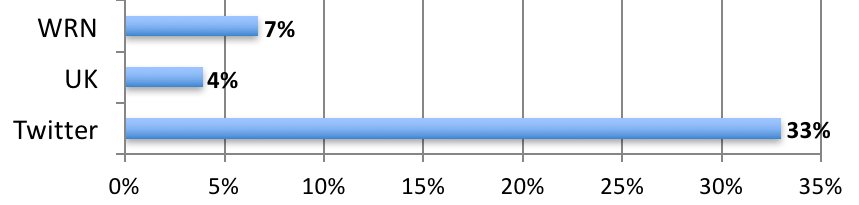}}
	\caption{Percentage of updated verteces in case of approximate PageRank in comparison to an exact one.}
	\label{tol_vs_itr_PR}
\end{figure}

\iftechreport

\begin{figure*}[th!]
	\centering
	\resizebox{15cm}{!}{\input{TekziStackGroupWorkload-khop-Gelly}}

	\caption{K-hop query results}
	\label{KHOP}
	
\end{figure*}
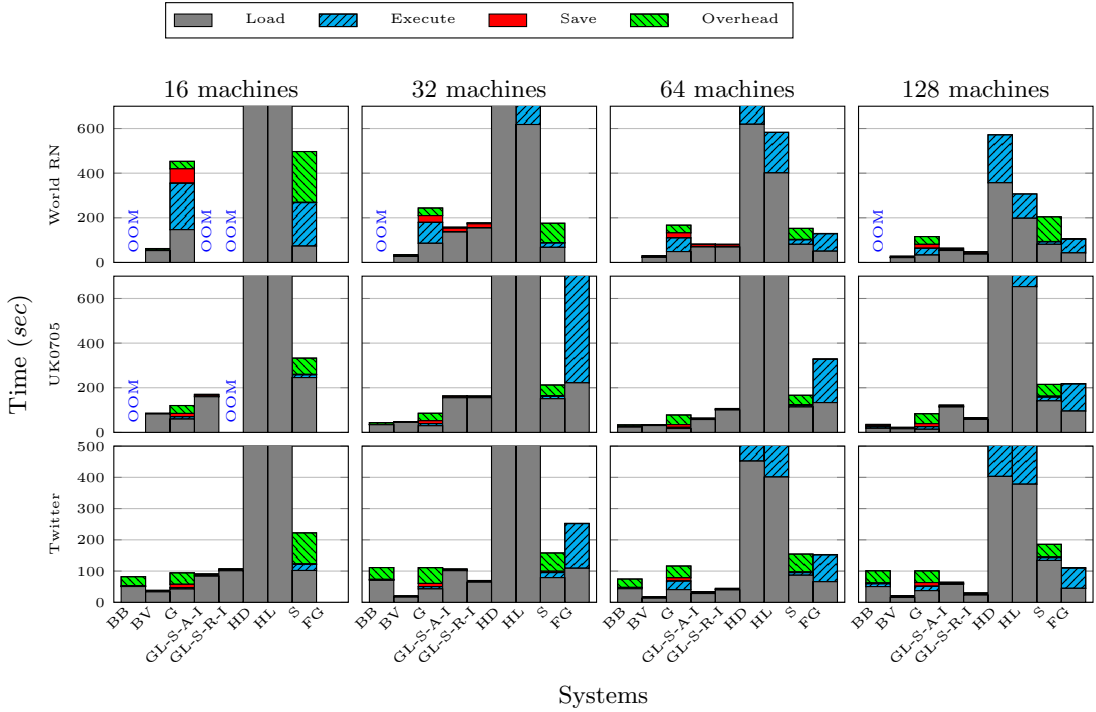

\begin{figure*}[th!]
	\centering
	\resizebox{15cm}{!}{\input{TekziStackGroupWorkload-sssp-Gelly}}

	\caption{SSSP  query results}
	\label{SSSP}
	
\end{figure*}
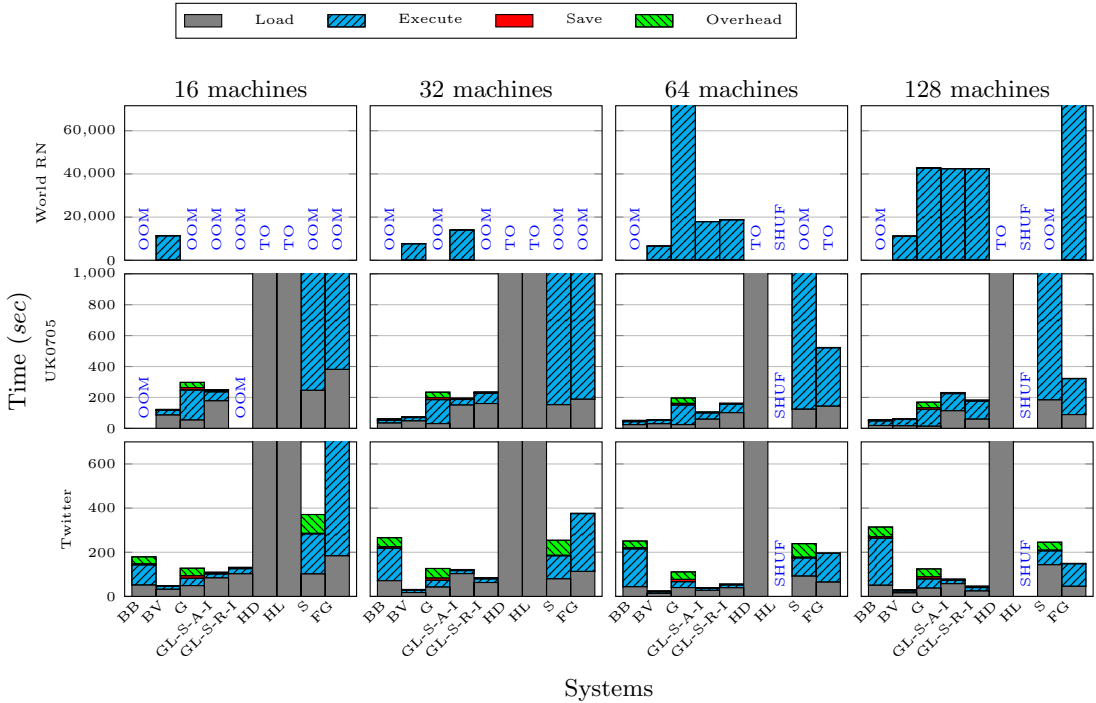

\begin{figure*}[th!]
	\centering
	\resizebox{15cm}{!}{\input{TekziStackGroupWorkload-wcc-Gelly}}

	\caption{WCC  query results}
	\label{WCC}
	
\end{figure*}
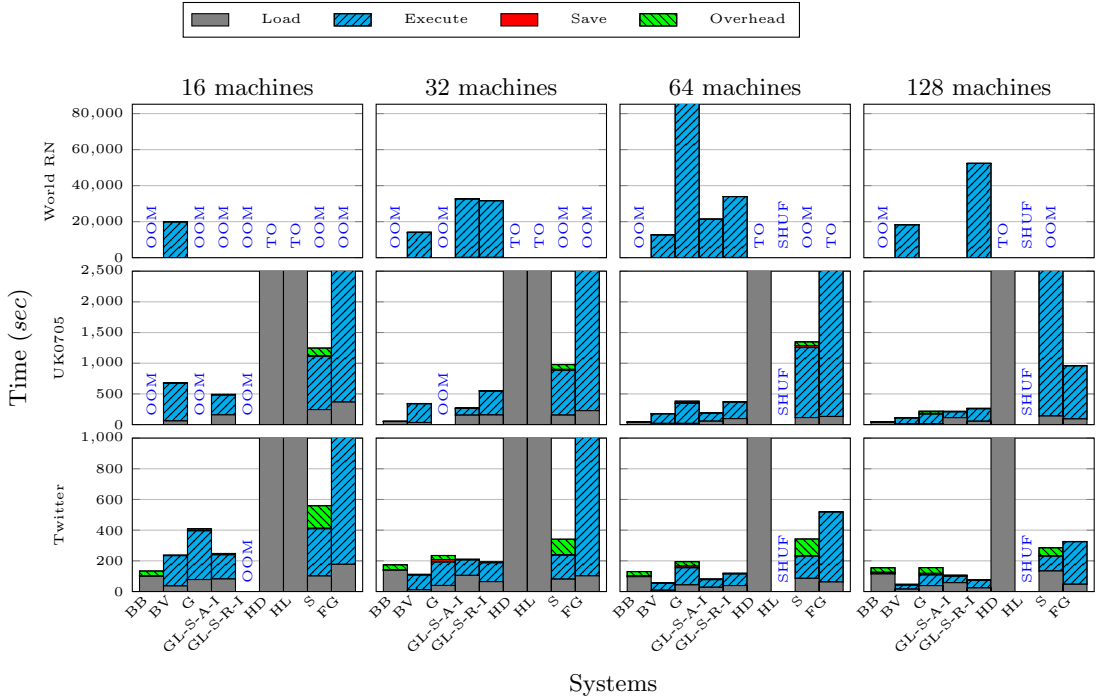

\fi

\subsection{Exact vs. Approximate PageRank}

The exact PageRank computation assumes that all vertices participate in the computation during all iterations. The approximate version allows vertices whose PageRank values have not changed (or changed by an amount smaller than a threshold) to become inactive and not participate in further computations. GraphLab is the only system that facilitates the latter because active vertices can gather the ranks of their in-neighbors even if these neighbors are not active. However, this also increases its memory overhead. Therefore, GraphLab fails with OOM error when using random partitioning for the UK0705 dataset over 16 machines. GraphLab also fails to load the WRN dataset to memory in the 16 machine configuration, regardless of the partitioning algorithm. Approximate PageRank in GraphLab is the only implementation that outperforms Blogel exact implementation. Approximate answers are less expensive than exact ones because many vertices converge in the first few iterations. For example, Figure~\ref{tol_vs_itr_PR} shows the ratio between the number of vertex updates in the approximate and exact implementations for three different datasets.

\subsection{Synchronous vs. Asynchronous}
\label{Sec:SyncVSasyncPAGERANK}

GraphLab is the only system that offers an asynchronous computation mode. 
However, it initiates thousands of threads per worker and allocates them to process vertices,  leading to distributed lock contention as previously reported~\cite{han2014experimental}. Therefore, PageRank asynchronous computation is typically slower than synchronous counterparts. Moreover, we found that asynchronous computation is only suitable for specific cluster sizes. It is not clear how to determine the right cluster size without trying multiple cluster configurations\footnote{This was suggested by the GraphLab team in the official forum: \url{
		http://forum.turi.com/discussion/714/
		}}.

\begin{figure}
	\vspace{-15mm}
	\centering
	\begin{subfigure}[t]{0.25\textwidth}
		\centering
		\begin{tikzpicture}
		\draw (0, 0) node[inner sep=0] {\includegraphics[width=0.8\linewidth]{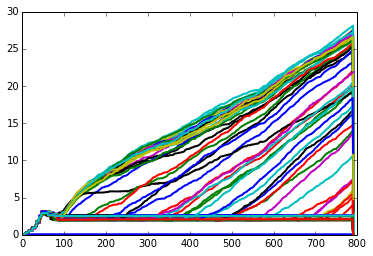}};
		\draw (-2, 0.5) node[text width=12em,rotate=90] {\scriptsize \color{annotationColor}Mem./worker (GB)};
		\end{tikzpicture}
		\vspace{-5mm}
		\caption{Asynchronous mode}
		\label{GraphLab_OOM_issue_async}
	\end{subfigure}%
	~
	\begin{subfigure}[t]{0.25\textwidth}
		\begin{tikzpicture}
		\draw (-3, 0) node[inner sep=0] {\includegraphics[width=0.8\linewidth]{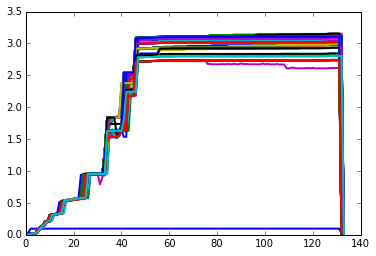}};
		\draw (0, 0.5) node[text width=12em,rotate=90] {\scriptsize \color{annotationColor} };
		\end{tikzpicture}
		\vspace{-9mm}
		\caption{Synchronous mode}
	\end{subfigure}
	\caption{Memory usage in GraphLab for PageRank of WRN using 128 machines. Each line represents the memory usage per worker per second; the X-axis is the time line for the computation in seconds. In the asynchronous mode, thousands of threads were created and allocated memory for vertices without releasing them quickly distributed locking which cause several machines to allocate large amount of memory, before the computation fails. }
	\label{GraphLab_OOM_issue}
\end{figure}

Unexpectedly, GraphLab asynchronous implementation fail\-ed with OOM while computing PageRank for the road network dataset using $128$ machines. Further analysis indicates that this is due to distributed locking. Figure~\ref{GraphLab_OOM_issue} shows  memory usage for synchronous and asynchronous; each line represents a machine. Data load overhead (time and memory) is the same for both modes. While the synchronous mode finished within a reasonable time with reasonable memory usage, in the asynchronous mode, many vertices started to allocate more memory (without releasing them due to distributed locking), which slowed down the computation performance and caused the failure.

\subsection{GraphLab partitioning optimization}
\label{results:graphlab_partitioning}

GraphLab has two partitioning modes: ``Random'' and ``Auto'' (Section~\ref{graphLabPartitioning}). Auto chooses between Grid, PDS an Oblivious. While Grid and PDF are faster than Oblivious, the latter does not have any requirements on the number of machines. Therefore, it gives priority to PDS or Grid, then uses Oblivious if neither are usable. 

In our reports, load time includes reading and partitioning the dataset. Figure~\ref{group-PR} shows that load time for GraphLab-auto is significantly smaller when using 16 or 64 machines (Grid) than the load time using 32 and 128 machines (Oblivious). This means that increasing the number of machines may lead to reduction of GraphLab performance.

\subsection{Giraph vs. GraphLab}
\label{results:GraphlabGirap}
Giraph is very competitive with GraphLab when the latter uses random partitioning and runs a fixed number of iterations (similar to Giraph). In fact, Giraph was faster than GraphLab in the 16 and 32 clusters. However, as the cluster size grows, Giraph spends more time in requesting resources and releasing them because it uses MapReduce platform for resource allocation. Therefore, both systems perform similar in the 64 cluster, but GraphLab finally wins in the 128 cluster.

\subsection{GraphX is not efficient when large number of iterations are required}
\label{GraphXdiscussion}

GraphX/Spark is slower than all other systems in our study because it suffers from Spark overheads, such as data shuffling, long RDD lineages, and checkpointing. Previous publications~\cite{GraphX} show that GraphX is efficient because it uses a special Spark prototype version that includes in-memory shuffling. This feature is not available in the latest Spark release. 

GraphX failed to compute WCC for the WRN dataset due to memory or timeout errors in all cluster sizes. It turns out that Spark fault-tolerance mechanism of maintaining RDD lineages is the culprit of memory errors. When the number of iterations grows, these lineages become long  leading to high memory usage, and potential out of memory errors. Recent introduction of  GraphFrames\footnote{\url{https://github.com/graphframes/graphframes}} should be a more efficient option for GraphX. We investigated GraphFrame implementations and found that many of its algorithms convert the input graph to GraphX format and then run GraphX algorithms. We also found that most algorithms have a default maximum limit on number of iterations to reduce the potential overhead of long lineage in RDDs. For example, SSSP has a limit of 10, otherwise it starts to checkpoint to avoid long lineages. Moreover, the default implementation of WCC requires checkpointing every two iterations. Checkpointing prevents lineage from being very long, but it leads to expensive I/O interactions with disk, which then leads to a timeout error. 
GraphFrames offers a version of the hash-min~\cite{Pegasus} algorithm originally used to compute WCC, called hash-to-min~\cite{MRcc2014} that uses fewer iterations. We tested this implementation as well and found that it was competitive with hash-min in Blogel.

\begin{figure}
	\centering
	\includegraphics[width=0.75\linewidth]{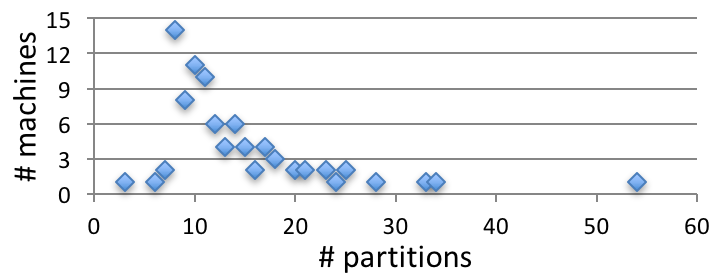}
	\caption{In a 128-machines cluster, GraphX does not balance number of partitions ($1200$) evenly to machines. A balanced distribution of workload would assign $1200/128=9.4$ partitions to each machine. However, one machine has 54 partitions.}
	\label{GraphXloadBalancing}
	\vspace{-5mm}
\end{figure}

We noticed that Spark could not uniformly distribute partitions to workers. As depicted in Figure~\ref{GraphXloadBalancing} some machines were assigned a large number of partitions. In a synchronous computation model, stragglers form in this case slowing down the workers who finish their tasks. In future Spark releases, this problem could be solved by implementing a more efficient load balancer. 

In practice, the number of partitions should not be more than the number of blocks in the input file, because this forces Spark to read the same data block more than once. On the other hand, the number of partitions should not be less than the number of cores in the cluster, because this would lead to CPU under-utilization. In our experiments, we set the number of partitions equal to the number of blocks as long as it does not exceed twice the number of cores. This allows Spark to handle stragglers  by assigning them to another core. Unfortunately, this does not guarantee the best performance (Figure~\ref{GraphXpartitions}) because the workload assignment is not balanced (Figure~\ref{GraphXloadBalancing}).

\subsection{System Overhead}
\label{Overhead-discussion}
The computation overhead is significantly larger for Giraph and GraphX. Giraph uses Hadoop and GraphX uses Spark for resource management, scheduling, and fault tolerance. The cost of starting a job and closing it are high in Hadoop and Spark. Blogel and GraphLab use MPI for communication between machines, and therefore, do not have the overhead of an underlying infrastructure. They interact with HDFS using C++ libraries but they do not depend on the job or task tracker in Hadoop. 

Although the overhead time is small in Flink Gelly, we found that the system frequently fails after running a few jobs. It turns out that Flink does not reclaim all memory used by the system in between workload executions. This causes the system to eventually fail due to out-of-memory error. Thus, we had to restart Flink after each workload. 

\subsection{WCC Experiments}
\label{wcc}

The WCC workload needs special handling because it requires the processing system to handle edges in both directions. Therefore, Gelly, Blogel and Giraph have the overhead of pre-computing the in-neighbors before executing the algorithm. Moreover, Blogel and Giraph cannot benefit from the message combiner in this workload, because messages in the first iteration should not be combined, since they are used to discover in-neighbors, not to find the smallest vertex id in the connected component. Furthermore, computing the WCC requires more memory than other workloads, because each vertex needs to recognize its in- and out-neighbors. Giraph failed to load the UK0705 in the 16 and 32 machine clusters and failed to load the WRN in the 16 machine cluster. Giraph could not finish computation of WCC for the WRN dataset in the 32 machine cluster, but succeeded to compute the WCC in almost 24 hours using the 64 machine cluster.

Blogel-V is the only system that could compute WCC on the WRN dataset using the 16 machine cluster due to its low memory requirements. GraphLab with random partitioning failed to load UK0705 dataset in the 16 machine cluster. On the other hand, GraphLab auto partitioning significantly reduces the execution time in comparison to random partitioning. However, the loading time for auto partitioning is high when Grid and PDS algorithms are not applicable (e.g. in the 32 and 128 machines cluster). 

 GraphX performance for WCC using the UK0705 dataset over 128 machines was significantly worse than all other systems and also worse than GraphX performance over 64 machines. This is an example of the influence of the number of partitions (Table~\ref{ExperimentGraphXpartitions}) in Spark on the GraphX performance.
 
 Gelly successfully finished the execution of WCC for Twitter and UK0705 in all clusters. However, it failed with time-out error to compute WCC for the WRN dataset in the 16, 32, and 64 cluster. Gelly finished WCC for WRN in slightly less than 24 hours using 128 machines.

\begin{table}
	\centering
	\tiny
	\begin{tabular}{|c||c|c||c|c|}
		\hline   & \multicolumn{2}{|c||}{Giraph}& \multicolumn{2}{c|}{GraphX} \\ 
		\hline  & SSSP & WCC & SSSP & WCC \\ \hline	
		\hline 16 & 6 & OOM & 120 & 420 \\ 
		\hline 32 & 3 & 3.2  & 17 & 30 \\ 
		\hline 
	\end{tabular}
	\caption{The time, in seconds, used by each iteration for the WRN dataset. For SSSP and WCC to finish in 24 hours, the iteration time should be 2.4 and 1.8 respectively.} 
	\label{NotFinishedNumbers}
\end{table}

\begin{table}
	\centering	
	\tiny										
	\begin{tabular}{|l|c|c|c|c|}
		\hline \textbf{Workload} & \textbf{Read} & \textbf{Execute} & \textbf{Save} & \textbf{Others} \\
		\hline 
		\hline \textbf{PageRank} & 132.5 & 139.7 & 10.5 & 15.3 \\ 
		\hline \textbf{WCC} & 134.1 & 152.5 & 11.5 & 10.6 \\ 
		\hline \textbf{SSSP} & 158.3 & 89.3 & 2.2 &  20.7 \\ 
		\hline \textbf{K-hop} & 161.6 & 0.03 & 0.2 &  16.4 \\ 
		\hline 
	\end{tabular} 
	\caption{Blogel-V performance on the ClueWeb dataset using a cluster of 128 machines. Numbers represent number
of seconds used for each processing phase.}
	\vspace{-5mm}
	\label{ClueWeb-performance}
\end{table}

\subsection{ClueWeb experiments}
\label{ClueWeb_results}
ClueWeb represents a web graph, and has 42.5 billion edges and almost one billion vertices. The size of this dataset is 700GB (adjacency list) and 1.2 TB (edge list). Only the 128 cluster can hold it using its total 3 TB memory.

GraphLab could not load the dataset in memory. Although we do not know ClueWeb's replication factor (since it could not be loaded), the other web graph, UK0705, has replication factors of 3.6 and 4.5 for the 64 and 128 machine clusters, respectively. If we assume a similar replication factor for ClueWeb, the data is larger than available memory. 
For the same reason, Gelly and Giraph could not finish their computation. We found that total physical memory used by Giraph to process UK0705 (originally 32 GB) using the 128 machine cluster is 1322 GB. Table~\ref{GiraphMemory} summarizes Giraph memory consumption for all datasets.

\begin{table}[t]
	\centering	
	\tiny										
	\begin{tabular}{|l|c|c|c|c|}
		\hline \textbf{dataset (size in GB)} & \textbf{16} & \textbf{32} & \textbf{64} & \textbf{128} \\
		\hline \hline \textbf{Twitter (12.5)} & 191.5 & 323.6 & 606.4 & 923.5 \\ 
		\hline \textbf{UK0705 (31.9)} & 264.0 & 411.8 & 717.6 & 1322.6 \\ 
		\hline \textbf{WRN (13.6)} & 363.7 & 475.4 & 683.4 & 1054.1 \\ 
		\hline 
	\end{tabular} 
	\caption{Total Giraph Memory across the cluster. All numbers are in GB; first row shows cluster size.}
	\label{GiraphMemory}
\end{table}

Blogel-V, was the only system that could perform any workload on ClueWeb in the 128 cluster (Table~\ref{ClueWeb-performance}). This suggests that graph processing systems should be conscious of memory requirements, despite the common assumption that memory is available and cheap and most real graphs can fit in memory of current workstations. Most graph systems are optimized to decrease processing time at the expense of larger memory, but our results suggest caution. Finally, ClueWeb results also show that Hadoop MapReduce platform and distributed out-of-core systems may have a role; they are significantly slower than in-memory systems, but they can finish the task when memory is constrained or graph size is too large.

\subsection{Hadoop and HaLoop Experiments}
\label{HaLoop-results}
As noted earlier, it was expected that Hadoop and HaLoop would be slower than in-memory systems. 
As expected, HaLoop was faster than Hadoop due to its optimizations. However, our experiments do not show the  $2\times$ speedup that was reported in the HaLoop paper. Moreover, existing HaLoop  implementation has some issues. The loop management mechanism introduced by HaLoop eliminates the usability of some basic Hadoop features, such as counters. It is not possible to use custom counters during iteration management to check for convergence. 
Moreover, HaLoop suffers from a bug that occasionally causes mapper output to be deleted before all reducers use them, in large cluster sizes\footnote{It typically fails after a few iterations in the 64 and 128 machine  clusters.}.

CPU utilization is better in HaLoop than Hadoop, because in Hadoop CPUs spend a long time waiting for I/O operations. Since HaLoop tries to allocate the same mapper to the same data partitions, there is not too much data shuffling. It is interesting to note that both Hadoop and HaLoop use similar average physical memory in their workers. This identifies an opportunity for HaLoop: instead of caching files on local disks, HaLoop could have utilized the available memory to further reduce execution time.

\subsection{Vertica Experiments}
\label{sec-vertica-results}

\begin{figure}
	\centering
	\includegraphics[width=0.9\linewidth]{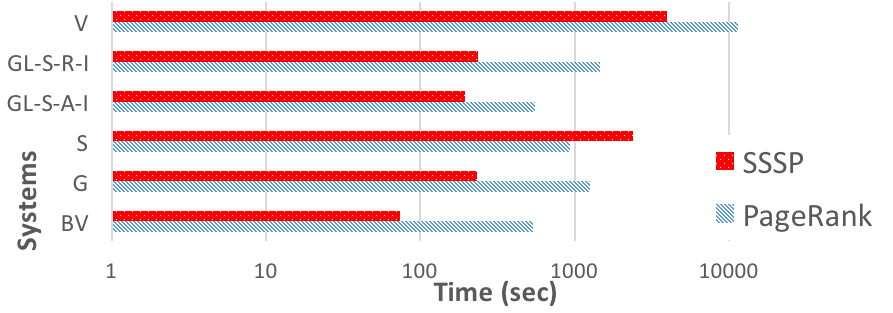}
	\caption{Computing SSSP (116 iterations) and 55 iterations of PageRank for the UK dataset using a cluster of 32 machines. }
	\label{verticaFigures}
\end{figure}

\begin{figure*}[t!]
	\centering
	\begin{subfigure}[t]{0.3\textwidth}
		\centering
		\includegraphics[width=0.9\linewidth]{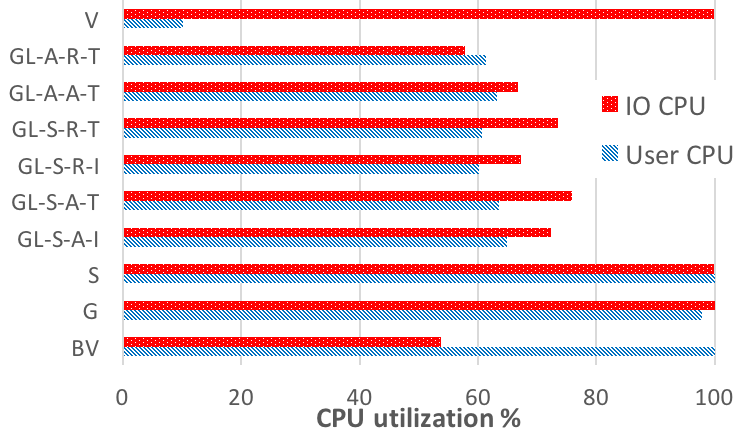}
		\caption{Maximum User and I/O CPU}
	\end{subfigure}%
	~
	\begin{subfigure}[t]{0.3\textwidth}
		\centering
		\includegraphics[width=0.9\linewidth]{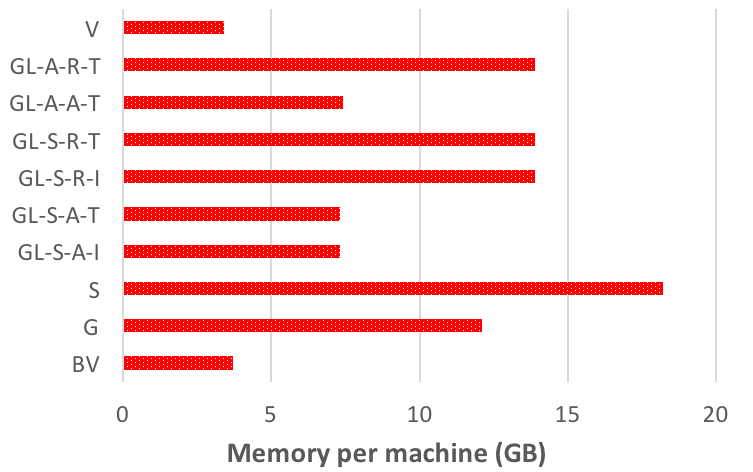}
		\caption{Memory footprint}
	\end{subfigure}
	~
	\begin{subfigure}[t]{0.3\textwidth}
		\centering
		\includegraphics[width=0.9\linewidth]{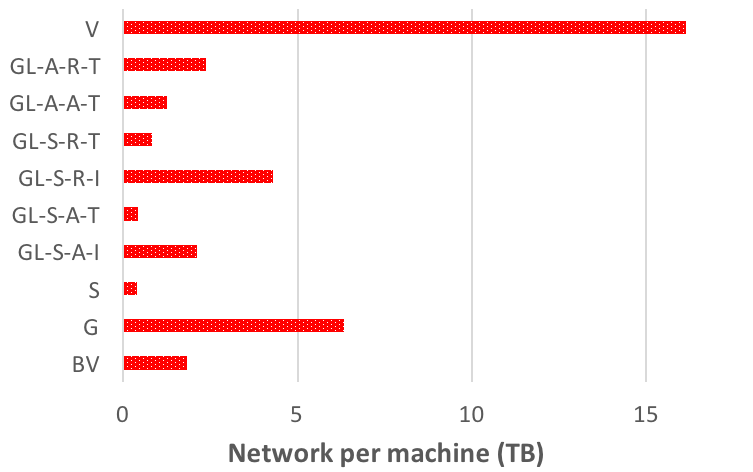}
		\caption{Network usage}
	\end{subfigure}
	
	\caption{Understanding how Vertica use its computing resources in comparison to other systems. All results were collected while computing 55 iterations of PageRank for the UK0705 dataset using a cluster of 64 machines.}
	\label{VerticaUtilization}

\end{figure*}

Although Vertica supports r3.4xlarge and r3.8xlarge instances only, we ran our experiments using r3.xlarge to make these results comparable to the rest of our experiments. We use similar SQL queries to the ones described in~\cite{Vertica4Graph}. These experiments are not as complete as others because we were allowed to use the system for a short trial period\footnote{The community edition of Vertica is restricted to 3 machines only.}. Nonetheless, we believe the results in Figures~\ref{verticaFigures} and~\ref{VerticaUtilization} present a good indication of its performance in large clusters.

Unlike previously reported results~\cite{Vertica4Graph}, Vertica is not competitive relative to native in-memory graph processing systems. As the cluster size increases, so does the gap between its performance and other systems. Previously reported experiments were conducted on only 4 machines and that may explain the competitive results.

The main reason behind Vertica's performance with large clusters is its requirement to create and delete new temporary tables during execution, because each table is partitioned across multiple machines. Moreover, self-join operation involves shuffling. The larger the cluster, the more expensive data shuffling becomes. Figure~\ref{VerticaUtilization} supports this argument. Although Vertica footprint is small, the I/O-wait time and network cost are significant. Increasing the cluster size, significantly adds to these overheads. On the other hand, distributed graph processing systems can utilize the computing power of larger clusters without significant I/O and network overhead.

There is a Vertica extension to avoid the intermediate disk I/O, but this extension is not yet publicly available. It works on multiple cores on a single machine, but it does not support shared-nothing parallel architectures.

\subsection{Scalability}
\label{scalability}

LDBC~\cite{LDBC} discusses two orthogonal types of scalability analysis: strong/weak  and horizontal/vertical scalability. In a strong scalability experiment the same dataset is used with different cluster sizes (horizontal scalability) or with one machine and different number of cores (vertical scalability). In a weak scalability experiment the data load (represented in graph size) of each machine in the cluster is fixed as we change the cluster size. For example, as we double the cluster size, we also double the graph size to keep the data load per machine constant. 

Our study does not include vertical scalability experiments because all our systems were introduced as parallel shared-nothing systems. We only consider real  datasets whose  sizes are fixed. Therefore, our scalability analysis is ``strong''.  

Blogel, Giraph, Gelly, and GraphLab show steady performance increase as the cluster size increases. GraphX and Vertica do not show the same scalability potential. GraphX suffers from load balancing issues: the higher the number of workers, the lower the balance between machines. Vertica, on the other hand, has to shuffle more data as the cluster gets larger.  That said, scalability was not noticeable during the execution of SSSP and K-hop workloads because most graph vertices do not participate in each iteration during SSSP and K-hop computation.

\subsection{COST Experiment}
\label{Cost-experiments}

COST~\cite{COST} stands for Configuration that Outperforms a Single Thread, and is used in the literature to evaluate the overhead of parallel algorithms and systems. The main idea is that parallel algorithms are often not optimal due to the special design considerations for parallel processing between machines. COST factor represents the response time of single thread divided by the response time of a parallel system. In the COST experiment, we used the single thread implementation of the GAP Benchmark Suite~\cite{GAP-benchmark} on a large machine with 512 GB memory. Table~\ref{COST_performance} summarizes the performance of a single-thread implementation (S) and the performance of the best parallel system (P) using 16 machines.
\iftechreport
Looking at Figures ~\ref{group-TW},~\ref{group-PR},~\ref{KHOP},~\ref{SSSP}, and ~\ref{WCC},
\fi 
it is clear that the performance of some systems using multiple machines is worse than their single thread performance. 

Some of the algorithms used in this experiment are different than ones described in Section~\ref{Sec-exp:workload}. The PR algorithm is similar to the one used by all systems. The SSSP algorithm\footnote{\url{
		https://tinyurl.com/y7rsgg9w
		%https://github.com/sbeamer/gapbs/blob/master/src/bfs.cc
		}} processes the workload from two directions and pre-computes each vertex degree in its initial phase~\cite{Direction-optimization-BFS}. 
The WCC algorithm implementation\footnote{\url{
		https://tinyurl.com/ydxhpfyx
		%https://github.com/sbeamer/gapbs/blob/master/src/cc.cc
		}} is based on Shiloach-Vishki algorithm with further optimizations \cite{bader2005architectural, shiloach1982logn, kothapalli2010fast}. 

\begin{table}
	\centering	
	\tiny										
	\begin{tabular}{|l||l|c||l|c||l|c|}
		\hline  
		& \multicolumn{2}{c||}{\textbf{PageRank}}  & \multicolumn{2}{c||}{\textbf{SSSP}} & \multicolumn{2}{c|}{\textbf{WCC}}\\ 
				   & P	& S & P	& S & P	& S\\ \hline 
		\hline \textbf{Twitter} & BV=260 & 490 & BV=48.3 & 422 & GL=248 & 452 \\ 
		\hline \textbf{UK0705} & BV=338.7 & 720 & BV=122.3 & 610 & GL=492.67 & 632  \\ 
		\hline \textbf{WRN} & BV=268.3 & 880 & BV=11295 & 455 & BV=19831 & 640  \\ 
		\hline 
	\end{tabular} 
	\caption{Time in seconds for a single thread algorithm (S) and best performing parallel system using 16 machines (P)}
	\vspace{-5mm}
	\label{COST_performance}
\end{table}

Although many systems, using 16 machines, have a COST $< 1$, meaning they perform worse than a single thread implementation,  best parallel systems perform better than the single thread in most cases. However, definitive conclusions using the COST factor are hard to reach. For PR, the cost factor is between $2$ and $3$ which means the 16-machines cluster performs two to three times faster than the single thread. For SSSP and WCC the cost factor is $0.5$ to $0.11$ for power-law datasets, but it is $0.04$ and $0.03$ respectively for WRN. This means that, for reachability-based workloads, the best parallel system could be two orders of magnitude slower than a single thread implementation. The large number of iterations leads to significant network overhead between machines. Of course, single-thread performance requires larger machines, e.g., running WCC on WRN using the single thread implementation uses $112$GB memory -- four times the memory available in the machines used in our experiments. 

The surprisingly bad COST factor of many parallel systems is due to three main factors:
\begin{itemize}
	\item Different algorithms: Parallel systems adopt simple algorithms that can scale well, while the single thread implementations include several optimizations. It is an interesting future work to study the possibility of parallelizing these optimizations.
	\item Replication: Parallel systems need to partition the dataset with significant replication factors (see Table \ref{replicationFactor}). This adds overhead to the overall dataset sizes.
	\item Network: Parallel systems incur network overhead, which, of course, is absent in single thread implementation.
\end{itemize}

%% file: TekziStackGroupDataset-twitter-annotated-Gelly.tex
\makeatletter
\pgfplotsset{
	groupplot xlabel/.initial={}, every tick label/.append style={font=	iny},
	every groupplot x label/.style={
		at={($({group c1r\pgfplots@group@rows.west}|-{group c1r\pgfplots@group@rows.outer south})!0.5!({group c\pgfplots@group@columns r\pgfplots@group@rows.east}|-{group c\pgfplots@group@columns r\pgfplots@group@rows.outer south})$)},
		anchor=north,
	},
	groupplot ylabel/.initial={},
	every groupplot y label/.style={
		rotate=90,
		at={($({group c1r1.north}-|{group c1r1.outer
				west})!0.5!({group c1r\pgfplots@group@rows.south}-|{group c1r\pgfplots@group@rows.outer west})$)},
		anchor=south
	},
	execute at end groupplot/.code={%
		\node [/pgfplots/every groupplot x label]
		{\pgfkeysvalueof{/pgfplots/groupplot xlabel}};  
		\node [/pgfplots/every groupplot y label] 
		{\pgfkeysvalueof{/pgfplots/groupplot ylabel}};
	},
	group/only outer labels/.style =
	{
		group/every plot/.code = {%
			\ifnum\pgfplots@group@current@row=\pgfplots@group@rows\else%
			\pgfkeys{xticklabels = {}, xlabel = {}}\fi%
			\ifnum\pgfplots@group@current@column=1\else%
			\pgfkeys{yticklabels = {}, ylabel = {}}\fi%
		}
	}
}
\def\endpgfplots@environment@groupplot{%
	\endpgfplots@environment@opt%
	\pgfkeys{/pgfplots/execute at end groupplot}%
	\endgroup%
}

\definecolor{RYB3}{RGB}{253, 180, 98}

\tikzset{nomorepostaction/.code=\let\tikz@postactions\pgfutil@empty}
\makeatother
\begin{tikzpicture}
\pgfplotsset{%
     tiny,samples=10,
	width=3cm,
	height=2cm,
	scale only axis,
	ymajorgrids,
	yminorgrids
}
   \begin{groupplot}[
         group style = {group size = 4 by 3, 
         	horizontal sep = 5pt,
         	vertical sep = 5pt}, 
         groupplot ylabel={Time ($sec$)},
         group/only outer labels, 
         ybar stacked, /pgf/bar width=1, /pgf/bar shift=0pt,
         area legend,  ymin=0,    xtick=data,xticklabels={BB, BV, G, GL-S-A-I, GL-S-R-I, HD, HL, S, FG},
         scaled ticks=false, xtick style={draw=none},  x tick label style={rotate=45,anchor=east}
]
      \nextgroupplot[ title = {16 machines}, 
      legend style = { column sep = 10pt, legend columns = 4, legend to name = grouplegend,}, ymax=800, ylabel={khop}] 

\node[font=\bf,white,above] at (axis cs: 6,0) {\rotatebox{90}{\tiny{1,164 sec}}}; 
\node[font=\bf,white,above] at (axis cs: 7,0) {\rotatebox{90}{\tiny{844 sec}}}; 

\addplot+[ybar] [black, fill=gray] coordinates{
   (1, 562.521629) 
   (2, 33.49368666666667)
   (3, 42.6)
   (4, 83.69456666666667)
   (5, 101.04786666666666)
   (6, 763.0)
   (7, 565.0)
   (8, 101.83999999999999)
   (9, 177.0)
};
\addlegendentry{Load}
\addplot+[ybar] [black, fill= cyan, postaction={nomorepostaction,pattern=north east lines}] coordinates{
   (1, 0.246264)
   (2, 0.17703233333333332)
   (3, 4.675333333333334)
   (4, 1.1582733333333335)
   (5, 1.2132533333333333)
   (6, 401.0)
   (7, 279.0)
   (8, 19.666666666666668)
   (9, 221.5)
};
\addlegendentry{Execute}
\addplot+[ybar] [black, fill=red] coordinates{
   (1, 0.679032)
   (2, 0.2055543333333333)
   (3, 10.342999999999998)
   (4, 2.567682000000002)
   (5, 2.587617000000001)
   (6, 0.0)
   (7, 0.0)
   (8, 0.9533906666666665)
   (9, 1.5)
};
\addlegendentry{Save}
\addplot+[ybar] [black, fill=green, postaction={nomorepostaction,pattern=north west lines}] coordinates{
   (1, 29.553075)
   (2, 4.45706)
   (3, 36.714999999999996)
   (4, 3.579478)
   (5, 2.484596333333333)
   (6, 2.0)
   (7, 2.0)
   (8, 99.87327599999999)
   (9, 0)
};
\addlegendentry{Overhead}
      \nextgroupplot[ title = {32 machines}, 
      ymax=800] 
      
\node[font=\bf,white,above] at (axis cs: 6,0) {\rotatebox{90}{\tiny{933 sec}}}; 
\node[font=\bf,white,above] at (axis cs: 7,0) {\rotatebox{90}{\tiny{773 sec}}}; 

\addplot+[ybar] [black, fill=gray] coordinates{
   (1, 575.3401685)  
   (2, 16.67743533333333)
   (3, 42.93142857142857)
   (4, 101.26638333333335)
   (5, 64.50108333333334)
   (6, 563)
   (7, 511)
   (8, 78.902)
   (9, 109.0)
};
\addplot+[ybar] [black, fill= cyan, postaction={nomorepostaction,pattern=north east lines}] coordinates{
   (1, 0.15432075)
   (2, 0.228992)
   (3, 7.282000000000001)
   (4, 1.0333333333333332)
   (5, 1.1666666666666667)
   (6, 370.0)
   (7, 262.0)
   (8, 16.333333333333332)
   (9, 143.0)
};
\addplot+[ybar] [black, fill=red] coordinates{
   (1, 0.9247397500000001)
   (2, 0.16912733333333332)
   (3, 10.178142857142856)
   (4, 2.04538)
   (5, 2.0039166666666666)
   (6,  0.0)
   (7,  0.0)
   (8, 4.056975666666666)
   (9, 1)
};
\addplot+[ybar] [black, fill=green, postaction={nomorepostaction,pattern=north west lines}] coordinates{
   (1, 37.830771)
   (2, 3.9244453333333333)
   (3, 50.608428571428576)
   (4, 2.488236666666667)
   (5, 2.1616666666666666)
   (6, 2.0)
   (7, 1.0)
   (8, 58.70769099999999)
   (9, 0)
};
      \nextgroupplot[ title = {64 machines}, 
      ymax=800]


\addplot+[ybar] [black, fill=gray] coordinates{
   (1, 381.434589200000005) 
   (2, 13.315258666666665)
   (3, 40.518750000000004)
   (4, 28.10118)
   (5, 39.68831)
   (6, 452)
   (7, 401)
   (8, 86.674)
   (9, 64.5)
};
\addplot+[ybar] [black, fill= cyan, postaction={nomorepostaction,pattern=north east lines}] coordinates{
   (1, 0.2346156)
   (2, 0.3958333333333333)
   (3, 27.659499999999998)
   (4, 1.1666666666666667)
   (5, 1.2333333333333334)
   (6, 318.0)
   (7, 241.0)
   (8, 8.5)
   (9, 86.0)
};
\addplot+[ybar] [black, fill=red] coordinates{
   (1, 1.0522358)
   (2, 0.16637433333333332)
   (3, 10.388500000000004)
   (4, 2.7830600000000003)
   (5, 1.5859766666666666)
   (6,  0.0)
   (7,  0.0)
   (8, 2.129772)
   (9, 1)
};
\addplot+[ybar] [black, fill=green, postaction={nomorepostaction,pattern=north west lines}] coordinates{
   (1, 26.8785594)
   (2, 4.1225336666666665)
   (3, 37.68325)
   (4, 2.2824266666666664)
   (5, 2.159046666666667)
   (6, .0)
   (7, 0)
   (8, 57.196228000000005)
   (9, 0)
};
      \nextgroupplot[ title = {128 machines}, 
      ymax=800] 
      

\addplot+[ybar] [black, fill=gray] coordinates{
   (1, 429.333451333333336) 
   (2, 15.512934666666666)
   (3, 37.36533333333333)
   (4, 57.43666666666667)
   (5, 23.268906666666666)
   (6, 403)
   (7, 378)
   (8, 133.96699999999998)
   (9, 45.5)
};
\addplot+[ybar] [black, fill= cyan, postaction={nomorepostaction,pattern=north east lines}] coordinates{
   (1, 0.216966333333332)
   (2, 1.0954756666666665)
   (3, 14.134666666666668)
   (4, 1.6580366666666666)
   (5, 1.7023133333333333)
   (6, 345.0)
   (7, 244.0)
   (8, 9.0)
   (9, 65.0)
};
\addplot+[ybar] [black, fill=red] coordinates{
   (1, 2.6021959999999997)
   (2, 0.16714933333333334)
   (3, 11.749333333333334)
   (4, 1.9364133333333333)
   (5, 1.74069)
   (6,  0.0)
   (7,  0.0)
   (8, 2.371534666666667)
   (9, 1)
};
\addplot+[ybar] [black, fill=green, postaction={nomorepostaction,pattern=north west lines}] coordinates{
   (1, 38.84738633333333)
   (2, 4.891107000000001)
   (3, 37.41733333333333)
   (4, 3.302216666666667)
   (5, 3.621423333333333)
   (6, .0)
   (7, 0)
   (8, 40.328132000000004)
   (9, 0)
};

      \nextgroupplot[ 
      ymax=800, ylabel={wcc}] 
      
\node[blue,above] at (axis cs: 5,0) {\rotatebox{90}{\tiny{OOM}}};
\node[font=\bf,white,above] at (axis cs: 6,0) {\rotatebox{90}{\tiny{20,459 sec}}}; 
\node[font=\bf,white,above] at (axis cs: 7,0) {\rotatebox{90}{\tiny{14,390 sec}}}; 

\addplot+[ybar] [black, fill=gray] coordinates{
   (1, 610.704912) 
   (2, 36.403749)
   (3, 78.1285)
   (4, 81.7675)
   (5, 0)
   (6, 12432.0)
   (7, 8974.0)
   (8, 101.24433333333332)
   (9, 177.0)
};
\addplot+[ybar] [black, fill= cyan, postaction={nomorepostaction,pattern=north east lines}] coordinates{
   (1, 1.115131)
   (2, 197.90251933333334)
   (3, 318.38525000000004)
   (4, 158.341)
   (5, 0)
   (6, 8027.0)
   (7, 5416.0)
   (8, 308.3333333333333)
   (9, 5750.25)
};
\addplot+[ybar] [black, fill=red] coordinates{
   (1, 3.281586)
   (2, 0.307373)
   (3, 10.481999999999946)
   (4, 3.1398460000000057)
   (5, 0)
   (6, 0.0)
   (7, 0.0)
   (8, 2.5688973333333336)
   (9, 2.25)
};
\addplot+[ybar] [black, fill=green, postaction={nomorepostaction,pattern=north west lines}] coordinates{
   (1, 30.898371)
   (2, 2.7196919999999998)
   (3, 0.275)
   (4, 4.751654)
   (5, 0)
   (6, 1.0)
   (7, 0.0)
   (8, 147.18676933333333)
   (9, 0)
};
\draw [thick,dash pattern={on 7pt off 2pt on 1pt off 3pt}] (axis cs:1,452) -- (axis cs:9,452) ;

\nextgroupplot[ ymax=800]

\node[font=\bf,white,above] at (axis cs: 6,0) {\rotatebox{90}{\tiny{11,149 sec}}}; 
\node[font=\bf,white,above] at (axis cs: 7,0) {\rotatebox{90}{\tiny{8,177 sec}}}; 

\addplot+[ybar] [black, fill=gray] coordinates{
   (1, 643.21549875) 
   (2, 11.745129333333333)
   (3, 39.413000000000004)
   (4, 105.08064999999999)
   (5, 64.463775)
   (6, 6831)
   (7, 5482)
   (8, 80.71266666666666)
   (9, 109.0)
};
\addplot+[ybar] [black, fill= cyan, postaction={nomorepostaction,pattern=north east lines}] coordinates{
   (1, 0.8446334999999999)
   (2, 96.001414)
   (3, 153.64075000000003)
   (4, 101.2945)
   (5, 123.28550000000001)
   (6, 4318.0 )
   (7, 2795.0 )
   (8, 157.0)
   (9, 1518.0)
};
\addplot+[ybar] [black, fill=red] coordinates{
   (1, 3.63659175)
   (2, 0.33340400000000003)
   (3, 14.863999999999987)
   (4, 1.952395)
   (5, 2.157745)
   (6, 0)
   (7, 0)
   (8, 1.9724993333333334)
   (9, 2.3)
};
\addplot+[ybar] [black, fill=green, postaction={nomorepostaction,pattern=north west lines}] coordinates{
   (1, 31.803276)
   (2, 3.920052666666667)
   (3, 26.907999999999998)
   (4, 2.1724550000000002)
   (5, 2.34298)
   (6, 0.0)
   (7, 0.0)
   (8, 101.314834)
   (9, 0)
};
\draw [thick,dash pattern={on 7pt off 2pt on 1pt off 3pt}] (axis cs:1,452) -- (axis cs:9,452) ;

      \nextgroupplot[ 
      ymax=800]

\node[font=\bf,white,above] at (axis cs: 6,0) {\rotatebox{90}{\tiny{8,906 sec}}}; 
\node[blue,above] at (axis cs: 7,0) {\rotatebox{90}{\tiny{SHFL}}};

\addplot+[ybar] [black, fill=gray] coordinates{
   (1, 434.63973519999999) 
   (2, 8.389617666666666)
   (3, 43.301)
   (4, 27.708516666666668)
   (5, 39.024319999999996)
   (6, 5712)
   (7, 0)
   (8, 85.893)
   (9, 64.5)
};
\addplot+[ybar] [black, fill= cyan, postaction={nomorepostaction,pattern=north east lines}] coordinates{
   (1, 2.9366894)
   (2, 44.515558)
   (3, 112.25450000000001)
   (4, 49.9972)
   (5, 76.2194)
   (6, 3194.0 )
   (7, 0)
   (8, 145.0)
   (9, 456.0)
};
\addplot+[ybar] [black, fill=red] coordinates{
   (1, 3.1212565999999997)
   (2, 0.37105000000000005)
   (3, 10.258749999999997)
   (4, 1.4668033333333332)
   (5, 1.5944933333333333)
   (6, 0)
   (7, 0)
   (8, 3.048936)
   (9, 1.3)
};
\addplot+[ybar] [black, fill=green, postaction={nomorepostaction,pattern=north west lines}] coordinates{
   (1, 27.502318799999994)
   (2, 4.057107666666667)
   (3, 28.999999999999996)
   (4, 2.160813333333333)
   (5, 2.49512)
   (6, 0.0)
   (7, 0)
   (8, 108.558064)
   (9, 0)
};
\draw [thick,dash pattern={on 7pt off 2pt on 1pt off 3pt}] (axis cs:1,452) -- (axis cs:9,452) ;

      \nextgroupplot[ 
      ymax=800]

\node[font=\bf,white,above] at (axis cs: 6,0) {\rotatebox{90}{\tiny{6,517 sec}}}; 
\node[blue,above] at (axis cs: 7,0) {\rotatebox{90}{\tiny{SHFL}}};

\addplot+[ybar] [black, fill=gray] coordinates{
   (1, 493.51784300000001) 
   (2, 15.585055333333335)
   (3, 38.55166666666667)
   (4, 57.6684)
   (5, 23.722896666666667)
   (6, 4298)
   (7, 0)
   (8, 133.85333333333332)
   (9, 45.5)
};
\addplot+[ybar] [black, fill= cyan, postaction={nomorepostaction,pattern=north east lines}] coordinates{
   (1, 6.232482666666667)
   (2, 26.469053666666667)
   (3, 70.373)
   (4, 42.763000000000005)
   (5, 48.44906666666666)
   (6, 2219.0 )
   (7, 0)
   (8, 97.0)
   (9, 277.5)
};
\addplot+[ybar] [black, fill=red] coordinates{
   (1, 4.541096666666666)
   (2, 0.35387833333333335)
   (3, 9.294333333333332)
   (4, 2.8388733333333334)
   (5, 2.0901300000000003)
   (6, 0)
   (7, 0)
   (8, 2.8669473333333335)
   (9, 1.25)
};
\addplot+[ybar] [black, fill=green, postaction={nomorepostaction,pattern=north west lines}] coordinates{
   (1, 30.041911)
   (2, 4.925346)
   (3, 37.781)
   (4, 3.3963933333333336)
   (5, 3.737906666666667)
   (6, 0.0)
   (7, 0)
   (8, 51.61305266666667)
   (9, 0)
};

\draw [thick,dash pattern={on 7pt off 2pt on 1pt off 3pt}] (axis cs:1,452) -- (axis cs:9,452) ;

\nextgroupplot[ ymax=800, ylabel={sssp}]

\node[font=\bf,white,above] at (axis cs: 6,0) {\rotatebox{90}{\tiny{17,353 sec}}}; 
\node[font=\bf,white,above] at (axis cs: 7,0) {\rotatebox{90}{\tiny{12,117 sec}}}; 

\addplot+[ybar] [black, fill=gray] coordinates{
   (1, 563.149076666666666) 
   (2, 31.99116166666667)
   (3, 48.63)
   (4, 83.52375)
   (5, 102.31815)
   (6, 11270.0)
   (7, 7873.0)
   (8, 101.40175)
   (9, 177.0)
};
\addplot+[ybar] [black, fill= cyan, postaction={nomorepostaction,pattern=north east lines}] coordinates{
   (1, 10.49)
   (2, 13.654413)
   (3, 33.24666666666667)
   (4, 17.81005)
   (5, 23.16885)
   (6, 6083.0)
   (7, 4244.0)
   (8, 182.0)
   (9, 729.5)
};
\addplot+[ybar] [black, fill=red] coordinates{
   (1, 0.382205)
   (2, 0.37798)
   (3, 11.068333333333326)
   (4, 3.8131065000000035)
   (5, 3.5019834999999944)
   (6, 0.0)
   (7, 0.0)
   (8, 1.9787024999999998)
   (9, 2.5)
};
\addplot+[ybar] [black, fill=green, postaction={nomorepostaction,pattern=north west lines}] coordinates{
   (1, 2.147376333333334)
   (2, 2.3097786666666664)
   (3, 34.721666666666664)
   (4, 4.3530935)
   (5, 2.5110165)
   (6, 1.0)
   (7, 0.0)
   (8, 85.6195475)
   (9, 0)
};

\draw [thick,dash pattern={on 7pt off 2pt on 1pt off 3pt}] (axis cs:1,422) -- (axis cs:9,422) ;

      \nextgroupplot[  
      ymax=800] 
      
\node[font=\bf,white,above] at (axis cs: 6,0) {\rotatebox{90}{\tiny{9,815 sec}}}; 
\node[font=\bf,white,above] at (axis cs: 7,0) {\rotatebox{90}{\tiny{6,965 sec}}}; 

\addplot+[ybar] [black, fill=gray] coordinates{
   (1, 575.96889125) 
   (2, 17.505749333333334)
   (3, 42.033)
   (4, 102.87665)
   (5, 62.7951)
   (6, 5921)
   (7, 4048)
   (8, 79.314)
   (9, 109.0)
};
\addplot+[ybar] [black, fill= cyan, postaction={nomorepostaction,pattern=north east lines}] coordinates{
   (1, 7.34145275)
   (2, 9.408379333333334)
   (3, 30.581999999999997)
   (4, 13.914575)
   (5, 16.68165)
   (6, 3894.0)
   (7, 2917.0)
   (8, 104.66666666666667)
   (9, 262.0)
};
\addplot+[ybar] [black, fill=red] coordinates{
   (1, 0.41311125)
   (2, 0.3196576666666666)
   (3, 10.068750000000001)
   (4, 2.3974125)
   (5, 2.4645799999999998)
   (6,  00)
   (7,  00)
   (8, 1.891532)
   (9, 2)
};
\addplot+[ybar] [black, fill=green, postaction={nomorepostaction,pattern=north west lines}] coordinates{
   (1, 4.27654475)
   (2, 4.099547)
   (3, 43.81625)
   (4, 2.5613625)
   (5, 2.0586699999999998)
   (6, 1.0)
   (7, 2.0)
   (8, 68.46113466666667)
   (9, 0)
};

\draw [thick,dash pattern={on 7pt off 2pt on 1pt off 3pt}] (axis cs:1,422) -- (axis cs:9,422) ;

      \nextgroupplot[  
      ymax=800]

\node[font=\bf,white,above] at (axis cs: 6,0) {\rotatebox{90}{\tiny{8,352 sec}}}; 
\node[blue,above] at (axis cs: 7,0) {\rotatebox{90}{\tiny{SHFL}}};

\addplot+[ybar] [black, fill=gray] coordinates{
   (1, 381.4611718) 
   (2, 13.234490333333332)
   (3, 39.76733333333333)
   (4, 27.106973333333332)
   (5, 39.025133333333336)
   (6, 5381)
   (7, 0)
   (8, 91.6455)
   (9, 64.5)
};
\addplot+[ybar] [black, fill= cyan, postaction={nomorepostaction,pattern=north east lines}] coordinates{
   (1, 6.7166228)
   (2, 7.063118333333333)
   (3, 26.847333333333335)
   (4, 9.050566666666667)
   (5, 12.8531)
   (6, 2971.0)
   (7, 0)
   (8, 82.5)
   (9, 130.5)
};
\addplot+[ybar] [black, fill=red] coordinates{
   (1, .6108366)
   (2, 0.24287666666666663)
   (3, 10.187)
   (4, 1.6465933333333336)
   (5, 1.68006)
   (6,  00)
   (7, 0)
   (8, 3.0587495000000002)
   (9, 1.5)
};
\addplot+[ybar] [black, fill=green, postaction={nomorepostaction,pattern=north west lines}] coordinates{
   (1, 3.211368800000002)
   (2, 4.459514666666666)
   (3, 34.19833333333333)
   (4, 1.8625333333333334)
   (5, 2.4417066666666667)
   (6, .0)
   (7, 0)
   (8, 61.7957505)
   (9, 0)
};

\draw [thick,dash pattern={on 7pt off 2pt on 1pt off 3pt}] (axis cs:1,422) -- (axis cs:9,422) ;

      \nextgroupplot[  
      ymax=800]

\node[font=\bf,white,above] at (axis cs: 6,0) {\rotatebox{90}{\tiny{6,282 sec}}}; 
\node[blue,above] at (axis cs: 7,0) {\rotatebox{90}{\tiny{SHFL}}};

\addplot+[ybar] [black, fill=gray] coordinates{
   (1, 429.203940666666675) 
   (2, 15.824115333333333)
   (3, 37.17433333333333)
   (4, 57.39723333333334)
   (5, 24.27839)
   (6, 4187)
   (7, 0)
   (8, 143.2285)
   (9, 45.5)
};
\addplot+[ybar] [black, fill= cyan, postaction={nomorepostaction,pattern=north east lines}] coordinates{
   (1, 7.36288633333334)
   (2, 8.628659666666666)
   (3, 41.57899999999999)
   (4, 14.718566666666666)
   (5, 15.686)
   (6, 2095.0)
   (7, 0)
   (8, 63.0)
   (9, 102.3)
};
\addplot+[ybar] [black, fill=red] coordinates{
   (1, 0.752490666666667)
   (2, 0.22257933333333335)
   (3, 9.343333333333339)
   (4, 1.9928899999999998)
   (5, 2.0968233333333335)
   (6,  00)
   (7, 0)
   (8, 2.8490064999999998)
   (9, 2)
};
\addplot+[ybar] [black, fill=green, postaction={nomorepostaction,pattern=north west lines}] coordinates{
   (1, 4.014015666666666)
   (2, 5.324645666666666)
   (3, 36.57)
   (4, 4.224643333333333)
   (5, 4.27212)
   (6, .0)
   (7, 0)
   (8, 36.4224935)
   (9, 0)
};
\draw [thick,dash pattern={on 7pt off 2pt on 1pt off 3pt}] (axis cs:1,422) -- (axis cs:9,422) ;

\makeatletter
\end{groupplot}
\end{tikzpicture}

%% file: TekziStackGroupWorkload-pagerank-annotated-Gelly.tex
\makeatletter
\pgfplotsset{
	groupplot xlabel/.initial={}, every tick label/.append style={font=	iny},
	every groupplot x label/.style={
		at={($({group c1r\pgfplots@group@rows.west}|-{group c1r\pgfplots@group@rows.outer south})!0.5!({group c\pgfplots@group@columns r\pgfplots@group@rows.east}|-{group c\pgfplots@group@columns r\pgfplots@group@rows.outer south})$)},
		anchor=north,
	},
	groupplot ylabel/.initial={},
	every groupplot y label/.style={
		rotate=90,
		at={($({group c1r1.north}-|{group c1r1.outer
				west})!0.5!({group c1r\pgfplots@group@rows.south}-|{group c1r\pgfplots@group@rows.outer west})$)},
		anchor=south
	},
	execute at end groupplot/.code={%
		\node [/pgfplots/every groupplot x label]
		{\pgfkeysvalueof{/pgfplots/groupplot xlabel}};  
		\node [/pgfplots/every groupplot y label] 
		{\pgfkeysvalueof{/pgfplots/groupplot ylabel}};
	},
	group/only outer labels/.style =
	{
		group/every plot/.code = {%
			\ifnum\pgfplots@group@current@row=\pgfplots@group@rows\else%
			\pgfkeys{xticklabels = {}, xlabel = {}}\fi%
			\ifnum\pgfplots@group@current@column=1\else%
			\pgfkeys{yticklabels = {}, ylabel = {}}\fi%
		}
	}
}
\def\endpgfplots@environment@groupplot{%
	\endpgfplots@environment@opt%
	\pgfkeys{/pgfplots/execute at end groupplot}%
	\endgroup%
}
\tikzset{nomorepostaction/.code=\let\tikz@postactions\pgfutil@empty}
\makeatother
\begin{tikzpicture}
\pgfplotsset{%
     tiny,samples=10,
	width=3cm,
	height=2cm,
	scale only axis,
	ymajorgrids,
	yminorgrids
}
   \begin{groupplot}[
   clip mode=individual,
         group style = {group size = 4 by 3, 
         	horizontal sep = 5pt,
         	vertical sep = 5pt}, 
         groupplot ylabel={Time ($sec$)},
         group/only outer labels, 
         ybar stacked, /pgf/bar width=1, /pgf/bar shift=0pt,
         area legend,  ymin=0,    xtick=data,xticklabels={BB, BV, G, GL-A-A-T, GL-A-R-T, GL-S-A-I, GL-S-A-T, GL-S-R-I, GL-S-R-T, HD, HL, S, FG},
         scaled ticks=false, xtick style={draw=none},  x tick label style={rotate=45,anchor=east}
]
      \nextgroupplot[ title = {16 machines}, 
      legend style = { column sep = 10pt, legend columns = 4, legend to name = grouplegend,}, ymax=2500, ylabel={World RN}] 
\node[font=\bf,white,above] at (axis cs: 10,0) {\rotatebox{90}{\tiny{19,389 sec}}}; 
\node[font=\bf,white,above] at (axis cs: 11,0) {\rotatebox{90}{\tiny{13,291 sec}}}; 
\addplot+[ybar] [black, fill=gray] coordinates{
   (1, 0)
   (2, 42.876456999999995)
   (3, 156.01000000000002)
   (4, 0)
   (5, 0)
   (6, 0)
   (7, 0)
   (8, 0)
   (9, 0)
   (10, 15308)
   (11, 10774)
   (12, 73.399)
   (13, 100)
};
\node[blue,above] at (axis cs: 1,0) {\rotatebox{90}{\tiny{OOM}}};
\node[blue,above] at (axis cs: 4,0) {\rotatebox{90}{\tiny{OOM}}};
\node[blue,above] at (axis cs: 5,0) {\rotatebox{90}{\tiny{OOM}}};
\node[blue,above] at (axis cs: 6,0) {\rotatebox{90}{\tiny{OOM}}};
\node[blue,above] at (axis cs: 7,0) {\rotatebox{90}{\tiny{OOM}}};
\node[blue,above] at (axis cs: 8,0) {\rotatebox{90}{\tiny{OOM}}};
\node[blue,above] at (axis cs: 9,0) {\rotatebox{90}{\tiny{OOM}}};

\addlegendentry{Load}
\addplot+[ybar] [black, fill= cyan, postaction={nomorepostaction,pattern=north east lines}] coordinates{
   (1, 0)
   (2, 201.99833066666667)
   (3, 1292.52125)
   (4, 0)
   (5, 0)
   (6, 0)
   (7, 0)
   (8, 0)
   (9, 0)
   (10, 4081.0 )
   (11, 2517.0 )
   (12, 1793.0)
   (13, 1912)
};

\addlegendentry{Execute}
\addplot+[ybar] [black, fill=red] coordinates{
   (1, 0)
   (2, 6.570897666666667)
   (3, 123.57475000000001)
   (4, 0)
   (5, 0)
   (6, 0)
   (7, 0)
   (8, 0)
   (9, 0)
   (10, 0)
   (11, 0)
   (12, 104.470985)
   (13, 30)
};
\addlegendentry{Save}
\addplot+[ybar] [black, fill=green, postaction={nomorepostaction,pattern=north west lines}] coordinates{
   (1, 0)
   (2, 16.887648000000002)
   (3, 34.644)
   (4, 0)
   (5, 0)
   (6, 0)
   (7, 0)
   (8, 0)
   (9, 0)
   (10, 0)
   (11, 0.0)
   (12, 321.130015)
   (13, 0)
};
\addlegendentry{Overhead}

\draw [thick,dash pattern={on 7pt off 2pt on 1pt off 3pt}, on layer=groupplot foreground] (axis cs:1,880) -- (axis cs:13,880) ;

      \nextgroupplot[ title = {32 machines}, 
      ymax=2500] 
      
\node[font=\bf,white,above] at (axis cs: 10,0) {\rotatebox{90}{\tiny{10,069 sec}}}; 
\node[font=\bf,white,above] at (axis cs: 11,0) {\rotatebox{90}{\tiny{7,391 sec}}}; 

\addplot+[ybar] [black, fill=gray] coordinates{
   (1, 0)
   (2, 22.454529)
   (3, 87.38775)
   (4, 159.1331)
   (5, 160.21474000000003)
   (6, 124.93927500000001)
   (7, 114.10507142857143)
   (8, 167.01547999999997)
   (9, 160.43184000000002)
   (10, 8013)
   (11, 5819)
   (12, 70.66766666666668)
   (13, 70)
};
\node[blue,above] at (axis cs: 1,0) {\rotatebox{90}{\tiny{OOM}}};

\addplot+[ybar] [black, fill= cyan, postaction={nomorepostaction,pattern=north east lines}] coordinates{
   (1, 0)
   (2, 145.789831)
   (3, 645.5335)
   (4, 1810.5279999999998)
   (5, 1875.4032)
   (6, 277.69055000000003)
   (7, 58.783828571428565)
   (8, 395.01120000000003)
   (9, 79.96302)
   (10, 2056.0 )
   (11, 1572.0 )
   (12, 465.3333333333333)
   (13, 676)
};
\addplot+[ybar] [black, fill=red] coordinates{
   (1, 0)
   (2, 4.084883)
   (3, 78.39699999999998)
   (4, 27.05842)
   (5, 29.2351)
   (6, 31.4574755)
   (7, 31.477395714285713)
   (8, 47.728139999999996)
   (9, 49.55792)
   (10, 0)
   (11, 0)
   (12, 52.10415433333333)
   (13, 20)
};
\addplot+[ybar] [black, fill=green, postaction={nomorepostaction,pattern=north west lines}] coordinates{
   (1, 0)
   (2, 10.670757)
   (3, 38.68175)
   (4, 3.08048)
   (5, 2.54696)
   (6, 3.0376994999999996)
   (7, 3.205132857142857)
   (8, 3.4451799999999997)
   (9, 2.64722)
   (10, 0)
   (11, 0.0)
   (12, 121.89484566666665)
   (13, 0)
};

\draw [thick,dash pattern={on 7pt off 2pt on 1pt off 3pt}] (axis cs:1,880) -- (axis cs:13,880) ;

      \nextgroupplot[ title = {64 machines}, 
      ymax=2500] 
\node[blue,above] at (axis cs: 1,0) {\rotatebox{90}{\tiny{OOM}}};
\node[blue,above] at (axis cs: 11,0) {\rotatebox{90}{\tiny{SHFL}}};
\node[font=\bf,white,above] at (axis cs: 10,0) {\rotatebox{90}{\tiny{6,779 sec}}}; 

\addplot+[ybar] [black, fill=gray] coordinates{
   (1, 0)
   (2, 17.234002666666665)
   (3, 46.42175)
   (4, 74.508065)
   (5, 72.74065)
   (6, 72.41987400000001)
   (7, 71.63462799999999)
   (8, 72.383258)
   (9, 70.482502)
   (10, 5381)
   (11, 0)
   (12, 85.29533333333333)
   (13, 44.3)
};
\addplot+[ybar] [black, fill= cyan, postaction={nomorepostaction,pattern=north east lines}] coordinates{
   (1, 0)
   (2, 128.538271)
   (3, 245.41849999999997)
   (4, 1715.4924999999998)
   (5, 1103.4894)
   (6, 193.4956)
   (7, 45.597680000000004)
   (8, 190.5024)
   (9, 44.47984)
   (10, 1398.0 )
   (11, 0)
   (12, 265.0)
   (13, 497.0)
};
\addplot+[ybar] [black, fill=red] coordinates{
   (1, 0)
   (2, 2.3079673333333335)
   (3, 36.79650000000002)
   (4, 15.013375)
   (5, 14.769300000000001)
   (6, 15.16962)
   (7, 18.987280000000002)
   (8, 14.837700000000002)
   (9, 20.19794)
   (10, 0)
   (11, 0)
   (12, 26.525848666666665)
   (13, 14)
};
\addplot+[ybar] [black, fill=green, postaction={nomorepostaction,pattern=north west lines}] coordinates{
   (1, 0)
   (2, 10.586425666666665)
   (3, 41.61325000000001)
   (4, 2.48606)
   (5, 2.6006500000000004)
   (6, 2.3149059999999997)
   (7, 2.7804119999999997)
   (8, 2.476642)
   (9, 2.439718)
   (10, 0)
   (11, 0)
   (12, 74.84548466666666)
   (13, 0)
};

\draw [thick,dash pattern={on 7pt off 2pt on 1pt off 3pt}] (axis cs:1,880) -- (axis cs:13,880) ;

      \nextgroupplot[ title = {128 machines}, 
      ymax=2500] 
\node[blue,above] at (axis cs: 1,0) {\rotatebox{90}{\tiny{OOM}}};
\node[blue,above] at (axis cs: 4,0) {\rotatebox{90}{\tiny{OOM}}};
\node[blue,above] at (axis cs: 5,0) {\rotatebox{90}{\tiny{OOM}}};
\node[blue,above] at (axis cs: 11,0) {\rotatebox{90}{\tiny{SHFL}}};
\node[font=\bf,white,above] at (axis cs: 10,0) {\rotatebox{90}{\tiny{4,010 sec}}}; 

\addplot+[ybar] [black, fill=gray] coordinates{
   (1, 0)
   (2, 17.952579999999998)
   (3, 21.951000000000004)
   (4, 0)
   (5, 0)
   (6, 48.516466666666666)
   (7, 48.88613333333333)
   (8, 37.50737333333333)
   (9, 37.855513333333334)
   (10, 3119)
   (11, 0)
   (12, 117.429)
   (13, 42.3)
};
\addplot+[ybar] [black, fill= cyan, postaction={nomorepostaction,pattern=north east lines}] coordinates{
   (1, 0)
   (2, 114.8165255)
   (3, 137.86800000000002)
   (4, 0)
   (5, 0)
   (6, 82.54623333333335)
   (7, 70.05566666666667)
   (8, 103.16699999999999)
   (9, 73.9499)
   (10, 891.0 )
   (11, 0)
   (12, 289.0)
   (13, 309)
};
\addplot+[ybar] [black, fill=red] coordinates{
   (1, 0)
   (2, 0.99722)
   (3, 23.45599999999999)
   (4, 0)
   (5, 0)
   (6, 8.44037)
   (7, 8.889416666666667)
   (8, 8.489743333333335)
   (9, 8.850746666666668)
   (10, 0)
   (11, 0)
   (12, 24.5972675)
   (13, 8)
};
\addplot+[ybar] [black, fill=green, postaction={nomorepostaction,pattern=north west lines}] coordinates{
   (1, 0)
   (2, 8.2336745)
   (3, 38.391666666666666)
   (4, 0)
   (5, 0)
   (6, 3.4969300000000003)
   (7, 4.168783333333334)
   (8, 4.169216666666667)
   (9, 4.34384)
   (10, 0)
   (11, 0)
   (12, 94.4737325)
   (13, 0)
};

\draw [thick,dash pattern={on 7pt off 2pt on 1pt off 3pt}] (axis cs:1,880) -- (axis cs:13,880) ;

      \nextgroupplot[ 
      ymax=2600, ylabel={UK0705}] 
      
\node[blue,above] at (axis cs: 5,0) {\rotatebox{90}{\tiny{OOM}}};
\node[blue,above] at (axis cs: 8,0) {\rotatebox{90}{\tiny{OOM}}};
\node[blue,above] at (axis cs: 9,0) {\rotatebox{90}{\tiny{OOM}}};
\node[font=\bf,white,above] at (axis cs: 10,0) {\rotatebox{90}{\tiny{83,897 sec}}}; 
\node[font=\bf,white,above] at (axis cs: 11,0) {\rotatebox{90}{\tiny{37,231 sec}}}; 

\addplot+[ybar] [black, fill=gray] coordinates{
   (1, 769.2)
   (2, 52.58375433333333)
   (3, 50.675666666666665)
   (4, 167.86294000000004)
   (5, 0)
   (6, 161.16060000000002)
   (7, 159.34713333333335)
   (8, 0)
   (9, 0)
   (10, 76816.0)
   (11, 32753.0)
   (12, 239.87766666666667)
   (13, 367.5)
};
\addplot+[ybar] [black, fill= cyan, postaction={nomorepostaction,pattern=north east lines}] coordinates{
   (1, 793.475068)
   (2, 795.3170373333334)
   (3, 2316.6593333333335)
   (4, 1069.2074)
   (5, 0)
   (6, 1118.42)
   (7, 164.27366666666668)
   (8, 0)
   (9, 0)
   (10, 7081.0)
   (11, 4478.0)
   (12, 984.6666666666666)
   (13, 11612)
};
\addplot+[ybar] [black, fill=red] coordinates{
   (1, 82.9)
   (2, 1.4515416666666667)
   (3, 26.55200000000016)
   (4, 8.743565600000002)
   (5, 0)
   (6, 10.022412000000001)
   (7, 9.091480333333356)
   (8, 0)
   (9, 0)
   (10, 0.0)
   (11, 0.0)
   (12, 32.60937066666666)
   (13, 8)
};
\addplot+[ybar] [black, fill=green, postaction={nomorepostaction,pattern=north west lines}] coordinates{
   (1, 0)
   (2, 3.981)
   (3, 34.779666666666664)
   (4, 3.5860944000000003)
   (5, 0)
   (6, 3.0636546666666664)
   (7, 5.954386333333333)
   (8, 0)
   (9, 0)
   (10, 2.0)
   (11, 1.0)
   (12, 89.17962933333332)
   (13, 0)
};

\draw [thick,dash pattern={on 7pt off 2pt on 1pt off 3pt}] (axis cs:1,720) -- (axis cs:13,720) ;

      \nextgroupplot[ 
      ymax=2600] 
      
\node[font=\bf,white,above] at (axis cs: 10,0) {\rotatebox{90}{\tiny{44,969 sec}}}; 
\node[font=\bf,white,above] at (axis cs: 11,0) {\rotatebox{90}{\tiny{21,410 sec}}}; 

\addplot+[ybar] [black, fill=gray] coordinates{
   (1, 615.4291538888889) 
   (2, 30.787167)
   (3, 29.558333333333334)
   (4, 153.4014)
   (5, 159.344425)
   (6, 153.06422500000002)
   (7, 152.60017499999998)
   (8, 158.417475)
   (9, 157.2022)
   (10, 40120)
   (11, 18830)
   (12, 156.037)
   (13, 188)
};
\addplot+[ybar] [black, fill= cyan, postaction={nomorepostaction,pattern=north east lines}] coordinates{
   (1, 608.0155966666666)
   (2, 500.781189)
   (3, 1178.0516666666665)
   (4, 393.912)
   (5, 756.46875)
   (6, 389.047)
   (7, 87.425)
   (8, 1294.0875)
   (9, 200.75225)
   (10, 4849.0 )
   (11, 2580.0 )
   (12, 707.0)
   (13, 3227)
};
\addplot+[ybar] [black, fill=red] coordinates{
   (1, 1.526111555555555)
   (2, 0.8560043333333333)
   (3, 18.194999999999908)
   (4, 5.7274175)
   (5, 7.657855)
   (6, 9.255165)
   (7, 5.640040000000001)
   (8, 5.2674025)
   (9, 5.1494925)
   (10, 0)
   (11, 0)
   (12, 11.118300666666668)
   (13, 5)
};
\addplot+[ybar] [black, fill=green, postaction={nomorepostaction,pattern=north west lines}] coordinates{
   (1, 20.695804555555554)
   (2, 2.908973)
   (3, 41.195)
   (4, 2.4591825)
   (5, 3.02897)
   (6, 2.63361)
   (7, 3.084785)
   (8, 2.9776225)
   (9, 2.8960575)
   (10, 0.0)
   (11, 0.0)
   (12, 61.17803266666667)
   (13, 0)
};

\draw [thick,dash pattern={on 7pt off 2pt on 1pt off 3pt}] (axis cs:1,720) -- (axis cs:13,720) ;

      \nextgroupplot[ 
      ymax=2600] 

\node[font=\bf,white,above] at (axis cs: 10,0) {\rotatebox{90}{\tiny{35,343 sec}}}; 
\node[blue,above] at (axis cs: 11,0) {\rotatebox{90}{\tiny{SHFL}}};

\addplot+[ybar] [black, fill=gray] coordinates{
   (1, 445.51358242857145) 
   (2, 20.073345333333332)
   (3, 26.93133333333333)
   (4, 58.004999999999995)
   (5, 106.1057)
   (6, 57.1222)
   (7, 58.95193333333333)
   (8, 104.49526666666667)
   (9, 102.31850000000001)
   (10, 31842)
   (11, 0)
   (12, 117.81299999999999)
   (13, 133.3)
};
\addplot+[ybar] [black, fill= cyan, postaction={nomorepostaction,pattern=north east lines}] coordinates{
   (1, 418.7402248571429)
   (2, 372.78366933333336)
   (3, 646.5736666666668)
   (4, 231.38899999999998)
   (5, 413.21166666666664)
   (6, 451.297)
   (7, 92.39913333333334)
   (8, 841.34)
   (9, 153.28)
   (10, 3501.0 )
   (11, 0)
   (12, 452.3333333333333)
   (13, 2013.33)
};
\addplot+[ybar] [black, fill=red] coordinates{
   (1, 2.1399642857142855)
   (2, 0.5492076666666666)
   (3, 13.722666666666484)
   (4, 3.090616666666667)
   (5, 3.454446666666667)
   (6, 3.0401799999999994)
   (7, 3.1367933333333333)
   (8, 3.29045)
   (9, 3.390363333333333)
   (10, 0)
   (11, 0)
   (12, 8.004201333333333)
   (13, 3.3)
};
\addplot+[ybar] [black, fill=green, postaction={nomorepostaction,pattern=north west lines}] coordinates{
   (1, 11.891942714285713)
   (2, 6.927111)
   (3, 53.439)
   (4, 2.5153833333333333)
   (5, 2.56152)
   (6, 2.54062)
   (7, 2.51214)
   (8, 2.874283333333333)
   (9, 2.34447)
   (10, 0.0)
   (11, 0)
   (12, 55.516132)
   (13, 0)
};

\draw [thick,dash pattern={on 7pt off 2pt on 1pt off 3pt}] (axis cs:1,720) -- (axis cs:13,720) ;

      \nextgroupplot[ 
      ymax=2600] 
      
\node[font=\bf,white,above] at (axis cs: 10,0) {\rotatebox{90}{\tiny{20,935 sec}}}; 
\node[blue,above] at (axis cs: 11,0) {\rotatebox{90}{\tiny{SHFL}}};

\addplot+[ybar] [black, fill=gray] coordinates{
   (1, 458.653153) 
   (2, 13.018840333333335)
   (3, 14.657000000000002)
   (4, 113.62219999999998)
   (5, 58.14591333333334)
   (6, 112.93256666666667)
   (7, 111.78643333333332)
   (8, 59.74666333333332)
   (9, 58.40942999999999)
   (10, 18042)
   (11, 0)
   (12, 152.793)
   (13, 93.0)
};
\addplot+[ybar] [black, fill= cyan, postaction={nomorepostaction,pattern=north east lines}] coordinates{
   (1, 337.146361)
   (2, 290.9291876666667)
   (3, 351.92799999999994)
   (4, 522.938)
   (5, 495.61433333333326)
   (6, 143.29833333333332)
   (7, 188.191)
   (8, 528.8543333333333)
   (9, 231.62199999999999)
   (10, 2893.0 )
   (11, 0)
   (12, 598.3333333333334)
   (13, 1272.0)
};
\addplot+[ybar] [black, fill=red] coordinates{
   (1, 0.4688840000000001)
   (2, 0.32955399999999996)
   (3, 11.480000000000032)
   (4, 2.1029366666666665)
   (5, 2.4839566666666664)
   (6, 2.230223333333333)
   (7, 2.78669)
   (8, 2.4832733333333334)
   (9, 2.8952799999999996)
   (10, 0)
   (11, 0)
   (12, 13.001398666666667)
   (13, 3)
};
\addplot+[ybar] [black, fill=green, postaction={nomorepostaction,pattern=north west lines}] coordinates{
   (1, 0.06493533333333329)
   (2, 5.722418)
   (3, 36.935)
   (4, 4.6701966666666666)
   (5, 5.755796666666666)
   (6, 4.205543333333334)
   (7, 3.9025433333333335)
   (8, 4.582396666666667)
   (9, 4.07329)
   (10, 0.0)
   (11, 0)
   (12, 84.53893466666666)
   (13, 0)
};

\draw [thick,dash pattern={on 7pt off 2pt on 1pt off 3pt}] (axis cs:1,720) -- (axis cs:13,720) ;

      \nextgroupplot[ 
      ymax=2500, ylabel={Twitter}] 
      
\node[font=\bf,white,above] at (axis cs: 10,0) {\rotatebox{90}{\tiny{23,239 sec}}}; 
\node[font=\bf,white,above] at (axis cs: 11,0) {\rotatebox{90}{\tiny{17,397 sec}}}; 

\addplot+[ybar] [black, fill=gray] coordinates{
   (1, 710.7718575) 
   (2, 23.868796)
   (3, 48.7435)
   (4, 87.834075)
   (5, 111.3424)
   (6, 88.030525)
   (7, 87.80665)
   (8, 108.29599999999999)
   (9, 113.45112499999999)
   (10, 15706.0)
   (11, 11974.0)
   (12, 105.94825)
   (13, 177.0)
   
};
\addplot+[ybar] [black, fill= cyan, postaction={nomorepostaction,pattern=north east lines}] coordinates{
   (1, 250.241928)
   (2, 232.79251733333334)
   (3, 471.42650000000003)
   (4, 503.49675)
   (5, 983.601)
   (6, 414.4375)
   (7, 314.63924999999995)
   (8, 574.30375)
   (9, 465.22225000000003)
   (10, 7533.0)
   (11, 5423.0)
   (12, 1052.25)
   (13, 1650)
};
\addplot+[ybar] [black, fill=red] coordinates{
   (1, 7.132467500000001)
   (2, 0.705127)
   (3, 15.836499999999901)
   (4, 4.134636749999998)
   (5, 3.9055917500000135)
   (6, 4.520130749999984)
   (7, 4.0574634999999795)
   (8, 4.0202755)
   (9, 4.197302500000002)
   (10, 0.0)
   (11, 0.0)
   (12, 17.40264775)
   (13, 2)
};
\addplot+[ybar] [black, fill=green, postaction={nomorepostaction,pattern=north west lines}] coordinates{
   (1, 13.8272425)
   (2, 2.6335596666666667)
   (3, 38.4935)
   (4, 2.2845382499999998)
   (5, 2.1510082500000003)
   (6, 2.7618442500000002)
   (7, 4.4966365)
   (8, 2.8799745)
   (9, 2.8793225)
   (10, 2.0)
   (11, 2.0)
   (12, 113.39910225)
   (13, 0)
};

\draw [thick,dash pattern={on 7pt off 2pt on 1pt off 3pt}] (axis cs:1,490) -- (axis cs:13,490) ;

      \nextgroupplot[  
      ymax=2500] 
      
\node[font=\bf,white,above] at (axis cs: 10,0) {\rotatebox{90}{\tiny{14,127 sec}}}; 
\node[font=\bf,white,above] at (axis cs: 11,0) {\rotatebox{90}{\tiny{11,014 sec}}}; 

\addplot+[ybar] [black, fill=gray] coordinates{
   (1, 647.7406784285714) 
   (2, 12.363546)
   (3, 33.096333333333334)
   (4, 102.78793333333333)
   (5, 65.81556666666667)
   (6, 115.46188)
   (7, 104.70033333333333)
   (8, 65.09338)
   (9, 64.58872333333333)
   (10, 8659.0)
   (11, 6830.0)
   (12, 84.00500000000001)
   (13, 109.0)
};
\addplot+[ybar] [black, fill= cyan, postaction={nomorepostaction,pattern=north east lines}] coordinates{
   (1, 180.0561952857145)
   (2, 151.69337699999997)
   (3, 233.11966666666663)
   (4, 449.6235)
   (5, 553.2728333333333)
   (6, 277.90700000000004)
   (7, 235.63116666666667)
   (8, 345.7946)
   (9, 286.5486666666667)
   (10, 5468.0 )
   (11, 4184.0 )
   (12, 640.3333333333334)
   (13, 998)
};
\addplot+[ybar] [black, fill=red] coordinates{
   (1, 2.5385127142857145)
   (2, 0.39395766666666665)
   (3, 12.581666666666672)
   (4, 2.775348333333333)
   (5, 2.748555)
   (6, 3.249684)
   (7, 2.703046666666667)
   (8, 3.014064)
   (9, 2.6332083333333336)
   (10, 0)
   (11, 0)
   (12, 9.924600666666668)
   (13, 2)
};
\addplot+[ybar] [black, fill=green, postaction={nomorepostaction,pattern=north west lines}] coordinates{
   (1, 10.09318499999998)
   (2, 3.882452666666667)
   (3, 40.869)
   (4, 2.3132183333333334)
   (5, 2.3297116666666664)
   (6, 2.981436)
   (7, 2.63212)
   (8, 2.6979560000000005)
   (9, 2.2294016666666665)
   (10, 0.0)
   (11, 0.0)
   (12, 70.73706600000001)
   (13, 0)
};

\draw [thick,dash pattern={on 7pt off 2pt on 1pt off 3pt}] (axis cs:1,490) -- (axis cs:13,490) ;

      \nextgroupplot[  
      ymax=2500] 
      
\node[font=\bf,white,above] at (axis cs: 10,0) {\rotatebox{90}{\tiny{10,857 sec}}}; 
\node[blue,above] at (axis cs: 11,0) {\rotatebox{90}{\tiny{SHFL}}};

\addplot+[ybar] [black, fill=gray] coordinates{
   (1, 479.04227275) 
   (2, 8.684023999999999)
   (3, 47.304500000000004)
   (4, 27.13949)
   (5, 39.16191666666667)
   (6, 26.947706666666665)
   (7, 27.328793333333333)
   (8, 38.90477666666667)
   (9, 38.515209999999996)
   (10, 6638)
   (11, 0)
   (12, 88.255)
   (13, 64.5)
};
\addplot+[ybar] [black, fill= cyan, postaction={nomorepostaction,pattern=north east lines}] coordinates{
   (1, 104.271396)
   (2, 93.451431)
   (3, 158.73825)
   (4, 228.30933333333334)
   (5, 362.6023333333333)
   (6, 140.78366666666668)
   (7, 110.19)
   (8, 219.23)
   (9, 194.27966666666666)
   (10, 4219.0 )
   (11, 0)
   (12, 574.0)
   (13, 708.5)
};
\addplot+[ybar] [black, fill=red] coordinates{
   (1, 6.16385075)
   (2, 0.29879066666666665)
   (3, 10.985750000000017)
   (4, 2.191613333333333)
   (5, 1.83362)
   (6, 1.7566199999999998)
   (7, 1.8400766666666666)
   (8, 1.8168266666666666)
   (9, 1.81686)
   (10, 0)
   (11, 0)
   (12, 4.0644725)
   (13, 1.5)
};
\addplot+[ybar] [black, fill=green, postaction={nomorepostaction,pattern=north west lines}] coordinates{
   (1, 10.0224805)
   (2, 8.232421)
   (3, 42.2215)
   (4, 2.6928966666666665)
   (5, 2.735463333333333)
   (6, 1.8453400000000002)
   (7, 3.6411300000000004)
   (8, 2.0483966666666666)
   (9, 2.05493)
   (10, 0.0)
   (11, 0)
   (12, 60.1805275)
   (13, 0)
};

\draw [thick,dash pattern={on 7pt off 2pt on 1pt off 3pt}] (axis cs:1,490) -- (axis cs:13,490) ;

      \nextgroupplot[  
      ymax=2500] 
      
\node[font=\bf,white,above] at (axis cs: 10,0) {\rotatebox{90}{\tiny{8,189 sec}}}; 
\node[blue,above] at (axis cs: 11,0) {\rotatebox{90}{\tiny{SHFL}}};

\addplot+[ybar] [black, fill=gray] coordinates{
   (1, 459.34048200000004) 
   (2, 14.830438000000001)
   (3, 48.029666666666664)
   (4, 57.72013333333334)
   (5, 24.78904)
   (6, 57.21983333333333)
   (7, 57.348733333333335)
   (8, 24.55156)
   (9, 24.539925)
   (10, 4671)
   (11, 0)
   (12, 155.801)
   (13, 45.5)
};
\addplot+[ybar] [black, fill= cyan, postaction={nomorepostaction,pattern=north east lines}] coordinates{
   (1, 80.62864133333333)
   (2, 78.158876)
   (3, 121.31600000000002)
   (4, 333.99033333333335)
   (5, 411.2486666666667)
   (6, 105.966)
   (7, 102.79133333333334)
   (8, 131.61800000000002)
   (9, 126.219)
   (10, 3518.0 )
   (11, 0)
   (12, 285.0)
   (13, 503.0)
};
\addplot+[ybar] [black, fill=red] coordinates{
   (1, 0.414048)
   (2, 0.21000300000000002)
   (3, 10.271333333333315)
   (4, 1.91648)
   (5, 2.1427333333333336)
   (6, 1.6429)
   (7, 1.74053)
   (8, 1.7429499999999998)
   (9, 1.69318)
   (10, 0)
   (11, 0)
   (12, 4.144316)
   (13, 2)
};
\addplot+[ybar] [black, fill=green, postaction={nomorepostaction,pattern=north west lines}] coordinates{
   (1, 12.28349533333333)
   (2, 5.467349666666666)
   (3, 44.383)
   (4, 4.373053333333334)
   (5, 3.486226666666667)
   (6, 3.5046)
   (7, 6.452736666666667)
   (8, 3.4208233333333333)
   (9, 3.547895)
   (10, 0.0)
   (11, 0)
   (12, 40.054684)
   (13, 0)
};
\draw [thick,dash pattern={on 7pt off 2pt on 1pt off 3pt}] (axis cs:1,490) -- (axis cs:13,490) ;

\makeatletter
\end{groupplot}
\node at ($(group c2r1) + (0,2.1cm)$) {\ref{grouplegend}};
\end{tikzpicture}

%% file: TekziStackGroupWorkload-khop-Gelly.tex
\makeatletter
\pgfplotsset{
	groupplot xlabel/.initial={}, every tick label/.append style={font=	iny},
	every groupplot x label/.style={
		at={($({group c1r\pgfplots@group@rows.west}|-{group c1r\pgfplots@group@rows.outer south})!0.5!({group c\pgfplots@group@columns r\pgfplots@group@rows.east}|-{group c\pgfplots@group@columns r\pgfplots@group@rows.outer south})$)},
		anchor=north,
	},
	groupplot ylabel/.initial={},
	every groupplot y label/.style={
		rotate=90,
		at={($({group c1r1.north}-|{group c1r1.outer
				west})!0.5!({group c1r\pgfplots@group@rows.south}-|{group c1r\pgfplots@group@rows.outer west})$)},
		anchor=south
	},
	execute at end groupplot/.code={%
		\node [/pgfplots/every groupplot x label]
		{\pgfkeysvalueof{/pgfplots/groupplot xlabel}};  
		\node [/pgfplots/every groupplot y label] 
		{\pgfkeysvalueof{/pgfplots/groupplot ylabel}};
	},
	group/only outer labels/.style =
	{
		group/every plot/.code = {%
			\ifnum\pgfplots@group@current@row=\pgfplots@group@rows\else%
			\pgfkeys{xticklabels = {}, xlabel = {}}\fi%
			\ifnum\pgfplots@group@current@column=1\else%
			\pgfkeys{yticklabels = {}, ylabel = {}}\fi%
		}
	}
}
\def\endpgfplots@environment@groupplot{%
	\endpgfplots@environment@opt%
	\pgfkeys{/pgfplots/execute at end groupplot}%
	\endgroup%
}
\tikzset{nomorepostaction/.code=\let\tikz@postactions\pgfutil@empty}
\makeatother
\begin{tikzpicture}
\pgfplotsset{%
     tiny,samples=10,
	width=3cm,
	height=2cm,
	scale only axis,
	ymajorgrids,
	yminorgrids
}
   \begin{groupplot}[
         group style = {group size = 4 by 3, 
         	horizontal sep = 5pt,
         	vertical sep = 5pt}, 
         groupplot ylabel={Time ($sec$)},
         groupplot xlabel={Systems},
         group/only outer labels, 
         ybar stacked, /pgf/bar width=1, /pgf/bar shift=0pt,
         area legend,  ymin=0,    xtick=data,xticklabels={BB, BV, G, GL-S-A-I, GL-S-R-I, HD, HL, S, FG},
         scaled ticks=false, xtick style={draw=none},  x tick label style={rotate=45,anchor=east}
]
      \nextgroupplot[ title = {16 machines}, 
      legend style = { column sep = 10pt, legend columns = 4, legend to name = grouplegend,}, ymax=700, ylabel={World RN}] 
\addplot+[ybar] [black, fill=gray] coordinates{
   (1, 0)
   (2, 53.08883599999999)
   (3, 146.822)
   (4, 0)
   (5, 0)
   (6, 1034)
   (7, 978)
   (8, 73.198)
   (9, 0)
};
\node[blue,above] at (axis cs: 1,0) {\rotatebox{90}{\tiny{OOM}}};
\node[blue,above] at (axis cs: 4,0) {\rotatebox{90}{\tiny{OOM}}};
\node[blue,above] at (axis cs: 5,0) {\rotatebox{90}{\tiny{OOM}}};

\addlegendentry{Load}
\addplot+[ybar] [black, fill= cyan, postaction={nomorepostaction,pattern=north east lines}] coordinates{
   (1, 0)
   (2, 1.2661616666666664)
   (3, 208.38433333333333)
   (4, 0)
   (5, 0)
   (6, 711.0 )
   (7, 418.0 )
   (8, 195.0)
   (9, 0)
};
\addlegendentry{Execute}
\addplot+[ybar] [black, fill=red] coordinates{
   (1, 0)
   (2, 0.26296933333333333)
   (3, 64.22133333333333)
   (4, 0)
   (5, 0)
   (6, 0)
   (7, 0)
   (8, 1.02995)
   (9, 0)
};
\addlegendentry{Save}
\addplot+[ybar] [black, fill=green, postaction={nomorepostaction,pattern=north west lines}] coordinates{
   (1, 0)
   (2, 7.715366333333333)
   (3, 33.57233333333334)
   (4, 0)
   (5, 0)
   (6, 0)
   (7, 0.0)
   (8, 227.77205)
   (9, 0)
};
\addlegendentry{Overhead}
      \nextgroupplot[ title = {32 machines}, 
      ymax=700] 
\addplot+[ybar] [black, fill=gray] coordinates{
   (1, 0)
   (2, 26.815555)
   (3, 85.63100000000001)
   (4, 136.0741)
   (5, 154.19743333333335)
   (6, 834)
   (7, 618)
   (8, 67.48566666666666)
   (9, 0)
};
\addplot+[ybar] [black, fill= cyan, postaction={nomorepostaction,pattern=north east lines}] coordinates{
   (1, 0)
   (2, 0.907568)
   (3, 93.88833333333334)
   (4, 1.52931)
   (5, 1.7999966666666667)
   (6, 513.0 )
   (7, 301.0 )
   (8, 20.0)
   (9, 0)
};
\node[blue,above] at (axis cs: 1,0) {\rotatebox{90}{\tiny{OOM}}};

\addplot+[ybar] [black, fill=red] coordinates{
   (1, 0)
   (2, 0.22325)
   (3, 30.24666666666666)
   (4, 15.725123000000002)
   (5, 17.04977)
   (6, 0)
   (7, 0)
   (8, 0.881764)
   (9, 0)
};
\addplot+[ybar] [black, fill=green, postaction={nomorepostaction,pattern=north west lines}] coordinates{
   (1, 0)
   (2, 7.053627)
   (3, 34.234)
   (4, 4.671467)
   (5, 4.286133333333333)
   (6, 0)
   (7, 0.0)
   (8, 87.29923600000001)
   (9, 0)
};
      \nextgroupplot[ title = {64 machines}, 
      ymax=700] 
\addplot+[ybar] [black, fill=gray] coordinates{
   (1, 0)
   (2, 22.678047333333335)
   (3, 48.57983333333333)
   (4, 69.384138)
   (5, 69.01358200000001)
   (6, 619)
   (7, 401)
   (8, 81.10533333333333)
   (9, 50.333333333333336)
};
\addplot+[ybar] [black, fill= cyan, postaction={nomorepostaction,pattern=north east lines}] coordinates{
   (1, 0)
   (2, 0.7557823333333333)
   (3, 61.791333333333334)
   (4, 1.3799999999999997)
   (5, 1.399996)
   (6, 340.0 )
   (7, 182.0 )
   (8, 20.666666666666668)
   (9, 78.33333333333333)
};
\addplot+[ybar] [black, fill=red] coordinates{
   (1, 0)
   (2, 0.159484)
   (3, 22.21749999999999)
   (4, 9.053749999999999)
   (5, 8.360838)
   (6, 0)
   (7, 0)
   (8, 1.2421533333333332)
   (9, 1.3333333333333333)
};
\addplot+[ybar] [black, fill=green, postaction={nomorepostaction,pattern=north west lines}] coordinates{
   (1, 0)
   (2, 6.740019666666666)
   (3, 34.41133333333333)
   (4, 2.3821120000000002)
   (5, 2.625584)
   (6, 0)
   (7, 0.0)
   (8, 49.65251333333333)
   (9, 0.0)
};
      \nextgroupplot[ title = {128 machines}, 
      ymax=700] 
\addplot+[ybar] [black, fill=gray] coordinates{
   (1, 0)
   (2, 21.502929)
   (3, 33.068)
   (4, 53.425399999999996)
   (5, 37.30454)
   (6, 357)
   (7, 198)
   (8, 81.0945)
   (9, 42.666666666666664)
};
\node[blue,above] at (axis cs: 1,0) {\rotatebox{90}{\tiny{OOM}}};

\addplot+[ybar] [black, fill= cyan, postaction={nomorepostaction,pattern=north east lines}] coordinates{
   (1, 0)
   (2, 1.044827)
   (3, 30.567999999999998)
   (4, 1.4)
   (5, 1.5)
   (6, 215.0 )
   (7, 109.0 )
   (8, 11.5)
   (9, 61.666666666666664)
};
\addplot+[ybar] [black, fill=red] coordinates{
   (1, 0)
   (2, 0.146725)
   (3, 17.336)
   (4, 4.824199999999999)
   (5, 4.59719)
   (6, 0)
   (7, 0)
   (8, 1.1710034999999999)
   (9, 1.3333333333333333)
};
\addplot+[ybar] [black, fill=green, postaction={nomorepostaction,pattern=north west lines}] coordinates{
   (1, 0)
   (2, 5.305519)
   (3, 35.028)
   (4, 4.3504)
   (5, 4.59827)
   (6, 0)
   (7, 0.0)
   (8, 110.7344965)
   (9, 0.0)
};
      \nextgroupplot[ 
      ymax=700, ylabel={UK0705}] 
\addplot+[ybar] [black, fill=gray] coordinates{
   (1, 0)
   (2, 82.40520933333333)
   (3, 59.57966666666667)
   (4, 159.61929999999998)
   (5, 0)
   (6, 1920.0)
   (7, 1526.0)
   (8, 245.4558)
   (9, 0)
};
\node[blue,above] at (axis cs: 1,0) {\rotatebox{90}{\tiny{OOM}}};
\node[blue,above] at (axis cs: 5,0) {\rotatebox{90}{\tiny{OOM}}};

\addplot+[ybar] [black, fill= cyan, postaction={nomorepostaction,pattern=north east lines}] coordinates{
   (1, 0)
   (2, 0.3427733333333333)
   (3, 10.762666666666666)
   (4, 1.4250866666666668)
   (5, 0)
   (6, 871.0)
   (7, 498.0)
   (8, 12.8)
   (9, 0)
};
\addplot+[ybar] [black, fill=red] coordinates{
   (1, 0)
   (2, 0.3672843333333333)
   (3, 14.403333333333327)
   (4, 5.578797000000004)
   (5, 0)
   (6, 0.0)
   (7, 0.0)
   (8, 1.6049156)
   (9, 0)
};
\addplot+[ybar] [black, fill=green, postaction={nomorepostaction,pattern=north west lines}] coordinates{
   (1, 0)
   (2, 2.5513996666666667)
   (3, 35.58766666666667)
   (4, 3.7101496666666667)
   (5, 0)
   (6, 2.0)
   (7, 1.0)
   (8, 72.93928439999999)
   (9, 0)
};
      \nextgroupplot[ 
      ymax=700] 
\addplot+[ybar] [black, fill=gray] coordinates{
   (1, 34.194781666666664)
   (2, 44.66458866666667)
   (3, 29.71666666666667)
   (4, 155.05622499999998)
   (5, 155.313025)
   (6, 1610.0)
   (7, 1246.0)
   (8, 150.81133333333335)
   (9, 222.0)
};
\addplot+[ybar] [black, fill= cyan, postaction={nomorepostaction,pattern=north east lines}] coordinates{
   (1, 0.8516813333333334)
   (2, 0.3777166666666667)
   (3, 9.595666666666666)
   (4, 1.0904125)
   (5, 1.5311074999999998)
   (6, 662.0)
   (7, 378.0)
   (8, 11.0)
   (9, 692.5)
};
\addplot+[ybar] [black, fill=red] coordinates{
   (1, 0.8285696666666666)
   (2, 0.35519233333333333)
   (3, 12.966666666666663)
   (4, 3.486653249999997)
   (5, 3.463965250000001)
   (6, 0.0)
   (7, 0.0)
   (8, 1.5291086666666664)
   (9, 0.5)
};
\addplot+[ybar] [black, fill=green, postaction={nomorepostaction,pattern=north west lines}] coordinates{
   (1, 8.124967333333332)
   (2, 2.6025023333333333)
   (3, 33.38766666666667)
   (4, 3.8667092499999995)
   (5, 2.94190225)
   (6, 2.0)
   (7, 1.0)
   (8, 48.99289133333334)
   (9, 0.0)
};
      \nextgroupplot[ 
      ymax=700] 
\addplot+[ybar] [black, fill=gray] coordinates{
   (1, 23.60583516666667)
   (2, 30.58771133333333)
   (3, 17.022666666666666)
   (4, 57.5288)
   (5, 100.26846666666667)
   (6, 1223.0)
   (7, 819.0)
   (8, 113.81466666666667)
   (9, 133.33333333333334)
};
\addplot+[ybar] [black, fill= cyan, postaction={nomorepostaction,pattern=north east lines}] coordinates{
   (1, 2.7966496666666667)
   (2, 0.4522353333333333)
   (3, 5.423333333333333)
   (4, 1.2)
   (5, 1.3666666666666665)
   (6, 418.0)
   (7, 218.0)
   (8, 7.666666666666667)
   (9, 194.33333333333334)
};
\addplot+[ybar] [black, fill=red] coordinates{
   (1, 1.0316885)
   (2, 0.24839766666666666)
   (3, 12.502333333333333)
   (4, 2.2903133333333336)
   (5, 2.3804200000000004)
   (6, 0.0)
   (7, 0.0)
   (8, 1.771869)
   (9, 1.0)
};
\addplot+[ybar] [black, fill=green, postaction={nomorepostaction,pattern=north west lines}] coordinates{
   (1, 6.399159999999999)
   (2, 3.711655666666667)
   (3, 43.05166666666667)
   (4, 2.9808866666666667)
   (5, 2.984446666666667)
   (6, 2.0)
   (7, 1.0)
   (8, 43.080131)
   (9, 0.0)
};
      \nextgroupplot[ 
      ymax=700] 
\addplot+[ybar] [black, fill=gray] coordinates{
   (1, 18.196128666666667)
   (2, 16.075389333333334)
   (3, 13.852666666666666)
   (4, 113.6961)
   (5, 58.16684)
   (6, 921.0)
   (7, 653.0)
   (8, 141.26850000000002)
   (9, 95.4)
};
\addplot+[ybar] [black, fill= cyan, postaction={nomorepostaction,pattern=north east lines}] coordinates{
   (1, 6.896332)
   (2, 0.9674346666666667)
   (3, 11.938333333333334)
   (4, 1.6463433333333333)
   (5, 1.8716266666666665)
   (6, 279.0)
   (7, 158.0)
   (8, 17.0)
   (9, 121.8)
};
\addplot+[ybar] [black, fill=red] coordinates{
   (1, 4.625543)
   (2, 0.18307833333333334)
   (3, 13.408666666666667)
   (4, 2.5614399999999997)
   (5, 1.7015233333333333)
   (6, 0.0)
   (7, 0.0)
   (8, 3.7695755)
   (9, 1.2)
};
\addplot+[ybar] [black, fill=green, postaction={nomorepostaction,pattern=north west lines}] coordinates{
   (1, 5.948663)
   (2, 5.440764333333334)
   (3, 44.467000000000006)
   (4, 4.42945)
   (5, 4.26001)
   (6, 2.0)
   (7, 1.0)
   (8, 52.961924499999995)
   (9, 0.0)
};
      \nextgroupplot[ 
      ymax=500, ylabel={Twitter}] 
\addplot+[ybar] [black, fill=gray] coordinates{
   (1, 50.521629)
   (2, 33.49368666666667)
   (3, 42.6)
   (4, 83.69456666666667)
   (5, 101.04786666666666)
   (6, 763.0)
   (7, 565.0)
   (8, 101.83999999999999)
   (9, 0)
};
\addplot+[ybar] [black, fill= cyan, postaction={nomorepostaction,pattern=north east lines}] coordinates{
   (1, 1.246264)
   (2, 0.17703233333333332)
   (3, 4.675333333333334)
   (4, 1.1582733333333335)
   (5, 1.2132533333333333)
   (6, 401.0)
   (7, 279.0)
   (8, 19.666666666666668)
   (9, 0)
};
\addplot+[ybar] [black, fill=red] coordinates{
   (1, 0.679032)
   (2, 0.2055543333333333)
   (3, 10.342999999999998)
   (4, 2.567682000000002)
   (5, 2.587617000000001)
   (6, 0.0)
   (7, 0.0)
   (8, 0.9533906666666665)
   (9, 0)
};
\addplot+[ybar] [black, fill=green, postaction={nomorepostaction,pattern=north west lines}] coordinates{
   (1, 29.553075)
   (2, 4.45706)
   (3, 36.714999999999996)
   (4, 3.579478)
   (5, 2.484596333333333)
   (6, 2.0)
   (7, 2.0)
   (8, 99.87327599999999)
   (9, 0)
};
      \nextgroupplot[  
      ymax=500] 
\addplot+[ybar] [black, fill=gray] coordinates{
   (1, 70.3401685)
   (2, 16.67743533333333)
   (3, 42.93142857142857)
   (4, 101.26638333333335)
   (5, 64.50108333333334)
   (6, 563)
   (7, 511)
   (8, 78.902)
   (9, 109.0)
};
\addplot+[ybar] [black, fill= cyan, postaction={nomorepostaction,pattern=north east lines}] coordinates{
   (1, 2.15432075)
   (2, 0.228992)
   (3, 7.282000000000001)
   (4, 1.0333333333333332)
   (5, 1.1666666666666667)
   (6, 370.0)
   (7, 262.0)
   (8, 16.333333333333332)
   (9, 143.0)
};
\addplot+[ybar] [black, fill=red] coordinates{
   (1, 0.9247397500000001)
   (2, 0.16912733333333332)
   (3, 10.178142857142856)
   (4, 2.04538)
   (5, 2.0039166666666666)
   (6,  0.0)
   (7,  0.0)
   (8, 4.056975666666666)
   (9, 1.0)
};
\addplot+[ybar] [black, fill=green, postaction={nomorepostaction,pattern=north west lines}] coordinates{
   (1, 37.830771)
   (2, 3.9244453333333333)
   (3, 50.608428571428576)
   (4, 2.488236666666667)
   (5, 2.1616666666666666)
   (6, 2.0)
   (7, 1.0)
   (8, 58.70769099999999)
   (9, 0.0)
};
      \nextgroupplot[  
      ymax=500] 
\addplot+[ybar] [black, fill=gray] coordinates{
   (1, 43.434589200000005)
   (2, 13.315258666666665)
   (3, 40.518750000000004)
   (4, 28.10118)
   (5, 39.68831)
   (6, 452)
   (7, 401)
   (8, 86.674)
   (9, 66.0)
};
\addplot+[ybar] [black, fill= cyan, postaction={nomorepostaction,pattern=north east lines}] coordinates{
   (1, 3.2346156)
   (2, 0.3958333333333333)
   (3, 27.659499999999998)
   (4, 1.1666666666666667)
   (5, 1.2333333333333334)
   (6, 318.0)
   (7, 241.0)
   (8, 8.5)
   (9, 86.0)
};
\addplot+[ybar] [black, fill=red] coordinates{
   (1, 1.0522358)
   (2, 0.16637433333333332)
   (3, 10.388500000000004)
   (4, 2.7830600000000003)
   (5, 1.5859766666666666)
   (6,  0.0)
   (7,  0.0)
   (8, 2.129772)
   (9, 1.0)
};
\addplot+[ybar] [black, fill=green, postaction={nomorepostaction,pattern=north west lines}] coordinates{
   (1, 26.8785594)
   (2, 4.1225336666666665)
   (3, 37.68325)
   (4, 2.2824266666666664)
   (5, 2.159046666666667)
   (6, .0)
   (7, 0)
   (8, 57.196228000000005)
   (9, 0.0)
};
      \nextgroupplot[  
      ymax=500] 
\addplot+[ybar] [black, fill=gray] coordinates{
   (1, 50.333451333333336)
   (2, 15.512934666666666)
   (3, 37.36533333333333)
   (4, 57.43666666666667)
   (5, 23.268906666666666)
   (6, 403)
   (7, 378)
   (8, 133.96699999999998)
   (9, 44.75)
};
\addplot+[ybar] [black, fill= cyan, postaction={nomorepostaction,pattern=north east lines}] coordinates{
   (1, 9.216966333333332)
   (2, 1.0954756666666665)
   (3, 14.134666666666668)
   (4, 1.6580366666666666)
   (5, 1.7023133333333333)
   (6, 345.0)
   (7, 244.0)
   (8, 9.0)
   (9, 65.0)
};
\addplot+[ybar] [black, fill=red] coordinates{
   (1, 2.6021959999999997)
   (2, 0.16714933333333334)
   (3, 11.749333333333334)
   (4, 1.9364133333333333)
   (5, 1.74069)
   (6,  0.0)
   (7,  0.0)
   (8, 2.371534666666667)
   (9, 1.5)
};
\addplot+[ybar] [black, fill=green, postaction={nomorepostaction,pattern=north west lines}] coordinates{
   (1, 38.84738633333333)
   (2, 4.891107000000001)
   (3, 37.41733333333333)
   (4, 3.302216666666667)
   (5, 3.621423333333333)
   (6, .0)
   (7, 0)
   (8, 40.328132000000004)
   (9, 0.0)
};
\makeatletter
\end{groupplot}
\node at ($(group c2r1) + (0,2.1cm)$) {\ref{grouplegend}};
\end{tikzpicture}

%% file: TekziStackGroupWorkload-sssp-Gelly.tex
\makeatletter
\pgfplotsset{
	groupplot xlabel/.initial={}, every tick label/.append style={font=	iny},
	every groupplot x label/.style={
		at={($({group c1r\pgfplots@group@rows.west}|-{group c1r\pgfplots@group@rows.outer south})!0.5!({group c\pgfplots@group@columns r\pgfplots@group@rows.east}|-{group c\pgfplots@group@columns r\pgfplots@group@rows.outer south})$)},
		anchor=north,
	},
	groupplot ylabel/.initial={},
	every groupplot y label/.style={
		rotate=90,
		at={($({group c1r1.north}-|{group c1r1.outer
				west})!0.5!({group c1r\pgfplots@group@rows.south}-|{group c1r\pgfplots@group@rows.outer west})$)},
		anchor=south
	},
	execute at end groupplot/.code={%
		\node [/pgfplots/every groupplot x label]
		{\pgfkeysvalueof{/pgfplots/groupplot xlabel}};  
		\node [/pgfplots/every groupplot y label] 
		{\pgfkeysvalueof{/pgfplots/groupplot ylabel}};
	},
	group/only outer labels/.style =
	{
		group/every plot/.code = {%
			\ifnum\pgfplots@group@current@row=\pgfplots@group@rows\else%
			\pgfkeys{xticklabels = {}, xlabel = {}}\fi%
			\ifnum\pgfplots@group@current@column=1\else%
			\pgfkeys{yticklabels = {}, ylabel = {}}\fi%
		}
	}
}
\def\endpgfplots@environment@groupplot{%
	\endpgfplots@environment@opt%
	\pgfkeys{/pgfplots/execute at end groupplot}%
	\endgroup%
}
\tikzset{nomorepostaction/.code=\let\tikz@postactions\pgfutil@empty}
\makeatother
\begin{tikzpicture}
\pgfplotsset{%
     tiny,samples=10,
	width=3cm,
	height=2cm,
	scale only axis,
	ymajorgrids,
	yminorgrids
}
   \begin{groupplot}[
         group style = {group size = 4 by 3, 
         	horizontal sep = 5pt,
         	vertical sep = 5pt}, 
         groupplot ylabel={Time ($sec$)},
         groupplot xlabel={Systems},
         group/only outer labels, 
         ybar stacked, /pgf/bar width=1, /pgf/bar shift=0pt,
         area legend,  ymin=0,    xtick=data,xticklabels={BB, BV, G, GL-S-A-I, GL-S-R-I, HD, HL, S, FG},
         scaled ticks=false, xtick style={draw=none},  x tick label style={rotate=45,anchor=east}
]
      \nextgroupplot[ title = {16 machines}, 
      legend style = { column sep = 10pt, legend columns = 4, legend to name = grouplegend,}, ymax=71600, ylabel={World RN}] 
\addplot+[ybar] [black, fill=gray] coordinates{
   (1, 0)
   (2, 51.517534)
   (3, 0)
   (4, 0)
   (5, 0)
   (6, 0)
   (7, 0)
   (8, 0)
   (9, 0)
};
\node[blue,above] at (axis cs: 1,0) {\rotatebox{90}{\tiny{OOM}}};
\node[blue,above] at (axis cs: 3,0) {\rotatebox{90}{\tiny{OOM}}};
\node[blue,above] at (axis cs: 4,0) {\rotatebox{90}{\tiny{OOM}}};
\node[blue,above] at (axis cs: 5,0) {\rotatebox{90}{\tiny{OOM}}};
\node[blue,above] at (axis cs: 6,0) {\rotatebox{90}{\tiny{TO}}};
\node[blue,above] at (axis cs: 7,0) {\rotatebox{90}{\tiny{TO}}};
\node[blue,above] at (axis cs: 8,0) {\rotatebox{90}{\tiny{OOM}}};
\node[blue,above] at (axis cs: 9,0) {\rotatebox{90}{\tiny{OOM}}};

\addlegendentry{Load}
\addplot+[ybar] [black, fill= cyan, postaction={nomorepostaction,pattern=north east lines}] coordinates{
   (1, 0)
   (2, 11234.800596)
   (3, 0)
   (4, 0)
   (5, 0)
   (6, 0)
   (7, 0)
   (8, 0)
   (9, 0)
};
\addlegendentry{Execute}
\addplot+[ybar] [black, fill=red] coordinates{
   (1, 0)
   (2, 1.733151)
   (3, 0)
   (4, 0)
   (5, 0)
   (6, 0)
   (7, 0)
   (8, 0)
   (9, 0)
};
\addlegendentry{Save}
\addplot+[ybar] [black, fill=green, postaction={nomorepostaction,pattern=north west lines}] coordinates{
   (1, 0)
   (2, 6.948719)
   (3, 0)
   (4, 0)
   (5, 0)
   (6, 0)
   (7, 0)
   (8, 0)
   (9, 0)
};
\addlegendentry{Overhead}
      \nextgroupplot[ title = {32 machines}, 
      ymax=71600] 
\addplot+[ybar] [black, fill=gray] coordinates{
   (1, 0)
   (2, 25.376101)
   (3, 0)
   (4, 114.7777)
   (5, 0)
   (6, 0)
   (7, 0)
   (8, 0)
   (9, 0)
};
\node[blue,above] at (axis cs: 1,0) {\rotatebox{90}{\tiny{OOM}}};
\node[blue,above] at (axis cs: 3,0) {\rotatebox{90}{\tiny{OOM}}};
\node[blue,above] at (axis cs: 5,0) {\rotatebox{90}{\tiny{OOM}}};
\node[blue,above] at (axis cs: 6,0) {\rotatebox{90}{\tiny{TO}}};
\node[blue,above] at (axis cs: 7,0) {\rotatebox{90}{\tiny{TO}}};
\node[blue,above] at (axis cs: 8,0) {\rotatebox{90}{\tiny{OOM}}};
\node[blue,above] at (axis cs: 9,0) {\rotatebox{90}{\tiny{OOM}}};

\addplot+[ybar] [black, fill= cyan, postaction={nomorepostaction,pattern=north east lines}] coordinates{
   (1, 0)
   (2, 7576.420914)
   (3, 0)
   (4, 13784.9)
   (5, 0)
   (6, 0)
   (7, 0)
   (8, 0)
   (9, 0)
};
\addplot+[ybar] [black, fill=red] coordinates{
   (1, 0)
   (2, 0.979776)
   (3, 0)
   (4, 18.2502)
   (5, 0)
   (6, 0)
   (7, 0)
   (8, 0)
   (9, 0)
};
\addplot+[ybar] [black, fill=green, postaction={nomorepostaction,pattern=north west lines}] coordinates{
   (1, 0)
   (2, 10.223209)
   (3, 0)
   (4, 3.0721)
   (5, 0)
   (6, 0)
   (7, 0)
   (8, 0)
   (9, 0)
};
      \nextgroupplot[ title = {64 machines}, 
      ymax=71600] 
\addplot+[ybar] [black, fill=gray] coordinates{
   (1, 0)
   (2, 21.968046666666666)
   (3, 37.966)
   (4, 74.26422)
   (5, 70.90457)
   (6, 0)
   (7, 0)
   (8, 0)
   (9, 0)
};
\node[blue,above] at (axis cs: 1,0) {\rotatebox{90}{\tiny{OOM}}};
\node[blue,above] at (axis cs: 6,0) {\rotatebox{90}{\tiny{TO}}};
\node[blue,above] at (axis cs: 7,0) {\rotatebox{90}{\tiny{SHUF}}};
\node[blue,above] at (axis cs: 8,0) {\rotatebox{90}{\tiny{OOM}}};
\node[blue,above] at (axis cs: 9,0) {\rotatebox{90}{\tiny{TO}}};

\addplot+[ybar] [black, fill= cyan, postaction={nomorepostaction,pattern=north east lines}] coordinates{
   (1, 0)
   (2, 6446.395362666667)
   (3, 86036.72600000049)
   (4, 17801.3)
   (5, 18615.5)
   (6, 0)
   (7, 0)
   (8, 0)
   (9, 0)
};
\addplot+[ybar] [black, fill=red] coordinates{
   (1, 0)
   (2, 0.5311493333333334)
   (3, 24.204999999508843)
   (4, 9.68944)
   (5, 12.7519)
   (6, 0)
   (7, 0)
   (8, 0)
   (9, 0)
};
\addplot+[ybar] [black, fill=green, postaction={nomorepostaction,pattern=north west lines}] coordinates{
   (1, 0)
   (2, 5.772108)
   (3, 31.103)
   (4, 1.74634)
   (5, 2.84353)
   (6, 0)
   (7, 0)
   (8, 0)
   (9, 0)
};
      \nextgroupplot[ title = {128 machines}, 
      ymax=71600] 
\addplot+[ybar] [black, fill=gray] coordinates{
   (1, 0)
   (2, 21.7625265)
   (3, 21.547)
   (4, 49.5877)
   (5, 37.76253)
   (6, 0)
   (7, 0)
   (8, 0)
   (9, 45.0)
};
\node[blue,above] at (axis cs: 1,0) {\rotatebox{90}{\tiny{OOM}}};
\node[blue,above] at (axis cs: 6,0) {\rotatebox{90}{\tiny{TO}}};
\node[blue,above] at (axis cs: 7,0) {\rotatebox{90}{\tiny{SHUF}}};
\node[blue,above] at (axis cs: 8,0) {\rotatebox{90}{\tiny{OOM}}};
\addplot+[ybar] [black, fill= cyan, postaction={nomorepostaction,pattern=north east lines}] coordinates{
   (1, 0)
   (2, 11065.383858500001)
   (3, 42634.9360000001)
   (4, 42344.9)
   (5, 42348.9)
   (6, 0)
   (7, 0)
   (8, 0)
   (9, 84574.0)
};
\addplot+[ybar] [black, fill=red] coordinates{
   (1, 0)
   (2, 0.38576900000000003)
   (3, 19.52699999989837)
   (4, 6.13449)
   (5, 6.20012)
   (6, 0)
   (7, 0)
   (8, 0.0)
   (9, 4.0)
};
\addplot+[ybar] [black, fill=green, postaction={nomorepostaction,pattern=north west lines}] coordinates{
   (1, 0)
   (2, 8.467846)
   (3, 41.99)
   (4, 5.37781)
   (5, 6.13735)
   (6, 0)
   (7, 0)
   (8, 0)
   (9, 0.0)
};
      \nextgroupplot[ 
      ymax=1000, ylabel={UK0705}] 
\addplot+[ybar] [black, fill=gray] coordinates{
   (1, 0)
   (2, 85.70174533333333)
   (3, 53.312666666666665)
   (4, 177.6935)
   (5, 0)
   (6, 41644.0)
   (7, 25246.0)
   (8, 245.4645)
   (9, 381.0)
};
\node[blue,above] at (axis cs: 1,0) {\rotatebox{90}{\tiny{OOM}}};
\node[blue,above] at (axis cs: 5,0) {\rotatebox{90}{\tiny{OOM}}};

\addplot+[ybar] [black, fill= cyan, postaction={nomorepostaction,pattern=north east lines}] coordinates{
   (1, 0)
   (2, 30.750173999999998)
   (3, 194.442)
   (4, 58.3951)
   (5, 0)
   (6, 40703.0)
   (7, 20703.0)
   (8, 1602.0)
   (9, 8017.0)
};
\addplot+[ybar] [black, fill=red] coordinates{
   (1, 0)
   (2, 1.2171786666666666)
   (3, 14.891000000000012)
   (4, 8.75466)
   (5, 0)
   (6, 0.0)
   (7, 0.0)
   (8, 48.6211775)
   (9, 5.0)
};
\addplot+[ybar] [black, fill=green, postaction={nomorepostaction,pattern=north west lines}] coordinates{
   (1, 0)
   (2, 4.664235333333333)
   (3, 34.687666666666665)
   (4, 4.65674)
   (5, 0)
   (6, 0.0)
   (7, 0.0)
   (8, 311.9143225)
   (9, 0.0)
};
      \nextgroupplot[ 
      ymax=1000] 
\addplot+[ybar] [black, fill=gray] coordinates{
   (1, 33.79293542857143)
   (2, 47.68844266666667)
   (3, 30.13766666666667)
   (4, 150.30616666666666)
   (5, 160.18173333333334)
   (6, 33824.0)
   (7, 20284.0)
   (8, 152.6903333333333)
   (9, 188.0)
};

\addplot+[ybar] [black, fill= cyan, postaction={nomorepostaction,pattern=north east lines}] coordinates{
   (1, 18.447706285714283)
   (2, 21.80083433333333)
   (3, 155.295)
   (4, 36.22253333333333)
   (5, 66.304)
   (6, 37923.0)
   (7, 15826.0)
   (8, 2087.3333333333335)
   (9, 947.0)
};
\addplot+[ybar] [black, fill=red] coordinates{
   (1, 1.824533142857143)
   (2, 0.667582)
   (3, 11.851333333333335)
   (4, 5.113036666666667)
   (5, 5.350273333333334)
   (6, 0.0)
   (7, 0.0)
   (8, 55.04292366666667)
   (9, 3.0)
};
\addplot+[ybar] [black, fill=green, postaction={nomorepostaction,pattern=north west lines}] coordinates{
   (1, 8.22053942857143)
   (2, 4.176474333333333)
   (3, 36.716)
   (4, 3.691596666666667)
   (5, 3.1639933333333334)
   (6, 0.0)
   (7, 0.0)
   (8, 109.266743)
   (9, 0.0)
};
      \nextgroupplot[ 
      ymax=1000] 
\addplot+[ybar] [black, fill=gray] coordinates{
   (1, 23.767031)
   (2, 29.268938666666667)
   (3, 23.868666666666666)
   (4, 58.732499999999995)
   (5, 100.8051)
   (6, 24276.0)
   (7, 0)
   (8, 123.35966666666667)
   (9, 143.0)
};
\node[blue,above] at (axis cs: 7,0) {\rotatebox{90}{\tiny{SHUF}}};

\addplot+[ybar] [black, fill= cyan, postaction={nomorepostaction,pattern=north east lines}] coordinates{
   (1, 19.262377166666663)
   (2, 23.208988666666666)
   (3, 126.63900000000001)
   (4, 40.82333333333334)
   (5, 55.35296666666667)
   (6, 22100.0)
   (7, 0)
   (8, 3833.0)
   (9, 378.0)
};
\addplot+[ybar] [black, fill=red] coordinates{
   (1, 1.9125173333333334)
   (2, 0.4456693333333333)
   (3, 10.663666666666662)
   (4, 3.1498933333333334)
   (5, 3.5432966666666665)
   (6, 0.0)
   (7, 0)
   (8, 93.17010066666667)
   (9, 2.5)
};
\addplot+[ybar] [black, fill=green, postaction={nomorepostaction,pattern=north west lines}] coordinates{
   (1, 6.391407833333332)
   (2, 3.0764033333333334)
   (3, 35.162000000000006)
   (4, 2.6276066666666664)
   (5, 2.6319700000000004)
   (6, 0.0)
   (7, 0)
   (8, 113.803566)
   (9, 0.0)
};
      \nextgroupplot[ 
      ymax=1000] 
\addplot+[ybar] [black, fill=gray] coordinates{
   (1, 18.232899333333336)
   (2, 15.708079333333336)
   (3, 13.085)
   (4, 113.41286666666667)
   (5, 59.082629999999995)
   (6, 18234.0)
   (7, 0)
   (8, 183.516)
   (9, 88.66666666666667)
};
\node[blue,above] at (axis cs: 7,0) {\rotatebox{90}{\tiny{SHUF}}};
\addplot+[ybar] [black, fill= cyan, postaction={nomorepostaction,pattern=north east lines}] coordinates{
   (1, 29.344478)
   (2, 41.383217333333334)
   (3, 108.73299999999999)
   (4, 110.94033333333334)
   (5, 116.96233333333333)
   (6, 15071.0)
   (7, 0)
   (8, 2535.0)
   (9, 232.66666666666666)
};
\addplot+[ybar] [black, fill=red] coordinates{
   (1, 2.770113333333333)
   (2, 0.3268376666666667)
   (3, 9.934333333333356)
   (4, 2.9694866666666666)
   (5, 3.0371)
   (6, 0.0)
   (7, 0)
   (8, 0.0)
   (9, 2.0)
};
\addplot+[ybar] [black, fill=green, postaction={nomorepostaction,pattern=north west lines}] coordinates{
   (1, 5.9858426666666675)
   (2, 5.248532333333333)
   (3, 37.580999999999996)
   (4, 4.343980000000001)
   (5, 3.584603333333334)
   (6, 0.0)
   (7, 0)
   (8, -2718.516)
   (9, 0.0)
};
      \nextgroupplot[ 
      ymax=700, ylabel={Twitter}] 
\addplot+[ybar] [black, fill=gray] coordinates{
   (1, 51.149076666666666)
   (2, 31.99116166666667)
   (3, 48.63)
   (4, 83.52375)
   (5, 102.31815)
   (6, 11270.0)
   (7, 7873.0)
   (8, 101.40175)
   (9, 183.5)
};

\addplot+[ybar] [black, fill= cyan, postaction={nomorepostaction,pattern=north east lines}] coordinates{
   (1, 91.65467533333333)
   (2, 13.654413)
   (3, 33.24666666666667)
   (4, 17.81005)
   (5, 23.16885)
   (6, 6083.0)
   (7, 4244.0)
   (8, 182.0)
   (9, 729.5)
};
\addplot+[ybar] [black, fill=red] coordinates{
   (1, 4.382205)
   (2, 0.37798)
   (3, 11.068333333333326)
   (4, 3.8131065000000035)
   (5, 3.5019834999999944)
   (6, 0.0)
   (7, 0.0)
   (8, 1.9787024999999998)
   (9, 2.5)
};
\addplot+[ybar] [black, fill=green, postaction={nomorepostaction,pattern=north west lines}] coordinates{
   (1, 32.147376333333334)
   (2, 2.3097786666666664)
   (3, 34.721666666666664)
   (4, 4.3530935)
   (5, 2.5110165)
   (6, 1.0)
   (7, 0.0)
   (8, 85.6195475)
   (9, 0.0)
};
      \nextgroupplot[  
      ymax=700] 
\addplot+[ybar] [black, fill=gray] coordinates{
   (1, 70.96889125)
   (2, 17.505749333333334)
   (3, 42.033)
   (4, 102.87665)
   (5, 62.7951)
   (6, 5921)
   (7, 4048)
   (8, 79.314)
   (9, 113.0)
};
\addplot+[ybar] [black, fill= cyan, postaction={nomorepostaction,pattern=north east lines}] coordinates{
   (1, 147.34145275)
   (2, 9.408379333333334)
   (3, 30.581999999999997)
   (4, 13.914575)
   (5, 16.68165)
   (6, 3894.0)
   (7, 2917.0)
   (8, 104.66666666666667)
   (9, 262.0)
};
\addplot+[ybar] [black, fill=red] coordinates{
   (1, 6.41311125)
   (2, 0.3196576666666666)
   (3, 10.068750000000001)
   (4, 2.3974125)
   (5, 2.4645799999999998)
   (6,  00)
   (7,  00)
   (8, 1.891532)
   (9, 2.0)
};
\addplot+[ybar] [black, fill=green, postaction={nomorepostaction,pattern=north west lines}] coordinates{
   (1, 41.27654475)
   (2, 4.099547)
   (3, 43.81625)
   (4, 2.5613625)
   (5, 2.0586699999999998)
   (6, 1.0)
   (7, 2.0)
   (8, 68.46113466666667)
   (9, 0.0)
};
      \nextgroupplot[  
      ymax=700] 
\addplot+[ybar] [black, fill=gray] coordinates{
   (1, 43.4611718)
   (2, 13.234490333333332)
   (3, 39.76733333333333)
   (4, 27.106973333333332)
   (5, 39.025133333333336)
   (6, 5381)
   (7, 0)
   (8, 91.6455)
   (9, 64.5)
};
\node[blue,above] at (axis cs: 7,0) {\rotatebox{90}{\tiny{SHUF}}};
\addplot+[ybar] [black, fill= cyan, postaction={nomorepostaction,pattern=north east lines}] coordinates{
   (1, 171.7166228)
   (2, 7.063118333333333)
   (3, 26.847333333333335)
   (4, 9.050566666666667)
   (5, 12.8531)
   (6, 2971.0)
   (7, 0)
   (8, 82.5)
   (9, 130.5)
};
\addplot+[ybar] [black, fill=red] coordinates{
   (1, 5.6108366)
   (2, 0.24287666666666663)
   (3, 10.187)
   (4, 1.6465933333333336)
   (5, 1.68006)
   (6, 0)
   (7, 0)
   (8, 3.0587495000000002)
   (9, 1.5)
};
\addplot+[ybar] [black, fill=green, postaction={nomorepostaction,pattern=north west lines}] coordinates{
   (1, 30.211368800000002)
   (2, 4.459514666666666)
   (3, 34.19833333333333)
   (4, 1.8625333333333334)
   (5, 2.4417066666666667)
   (6, 0.0)
   (7, 0)
   (8, 61.7957505)
   (9, 0.0)
};
      \nextgroupplot[  
      ymax=700] 
\addplot+[ybar] [black, fill=gray] coordinates{
   (1, 50.203940666666675)
   (2, 15.824115333333333)
   (3, 37.17433333333333)
   (4, 57.39723333333334)
   (5, 24.27839)
   (6, 4187)
   (7, 0)
   (8, 143.2285)
   (9, 45.5)
};
\node[blue,above] at (axis cs: 7,0) {\rotatebox{90}{\tiny{SHUF}}};
\addplot+[ybar] [black, fill= cyan, postaction={nomorepostaction,pattern=north east lines}] coordinates{
   (1, 213.36288633333334)
   (2, 8.628659666666666)
   (3, 41.57899999999999)
   (4, 14.718566666666666)
   (5, 15.686)
   (6, 2095.0)
   (7, 0)
   (8, 63.0)
   (9, 102.33333333333333)
};
\addplot+[ybar] [black, fill=red] coordinates{
   (1, 6.752490666666667)
   (2, 0.22257933333333335)
   (3, 9.343333333333339)
   (4, 1.9928899999999998)
   (5, 2.0968233333333335)
   (6,  00)
   (7, 0)
   (8, 2.8490064999999998)
   (9, 2.0)
};
\addplot+[ybar] [black, fill=green, postaction={nomorepostaction,pattern=north west lines}] coordinates{
   (1, 44.014015666666666)
   (2, 5.324645666666666)
   (3, 36.57)
   (4, 4.224643333333333)
   (5, 4.27212)
   (6, .0)
   (7, 0)
   (8, 36.4224935)
   (9, 0.0)
};
\makeatletter
\end{groupplot}
\node at ($(group c2r1) + (0,2.1cm)$) {\ref{grouplegend}};
\end{tikzpicture}

%% file: TekziStackGroupWorkload-wcc-Gelly.tex
\makeatletter
\pgfplotsset{
	groupplot xlabel/.initial={}, every tick label/.append style={font=	iny},
	every groupplot x label/.style={
		at={($({group c1r\pgfplots@group@rows.west}|-{group c1r\pgfplots@group@rows.outer south})!0.5!({group c\pgfplots@group@columns r\pgfplots@group@rows.east}|-{group c\pgfplots@group@columns r\pgfplots@group@rows.outer south})$)},
		anchor=north,
	},
	groupplot ylabel/.initial={},
	every groupplot y label/.style={
		rotate=90,
		at={($({group c1r1.north}-|{group c1r1.outer
				west})!0.5!({group c1r\pgfplots@group@rows.south}-|{group c1r\pgfplots@group@rows.outer west})$)},
		anchor=south
	},
	execute at end groupplot/.code={%
		\node [/pgfplots/every groupplot x label]
		{\pgfkeysvalueof{/pgfplots/groupplot xlabel}};  
		\node [/pgfplots/every groupplot y label] 
		{\pgfkeysvalueof{/pgfplots/groupplot ylabel}};
	},
	group/only outer labels/.style =
	{
		group/every plot/.code = {%
			\ifnum\pgfplots@group@current@row=\pgfplots@group@rows\else%
			\pgfkeys{xticklabels = {}, xlabel = {}}\fi%
			\ifnum\pgfplots@group@current@column=1\else%
			\pgfkeys{yticklabels = {}, ylabel = {}}\fi%
		}
	}
}
\def\endpgfplots@environment@groupplot{%
	\endpgfplots@environment@opt%
	\pgfkeys{/pgfplots/execute at end groupplot}%
	\endgroup%
}
\tikzset{nomorepostaction/.code=\let\tikz@postactions\pgfutil@empty}
\makeatother
\begin{tikzpicture}
\pgfplotsset{%
     tiny,samples=10,
	width=3cm,
	height=2cm,
	scale only axis,
	ymajorgrids,
	yminorgrids
}
   \begin{groupplot}[
         group style = {group size = 4 by 3, 
         	horizontal sep = 5pt,
         	vertical sep = 5pt}, 
         groupplot ylabel={Time ($sec$)},
         groupplot xlabel={Systems},
         group/only outer labels, 
         ybar stacked, /pgf/bar width=1, /pgf/bar shift=0pt,
         area legend,  ymin=0,    xtick=data,xticklabels={BB, BV, G, GL-S-A-I, GL-S-R-I, HD, HL, S, FG},
         scaled ticks=false, xtick style={draw=none},  x tick label style={rotate=45,anchor=east}
]
      \nextgroupplot[ title = {16 machines}, 
      legend style = { column sep = 10pt, legend columns = 4, legend to name = grouplegend,}, ymax=85200, ylabel={World RN}] 
\addplot+[ybar] [black, fill=gray] coordinates{
   (1, 0)
   (2, 45.5773525)
   (3, 0)
   (4, 0)
   (5, 0)
   (6, 0)
   (7, 0)
   (8, 0)
   (9, 0)
};
\node[blue,above] at (axis cs: 1,0) {\rotatebox{90}{\tiny{OOM}}};
\node[blue,above] at (axis cs: 3,0) {\rotatebox{90}{\tiny{OOM}}};
\node[blue,above] at (axis cs: 4,0) {\rotatebox{90}{\tiny{OOM}}};
\node[blue,above] at (axis cs: 5,0) {\rotatebox{90}{\tiny{OOM}}};
\node[blue,above] at (axis cs: 6,0) {\rotatebox{90}{\tiny{TO}}};
\node[blue,above] at (axis cs: 7,0) {\rotatebox{90}{\tiny{TO}}};
\node[blue,above] at (axis cs: 8,0) {\rotatebox{90}{\tiny{OOM}}};
\node[blue,above] at (axis cs: 9,0) {\rotatebox{90}{\tiny{OOM}}};

\addlegendentry{Load}
\addplot+[ybar] [black, fill= cyan, postaction={nomorepostaction,pattern=north east lines}] coordinates{
   (1, 0)
   (2, 19769.162374500003)
   (3, 0)
   (4, 0)
   (5, 0)
   (6, 0)
   (7, 0)
   (8, 0)
   (9, 0)
};
\addlegendentry{Execute}
\addplot+[ybar] [black, fill=red] coordinates{
   (1, 0)
   (2, 4.7513784999999995)
   (3, 0)
   (4, 0)
   (5, 0)
   (6, 0)
   (7, 0)
   (8, 0)
   (9, 0)
};
\addlegendentry{Save}
\addplot+[ybar] [black, fill=green, postaction={nomorepostaction,pattern=north west lines}] coordinates{
   (1, 0)
   (2, 12.0088945)
   (3, 0)
   (4, 0)
   (5, 0)
   (6, 0)
   (7, 0)
   (8, 0)
   (9, 0)
};
\addlegendentry{Overhead}
      \nextgroupplot[ title = {32 machines}, 
      ymax=85200] 
\addplot+[ybar] [black, fill=gray] coordinates{
   (1, 0)
   (2, 21.694711)
   (3, 0)
   (4, 129.49136666666666)
   (5, 142.63799999999998)
   (6, 0)
   (7, 0)
   (8, 0)
   (9, 0)
};
\node[blue,above] at (axis cs: 1,0) {\rotatebox{90}{\tiny{OOM}}};
\node[blue,above] at (axis cs: 3,0) {\rotatebox{90}{\tiny{OOM}}};
\node[blue,above] at (axis cs: 6,0) {\rotatebox{90}{\tiny{TO}}};
\node[blue,above] at (axis cs: 7,0) {\rotatebox{90}{\tiny{TO}}};
\node[blue,above] at (axis cs: 8,0) {\rotatebox{90}{\tiny{OOM}}};
\node[blue,above] at (axis cs: 9,0) {\rotatebox{90}{\tiny{OOM}}};

\addplot+[ybar] [black, fill= cyan, postaction={nomorepostaction,pattern=north east lines}] coordinates{
   (1, 0)
   (2, 13991.736826)
   (3, 0)
   (4, 32390.93333333333)
   (5, 31459.199999999997)
   (6, 0)
   (7, 0)
   (8, 0)
   (9, 0)
};
\addplot+[ybar] [black, fill=red] coordinates{
   (1, 0)
   (2, 2.473514)
   (3, 0)
   (4, 21.546000000000003)
   (5, 20.9999)
   (6, 0)
   (7, 0)
   (8, 0)
   (9, 0)
};
\addplot+[ybar] [black, fill=green, postaction={nomorepostaction,pattern=north west lines}] coordinates{
   (1, 0)
   (2, 8.094949)
   (3, 0)
   (4, 2.695966666668667)
   (5, 2.6620999999999997)
   (6, 0)
   (7, 0)
   (8, 0)
   (9, 0)
};
      \nextgroupplot[ title = {64 machines}, 
      ymax=85200] 
\addplot+[ybar] [black, fill=gray] coordinates{
   (1, 0)
   (2, 17.25363366666667)
   (3, 63.7235)
   (4, 69.60108500000001)
   (5, 102.1639)
   (6, 0)
   (7, 0)
   (8, 0)
   (9, 0)
};
\node[blue,above] at (axis cs: 1,0) {\rotatebox{90}{\tiny{OOM}}};
\node[blue,above] at (axis cs: 6,0) {\rotatebox{90}{\tiny{TO}}};
\node[blue,above] at (axis cs: 7,0) {\rotatebox{90}{\tiny{SHUF}}};
\node[blue,above] at (axis cs: 8,0) {\rotatebox{90}{\tiny{OOM}}};
\node[blue,above] at (axis cs: 9,0) {\rotatebox{90}{\tiny{TO}}};
\addplot+[ybar] [black, fill= cyan, postaction={nomorepostaction,pattern=north east lines}] coordinates{
   (1, 0)
   (2, 12567.882173333333)
   (3, 101020.71950000004)
   (4, 21209.75)
   (5, 33813.2)
   (6, 0)
   (7, 0)
   (8, 0)
   (9, 0)
};
\addplot+[ybar] [black, fill=red] coordinates{
   (1, 0)
   (2, 1.5573750000000002)
   (3, 29.171499999956694)
   (4, 10.71445)
   (5, 11.3605)
   (6, 0)
   (7, 0)
   (8, 0.0)
   (9, 0)
};
\addplot+[ybar] [black, fill=green, postaction={nomorepostaction,pattern=north west lines}] coordinates{
   (1, 0)
   (2, 6.973484666666667)
   (3, 16.7044999999965)
   (4, 2.434465)
   (5, 4.2756)
   (6, 0)
   (7, 0)
   (8, 0)
   (9, 0)
};
      \nextgroupplot[ title = {128 machines}, 
      ymax=85200] 
\addplot+[ybar] [black, fill=gray] coordinates{
   (1, 0)
   (2, 18.5525605)
   (3, 0)
   (4, 0)
   (5, 39.963080000000005)
   (6, 0)
   (7, 0)
   (8, 0)
   (9, 0)
};
\node[blue,above] at (axis cs: 1,0) {\rotatebox{90}{\tiny{OOM}}};
\node[blue,above] at (axis cs: 6,0) {\rotatebox{90}{\tiny{TO}}};
\node[blue,above] at (axis cs: 7,0) {\rotatebox{90}{\tiny{SHUF}}};
\node[blue,above] at (axis cs: 8,0) {\rotatebox{90}{\tiny{OOM}}};

\addplot+[ybar] [black, fill= cyan, postaction={nomorepostaction,pattern=north east lines}] coordinates{
   (1, 0)
   (2, 18204.9472455)
   (3, 0)
   (4, 0)
   (5, 52267.7)
   (6, 0)
   (7, 0)
   (8, 0)
   (9, 0)
};
\addplot+[ybar] [black, fill=red] coordinates{
   (1, 0)
   (2, 0.8262860000000001)
   (3, 0)
   (4, 0)
   (5, 7.14988)
   (6, 0)
   (7, 0)
   (8, 0)
   (9, 0)
};
\addplot+[ybar] [black, fill=green, postaction={nomorepostaction,pattern=north west lines}] coordinates{
   (1, 0)
   (2, 6.673908)
   (3, 0)
   (4, 0)
   (5, 6.18704)
   (6, 0)
   (7, 0)
   (8, 0)
   (9, 0)
};
      \nextgroupplot[ 
      ymax=2500, ylabel={UK0705}] 
\addplot+[ybar] [black, fill=gray] coordinates{
   (1, 0)
   (2, 59.31099233333333)
   (3, 0)
   (4, 158.48593333333335)
   (5, 0)
   (6, 29213.0)
   (7, 16224.0)
   (8, 244.343)
   (9, 367.5)
};
\node[blue,above] at (axis cs: 1,0) {\rotatebox{90}{\tiny{OOM}}};
\node[blue,above] at (axis cs: 3,0) {\rotatebox{90}{\tiny{OOM}}};
\node[blue,above] at (axis cs: 5,0) {\rotatebox{90}{\tiny{OOM}}};

\addplot+[ybar] [black, fill= cyan, postaction={nomorepostaction,pattern=north east lines}] coordinates{
   (1, 0)
   (2, 613.3280663333334)
   (3, 0)
   (4, 324.0753333333333)
   (5, 0)
   (6, 19238.0)
   (7, 9274.0)
   (8, 866.3333333333334)
   (9, 22322.5)
};
\addplot+[ybar] [black, fill=red] coordinates{
   (1, 0)
   (2, 1.116326)
   (3, 0)
   (4, 6.654410666666652)
   (5, 0)
   (6, 0.0)
   (7, 0.0)
   (8, 10.105969)
   (9, 4.0)
};
\addplot+[ybar] [black, fill=green, postaction={nomorepostaction,pattern=north west lines}] coordinates{
   (1, 0)
   (2, 4.577948666666667)
   (3, 0)
   (4, 3.4509893333333337)
   (5, 0)
   (6, 0.0)
   (7, 1.0)
   (8, 126.21769766666667)
   (9, 0.0)
};
      \nextgroupplot[ 
      ymax=2500] 
\addplot+[ybar] [black, fill=gray] coordinates{
   (1, 48.627336400000004)
   (2, 34.187789)
   (3, 0)
   (4, 153.82656666666665)
   (5, 156.7987)
   (6, 23762.0)
   (7, 12297.0)
   (8, 154.55933333333334)
   (9, 227.0)
};
\node[blue,above] at (axis cs: 3,0) {\rotatebox{90}{\tiny{OOM}}};

\addplot+[ybar] [black, fill= cyan, postaction={nomorepostaction,pattern=north east lines}] coordinates{
   (1, 0.9075414)
   (2, 305.6723406666667)
   (3, 0)
   (4, 109.106)
   (5, 387.93399999999997)
   (6, 14138.0)
   (7, 7871.0)
   (8, 731.3333333333334)
   (9, 11508.5)
};
\addplot+[ybar] [black, fill=red] coordinates{
   (1, 1.3862074)
   (2, 0.6478433333333333)
   (3, 0)
   (4, 3.7400489999999937)
   (5, 3.9501100000000005)
   (6, 0.0)
   (7, 0.0)
   (8, 13.322171333333332)
   (9, 2.25)
};
\addplot+[ybar] [black, fill=green, postaction={nomorepostaction,pattern=north west lines}] coordinates{
   (1, 5.0789148)
   (2, 3.1586936666666667)
   (3, 0)
   (4, 2.9940510000000002)
   (5, 3.81719)
   (6, 0.0)
   (7, 1.0)
   (8, 78.785162)
   (9, 0.0)
};
      \nextgroupplot[ 
      ymax=2500] 
\addplot+[ybar] [black, fill=gray] coordinates{
   (1, 36.08626383333334)
   (2, 19.765246333333334)
   (3, 19.478)
   (4, 55.4957)
   (5, 97.01243333333332)
   (6, 19086.0)
   (7, 0)
   (8, 114.3075)
   (9, 132.5)
};
\node[blue,above] at (axis cs: 7,0) {\rotatebox{90}{\tiny{SHUF}}};

\addplot+[ybar] [black, fill= cyan, postaction={nomorepostaction,pattern=north east lines}] coordinates{
   (1, 2.821915166666667)
   (2, 155.28702166666665)
   (3, 325.60124999999994)
   (4, 126.423)
   (5, 269.40633333333335)
   (6, 11192.0)
   (7, 0)
   (8, 1141.5)
   (9, 3993.5)
};
\addplot+[ybar] [black, fill=red] coordinates{
   (1, 1.4476603333333333)
   (2, 0.42414)
   (3, 10.565750000000032)
   (4, 3.10364)
   (5, 2.766363333333333)
   (6, 0.0)
   (7, 0)
   (8, 27.459024499999998)
   (9, 3.0)
};
\addplot+[ybar] [black, fill=green, postaction={nomorepostaction,pattern=north west lines}] coordinates{
   (1, 4.144160666666667)
   (2, 3.5235920000000003)
   (3, 26.273750000000003)
   (4, 2.6443266666666667)
   (5, 2.1482033333333335)
   (6, 0.0)
   (7, 0)
   (8, 65.2334755)
   (9, 0.0)
};
      \nextgroupplot[ 
      ymax=2500] 
\addplot+[ybar] [black, fill=gray] coordinates{
   (1, 35.409181000000004)
   (2, 13.286182666666667)
   (3, 14.212666666666665)
   (4, 113.16516666666666)
   (5, 54.79320666666666)
   (6, 15236.0)
   (7, 0)
   (8, 139.899)
   (9, 92.5)
};
\node[blue,above] at (axis cs: 7,0) {\rotatebox{90}{\tiny{SHUF}}};
\addplot+[ybar] [black, fill= cyan, postaction={nomorepostaction,pattern=north east lines}] coordinates{
   (1, 6.701697666666667)
   (2, 94.17889733333334)
   (3, 158.72066666666663)
   (4, 96.15926666666667)
   (5, 204.07466666666667)
   (6, 9258.0)
   (7, 0)
   (8, 3452.0)
   (9, 865.5)
};
\addplot+[ybar] [black, fill=red] coordinates{
   (1, 2.391265666666667)
   (2, 0.4080626666666667)
   (3, 9.838666666666695)
   (4, 2.3223333333333334)
   (5, 2.5233266666666663)
   (6, 0.0)
   (7, 0)
   (8, 113.471982)
   (9, 2.0)
};
\addplot+[ybar] [black, fill=green, postaction={nomorepostaction,pattern=north west lines}] coordinates{
   (1, 4.164522333333333)
   (2, 4.793524000000001)
   (3, 35.56133333333333)
   (4, 4.686566666666667)
   (5, 3.942133333333333)
   (6, 0.0)
   (7, 0)
   (8, 74.629018)
   (9, 0.0)
};
      \nextgroupplot[ 
      ymax=1000, ylabel={Twitter}] 
\addplot+[ybar] [black, fill=gray] coordinates{
   (1, 98.704912)
   (2, 36.403749)
   (3, 78.1285)
   (4, 81.7675)
   (5, 0)
   (6, 12432.0)
   (7, 8974.0)
   (8, 101.24433333333332)
   (9, 177.0)
};
\node[blue,above] at (axis cs: 5,0) {\rotatebox{90}{\tiny{OOM}}};
\addplot+[ybar] [black, fill= cyan, postaction={nomorepostaction,pattern=north east lines}] coordinates{
   (1, 1.115131)
   (2, 197.90251933333334)
   (3, 318.38525000000004)
   (4, 158.341)
   (5, 0)
   (6, 8027.0)
   (7, 5416.0)
   (8, 308.3333333333333)
   (9, 5750.25)
};
\addplot+[ybar] [black, fill=red] coordinates{
   (1, 3.281586)
   (2, 0.307373)
   (3, 10.481999999999946)
   (4, 3.1398460000000057)
   (5, 0)
   (6, 0.0)
   (7, 0.0)
   (8, 2.5688973333333336)
   (9, 2.25)
};
\addplot+[ybar] [black, fill=green, postaction={nomorepostaction,pattern=north west lines}] coordinates{
   (1, 30.898371)
   (2, 2.7196919999999998)
   (3, 0.275)
   (4, 4.751654)
   (5, 0)
   (6, 1.0)
   (7, 0.0)
   (8, 147.18676933333333)
   (9, 0.0)
};
      \nextgroupplot[  
      ymax=1000] 
\addplot+[ybar] [black, fill=gray] coordinates{
   (1, 138.21549875)
   (2, 11.745129333333333)
   (3, 39.413000000000004)
   (4, 105.08064999999999)
   (5, 64.463775)
   (6, 6831)
   (7, 5982)
   (8, 80.71266666666666)
   (9, 102.33333333333333)
};
\addplot+[ybar] [black, fill= cyan, postaction={nomorepostaction,pattern=north east lines}] coordinates{
   (1, 0.8446334999999999)
   (2, 96.001414)
   (3, 153.64075000000003)
   (4, 101.2945)
   (5, 123.28550000000001)
   (6, 4318.0 )
   (7, 3195.0 )
   (8, 157.0)
   (9, 1518.0)
};
\addplot+[ybar] [black, fill=red] coordinates{
   (1, 3.63659175)
   (2, 0.33340400000000003)
   (3, 14.863999999999987)
   (4, 1.952395)
   (5, 2.157745)
   (6, 0)
   (7, 0)
   (8, 1.9724993333333334)
   (9, 2.3333333333333335)
};
\addplot+[ybar] [black, fill=green, postaction={nomorepostaction,pattern=north west lines}] coordinates{
   (1, 31.803276)
   (2, 3.920052666666667)
   (3, 26.907999999999998)
   (4, 2.1724550000000002)
   (5, 2.34298)
   (6, 0.0)
   (7, 0.0)
   (8, 101.314834)
   (9, 0.0)
};
      \nextgroupplot[  
      ymax=1000] 
\addplot+[ybar] [black, fill=gray] coordinates{
   (1, 96.63973519999999)
   (2, 8.389617666666666)
   (3, 43.301)
   (4, 27.708516666666668)
   (5, 39.024319999999996)
   (6, 5712)
   (7, 0)
   (8, 85.893)
   (9, 62.333333333333336)
};
\node[blue,above] at (axis cs: 7,0) {\rotatebox{90}{\tiny{SHUF}}};
\addplot+[ybar] [black, fill= cyan, postaction={nomorepostaction,pattern=north east lines}] coordinates{
   (1, 2.9366894)
   (2, 44.515558)
   (3, 112.25450000000001)
   (4, 49.9972)
   (5, 76.2194)
   (6, 3194.0 )
   (7, 0)
   (8, 145.0)
   (9, 456.0)
};
\addplot+[ybar] [black, fill=red] coordinates{
   (1, 3.1212565999999997)
   (2, 0.37105000000000005)
   (3, 10.258749999999997)
   (4, 1.4668033333333332)
   (5, 1.5944933333333333)
   (6, 0)
   (7, 0)
   (8, 3.048936)
   (9, 1.3333333333333333)
};
\addplot+[ybar] [black, fill=green, postaction={nomorepostaction,pattern=north west lines}] coordinates{
   (1, 27.502318799999994)
   (2, 4.057107666666667)
   (3, 28.999999999999996)
   (4, 2.160813333333333)
   (5, 2.49512)
   (6, 0.0)
   (7, 0)
   (8, 108.558064)
   (9, 0.0)
};
      \nextgroupplot[  
      ymax=1000] 
\addplot+[ybar] [black, fill=gray] coordinates{
   (1, 114.51784300000001)
   (2, 15.585055333333335)
   (3, 38.55166666666667)
   (4, 57.6684)
   (5, 23.722896666666667)
   (6, 4298)
   (7, 0)
   (8, 133.85333333333332)
   (9, 47.5)
};
\node[blue,above] at (axis cs: 7,0) {\rotatebox{90}{\tiny{SHUF}}};

\addplot+[ybar] [black, fill= cyan, postaction={nomorepostaction,pattern=north east lines}] coordinates{
   (1, 6.232482666666667)
   (2, 26.469053666666667)
   (3, 70.373)
   (4, 42.763000000000005)
   (5, 48.44906666666666)
   (6, 2219.0 )
   (7, 0)
   (8, 97.0)
   (9, 277.5)
};
\addplot+[ybar] [black, fill=red] coordinates{
   (1, 4.541096666666666)
   (2, 0.35387833333333335)
   (3, 9.294333333333332)
   (4, 2.8388733333333334)
   (5, 2.0901300000000003)
   (6, 0)
   (7, 0)
   (8, 2.8669473333333335)
   (9, 1.25)
};
\addplot+[ybar] [black, fill=green, postaction={nomorepostaction,pattern=north west lines}] coordinates{
   (1, 30.041911)
   (2, 4.925346)
   (3, 37.781)
   (4, 3.3963933333333336)
   (5, 3.737906666666667)
   (6, 0.0)
   (7, 0)
   (8, 51.61305266666667)
   (9, 0.0)
};
\makeatletter
\end{groupplot}
\node at ($(group c2r1) + (0,2.1cm)$) {\ref{grouplegend}};
\end{tikzpicture}

%% file: related.tex
\section{Related Work}
\label{relatedStudies}

Many studies of graph processing systems focus on Pregel-like systems \cite{han2014experimental, Sakr-GraphExperiments, experiments}. LDBC~\cite{LDBC} includes industry-driven systems such as GraphPad~\cite{graphpad} from Intel and PGX~\cite{PGXD} from Oracle. In contrast, our choice of systems under study follows a systematic classification (Section~\ref{Sec-exp:intro}). 

LDBC  is the only study that uses vertical/horizontal and strong/weak scalability. Our study does not include vertical scalability experiments for reasons discussed in Section~\ref{scalability}.

All studies consider power-law real and synthetic datasets. Many systems use graph data generators, such as DataGen~\cite{DataGen}, WGB~\cite{wgb-short}, RTG~\cite{RTG}, LUBM~\cite{LUBM}, and gMark~\cite{gMark} to create large datasets. The main objective of these graph generators is to represent real graphs and allow  the creation of graph datasets with different sizes. In this this study, we only use real graph datasets of varying sizes, some of which are larger than any other dataset in existing studies. We also include more diverse datasets (road network, social network, and web pages) than  others study. 

The following  highlight the major findings of previous studies and where our results differ:
	\begin{itemize}
		\item[~\cite{experiments, han2014experimental}] Giraph and GraphLab have similar performance when both use random partitioning. However, GraphLab has an auto mode that allows it to beat Giraph in certain cluster sizes.
		
		\item[~\cite{Blogel}] It is not fair to say that existing implementation of block-centric computation is faster than vertex-centric. The execution time is faster in block-centric, but the overall response time is slower due to overheads discussed in Section~\ref{BlogelDiscussion}. 
		\item[~\cite{Sakr-GraphExperiments}] It was previously reported that GraphX has the best performance across all systems~\cite{Sakr-GraphExperiments}. Our results contradicts this assertion. GraphX paper~\cite{GraphX} never claimed that it was more efficient than other systems, even though a special version of GraphX was used that includes multiple optimizations that are not yet available in the most recent Spark release\footnote{This was confirmed by GraphX/Spark team: \url{
				https://tinyurl.com/y9246wpp
				%http://apache-spark-user-list.1001560.n3.nabble.com/Is-Spark-or-GraphX-runs-fast-a-performance-comparison-on-Page-Rank-tp19710p19956.html
			}.}. Furthermore, the version of Spark (v 1.3.0) used in the study making this claim~\cite{Sakr-GraphExperiments} does not compute PageRank values accurately in all cases. We suspect this causes fast convergence because we had similar experience in our earlier results. 
		\item[~\cite{Vertica4Graph}] Vertica is not competitive to existing parallel graph processing systems. Section~\ref{sec-vertica} has further details about Vertica performance. In a nutshell, Vertica I/O and network overhead is significantly larger than graph systems. This overhead increases as the cluster size increases.
	\end{itemize}

Finally, a recent study~\cite{sahu2017ubiquity} reports a user survey to explain how graphs are used in real life. The paper aims to understand types of graphs, computations, and software users need when processing their graphs. The focus is different and perhaps complementary to our paper.

%% file: conclusions.tex
\section{Conclusion}
\label{Sec-exp:conclusion}

In this paper we present results from extensive experiments on eight distributed graph processing systems (Hadoop, HaLoop, Vertica, Giraph, GraphLab, GraphX, Flink Gelly, and Blogel) across four workloads (PageRank, WCC, SPSS, K-hop) over four very large datasets (Twitter, World Road Network, UK 200705, and ClueWeb). We focused on scale-out performance. Our experiments evaluate a wider set of systems that follow more varied  computation models, over a larger and more diverse datasets than previous studies \cite{Sakr-GraphExperiments, han2014experimental,  experiments}. Ours is the most extensive independent study of graph processing systems to date. Our results indicate that the best system varies according to workload and particular data graph that is used. In some cases our study confirms previously reported results, in other cases the results are different and the reasons for the divergence are explained.

%% file: arxiv-version.bbl
\begin{thebibliography}{10}

\bibitem{Flink}
Flink.
\newblock https://flink.apache.org/.

\bibitem{Gelly}
Gelly: Flink graph api.
\newblock https://ci.apache.org/projects/flink/flink-docs-stable/.

\bibitem{Giraph}
Giraph.
\newblock http://giraph.apache.org.

\bibitem{Hadoop}
Hadoop.
\newblock http://hadoop.apache.org.

\bibitem{Timely}
Timely data flow.
\newblock https://github.com/frankmcsherry/timely-dataflow.

\bibitem{BlogelCommunication}
Private correspondence with {B}logel team., 2015.

\bibitem{TensorFlow}
M.~Abadi, A.~Agarwal, P.~Barham, E.~Brevdo, Z.~Chen, C.~Citro, G.~S. Corrado,
  A.~Davis, J.~Dean, M.~Devin, et~al.
\newblock Tensorflow: Large-scale machine learning on heterogeneous distributed
  systems.
\newblock {\em arXiv preprint arXiv:1603.04467}, 2016.

\bibitem{RTG}
L.~Akoglu and C.~Faloutsos.
\newblock Rtg: a recursive realistic graph generator using random typing.
\newblock {\em Data Mining and Knowledge Discovery}, 19(2):194--209, 2009.

\bibitem{wgb-short}
K.~Ammar and M.~{\"O}zsu.
\newblock {WGB}: Towards a universal graph benchmark.
\newblock In e.~a. Rabl, Tilmann, editor, {\em Advancing Big Data Benchmarks},
  Lecture Notes in Computer Science, pages 58--72. Springer, 2014.

\bibitem{graphpad}
M.~J. Anderson, N.~Sundaram, N.~Satish, M.~M.~A. Patwary, T.~L. Willke, and
  P.~Dubey.
\newblock Graphpad: Optimized graph primitives for parallel and distributed
  platforms.
\newblock In {\em Proc. 30th Int. Parallel \& Distributed Processing Symp.},
  pages 313--322, 2016.

\bibitem{bader2005architectural}
D.~A. Bader, G.~Cong, and J.~Feo.
\newblock On the architectural requirements for efficient execution of graph
  algorithms.
\newblock In {\em Proc. of Parallel Processing}, pages 547--556, 2005.

\bibitem{gMark}
G.~Bagan, A.~Bonifati, R.~Ciucanu, G.~H. Fletcher, A.~Lemay, and N.~Advokaat.
\newblock gmark: schema-driven generation of graphs and queries.
\newblock {\em IEEE transactions on knowledge and data engineering},
  29(4):856--869, 2017.

\bibitem{Sakr-GraphExperiments}
O.~Batarfi, R.~Shawi, A.~Fayoumi, R.~Nouri, S.-M.-R. Beheshti, A.~Barnawi, and
  S.~Sakr.
\newblock Large scale graph processing systems: survey and an experimental
  evaluation.
\newblock {\em Cluster Computing}, 18(3):1189--1213, 2015.

\bibitem{Direction-optimization-BFS}
S.~Beamer, K.~Asanovi{\'c}, and D.~Patterson.
\newblock Direction-optimizing breadth-first search.
\newblock In {\em International Conference on High Performance Computing,
  Networking, Storage and Analysis}, volume~21, pages 12:1--12:10, 2013.

\bibitem{GAP-benchmark}
S.~Beamer, K.~Asanovi{\'c}, and D.~Patterson.
\newblock The gap benchmark suite.
\newblock {\em arXiv preprint arXiv:1508.03619}, 2015.

\bibitem{Haloop}
Y.~Bu, B.~Howe, M.~Balazinska, and M.~D. Ernst.
\newblock The {HaLoop} approach to large-scale iterative data analysis.
\newblock {\em VLDB J.}, 21(2):169--190, 2012.

\bibitem{ChapterLawsGenerators}
D.~Chakrabarti, C.~Faloutsos, and M.~McGlohon.
\newblock Graph {M}ining: Laws and {G}enerators.
\newblock In {\em Proc. Managing and Mining Graph Data}, pages 69--123, 2010.

\bibitem{FB_trillionGraph}
A.~Ching, S.~Edunov, M.~Kabiljo, D.~Logothetis, and S.~Muthukrishnan.
\newblock One trillion edges: Graph processing at facebook-scale.
\newblock {\em Proc. VLDB Endowment}, 8(12):1804--1815, 2015.

\bibitem{Dean2004}
J.~Dean and S.~Ghemawat.
\newblock {MapReduce}: Simplified data processing on large clusters.
\newblock In {\em Proc. 6th USENIX Symp. on Operating System Design and
  Implementation}, pages 137--149, 2004.

\bibitem{DataGen}
O.~Erling, A.~Averbuch, J.~Larriba-Pey, H.~Chafi, A.~Gubichev, A.~Prat, M.-D.
  Pham, and P.~Boncz.
\newblock The ldbc social network benchmark: Interactive workload.
\newblock In {\em Proc. ACM SIGMOD Int. Conf. on Management of Data}, pages
  619--630, 2015.

\bibitem{GVD}
M.~Erwig and F.~Hagen.
\newblock The graph voronoi diagram with applications.
\newblock {\em Networks}, 36:156--163, 2000.

\bibitem{againstGraphProcessing}
J.~Fan, A.~G.~S. Raj, and J.~M. Patel.
\newblock The case against specialized graph analytics engines.
\newblock In {\em Proc. 7th Biennial Conference on Innovative Data Systems
  Research}, pages 1--10, 2015.

\bibitem{powerGraph}
J.~E. Gonzalez, Y.~Low, H.~Gu, D.~Bickson, and C.~Guestrin.
\newblock Powergraph: Distributed graph-parallel computation on natural graphs.
\newblock In {\em Proc. 10th USENIX Symp. on Operating System Design and
  Implementation}, pages 17--30, 2012.

\bibitem{GraphX}
J.~E. Gonzalez, R.~S. Xin, A.~Dave, D.~Crankshaw, M.~J. Franklin, and
  I.~Stoica.
\newblock Graphx: Graph processing in a distributed dataflow framework.
\newblock In {\em Proc. 11th USENIX Symp. on Operating System Design and
  Implementation}, pages 599--613, 2014.

\bibitem{LUBM}
Y.~Guo, Z.~Pan, and J.~Heflin.
\newblock Lubm: A benchmark for owl knowledge base systems.
\newblock {\em Web Semantics: Science, Services and Agents on the World Wide
  Web}, 3(2-3):158--182, 2005.

\bibitem{han2014experimental}
M.~Han, K.~Daudjee, K.~Ammar, M.~T. \"{O}zsu, X.~Wang, and T.~Jin.
\newblock An experimental comparison of pregel-like graph processing systems.
\newblock {\em Proc. VLDB Endowment}, 7(12):1047--1058, 2014.

\bibitem{PGXD}
S.~Hong, S.~Depner, T.~Manhardt, J.~V.~D. Lugt, M.~Verstraaten, and H.~Chafi.
\newblock Pgx.d: a fast distributed graph processing engine.
\newblock In {\em Proc. of Int. Conf. for High Performance Computing,
  Networking, Storage and Analysis}, pages 1--12, 2015.

\bibitem{LDBC}
A.~Iosup, T.~Hegeman, W.~L. Ngai, S.~Heldens, A.~Prat-P{\'e}rez, T.~Manhardto,
  H.~Chafio, M.~Capot{\u{a}}, N.~Sundaram, M.~Anderson, et~al.
\newblock {LDBC} graphalytics: A benchmark for large-scale graph analysis on
  parallel and distributed platforms.
\newblock {\em Proc. VLDB Endowment}, 9(13):1317--1328, 2016.

\bibitem{Vertica4Graph}
A.~Jindal, S.~Madden, M.~Castellanos, and M.~Hsu.
\newblock Graph analytics using vertica relational database.
\newblock In {\em Proc. IEEE International Conference on Big Data}, pages
  1191--1200, 2015.

\bibitem{Pegasus}
U.~Kang, C.~E. Tsourakakis, and C.~Faloutsos.
\newblock {PEGASUS:} a peta-scale graph mining system implementation and
  observations.
\newblock In {\em Proc. 2009 IEEE Int. Conf. on Data Mining}, pages 229--238,
  2009.

\bibitem{MRcc2014}
R.~Kiveris, S.~Lattanzi, V.~Mirrokni, V.~Rastogi, and S.~Vassilvitskii.
\newblock Connected components in mapreduce and beyond.
\newblock In {\em Proc. 5nd ACM Symp. on Cloud Computing}, pages 18:1--18:13,
  2014.

\bibitem{kothapalli2010fast}
K.~Kothapalli, J.~Soman, and P.~Narayanan.
\newblock Fast {GPU} algorithms for graph connectivity.
\newblock In {\em Proc. Workshop on Large Scale Parallel Processing}, pages
  66--75, 2010.

\bibitem{DynamicGraphProperties}
J.~Leskovec, J.~Kleinberg, and C.~Faloutsos.
\newblock Graphs over time: densification laws, shrinking diameters and
  possible explanations.
\newblock In {\em Proc. 11th ACM SIGKDD Int. Conf. on Knowledge Discovery and
  Data Mining}, pages 177--187, 2005.

\bibitem{DynamicGraphsProperties}
J.~Leskovec, J.~Kleinberg, and C.~Faloutsos.
\newblock Graphs over time: Densification laws, shrinking diameters and
  possible explanations.
\newblock In {\em Proc. 11th ACM SIGKDD Int. Conf. on Knowledge Discovery and
  Data Mining}, pages 177--187, 2005.

\bibitem{GraphLab}
Y.~Low, J.~Gonzalez, A.~Kyrola, D.~Bickson, C.~Guestrin, and J.~M. Hellerstein.
\newblock Distributed {G}raph{L}ab: A framework for machine learning in the
  cloud.
\newblock {\em Proc. VLDB Endowment}, 5(8):716--727, 2012.

\bibitem{experiments}
Y.~Lu, J.~Cheng, D.~Yan, and H.~Wu.
\newblock Large-scale distributed graph computing systems: An experimental
  evaluation.
\newblock {\em Proc. VLDB Endowment}, 8(3):281--292, 2014.

\bibitem{pregel}
G.~Malewicz, M.~H. Austern, A.~J. Bik, J.~C. Dehnert, I.~Horn, N.~Leiser, and
  G.~Czajkowski.
\newblock Pregel: a system for large-scale graph processing.
\newblock In {\em Proc. ACM SIGMOD Int. Conf. on Management of Data}, pages
  135--146, 2010.

\bibitem{COST}
F.~McSherry, M.~Isard, and D.~G. Murray.
\newblock Scalability! but at what cost?
\newblock In {\em Proc of the 15th {USENIX} Conference on Hot Topics in
  Operating Systems}, 2015.

\bibitem{DifferentialDataFlow}
F.~McSherry, D.~G. Murray, R.~Isaacs, and M.~Isard.
\newblock Differential dataflow.
\newblock In {\em Proc. 6th Biennial Conference on Innovative Data Systems
  Research}, 2013.

\bibitem{naiad}
D.~G. Murray, F.~McSherry, R.~Isaacs, M.~Isard, P.~Barham, and M.~Abadi.
\newblock Naiad: A timely dataflow system.
\newblock In {\em Proc. 24th ACM Symp. on Operating System Principles}, pages
  439--455, 2013.

\bibitem{sahu2017ubiquity}
S.~Sahu, A.~Mhedhbi, S.~Salihoglu, J.~Lin, and M.~T. {\"O}zsu.
\newblock The ubiquity of large graphs and surprising challenges of graph
  processing.
\newblock {\em Proc. VLDB Endowment}, 11(4), 2017.

\bibitem{GPS}
S.~Salihoglu and J.~Widom.
\newblock {GPS}: A graph processing system.
\newblock In {\em Proc. 25th Int. Conf. on Scientific and Statistical Database
  Management}, pages 1--12, 2013.

\bibitem{shiloach1982logn}
Y.~Shiloach and U.~Vishkin.
\newblock An o (logn) parallel connectivity algorithm.
\newblock {\em Journal of Algorithms}, 3(1):57--67, 1982.

\bibitem{Giraph++}
Y.~Tian, A.~Balmin, S.~A. Corsten, S.~Tatikonda, and J.~McPherson.
\newblock From ``think like a vertex" to ``think like a graph".
\newblock {\em Proc. VLDB Endowment}, 7(3):193--204, 2013.

\bibitem{PartitioningStudy}
S.~Verma, L.~M. Leslie, Y.~Shin, and I.~Gupta.
\newblock An experimental comparison of partitioning strategies in distributed
  graph processing.
\newblock {\em Proc. VLDB Endowment}, 10(5):493--504, Jan. 2017.

\bibitem{Blogel}
D.~Yan, J.~Cheng, Y.~Lu, and W.~Ng.
\newblock Blogel: A block-centric framework for distributed computation on
  real-world graphs.
\newblock {\em Proc. VLDB Endowment}, 7(14):1981--1992, 2014.

\bibitem{Spark}
M.~Zaharia, M.~Chowdhury, T.~Das, A.~Dave, J.~Ma, M.~McCauly, M.~J. Franklin,
  S.~Shenker, and I.~Stoica.
\newblock Resilient distributed datasets: {A} fault-tolerant abstraction for
  in-memory cluster computing.
\newblock In {\em Proc. 9th {USENIX} Symp. on Networked Systems Design and
  Implementation}, pages 15--28, 2012.

\end{thebibliography}
